\documentclass{aastex63}
\def\actaa{Acta Astronomica}

\usepackage{lineno}
\hypersetup{linkcolor=red,citecolor=blue,filecolor=green,urlcolor=magenta}

\begin{document}

\shorttitle{PL and PLC Relations for Contact Binaries}
\shortauthors{Ngeow et al.}

\title{Zwicky Transient Facility and Globular Clusters: the Period-Luminosity and Period-Luminosity-Color Relations for Late-Type Contact Binaries}

\correspondingauthor{C.-C. Ngeow}
\email{cngeow@astro.ncu.edu.tw}

\author[0000-0001-8771-7554]{Chow-Choong Ngeow}
%\author{Chow-Choong Ngeow}
\affil{Graduate Institute of Astronomy, National Central University, 300 Jhongda Road, 32001 Jhongli, Taiwan}

\author{Szu-Han Liao}
\affil{Department of Physics, National Central University, 300 Jhongda Road, 32001 Jhongli, Taiwan}

\author[0000-0001-8018-5348]{Eric C. Bellm}
%\author{Eric C. Bellm}
\affiliation{DIRAC Institute, Department of Astronomy, University of Washington, 3910 15th Avenue NE, Seattle, WA 98195, USA}

\author[0000-0001-5060-8733]{Dmitry A. Duev} 
%\author{Dmitry A. Duev} 
\affiliation{Division of Physics, Mathematics, and Astronomy, California Institute of Technology, Pasadena, CA 91125, USA}  

\author[0000-0002-3168-0139]{Matthew J. Graham}
%\author{Matthew J. Graham}
\affiliation{Division of Physics, Mathematics, and Astronomy, California Institute of Technology, Pasadena, CA 91125, USA}

\author[0000-0003-2242-0244]{Ashish~A.~Mahabal}
%\author{Ashish A. Mahabal}
\affiliation{Division of Physics, Mathematics and Astronomy, California Institute of Technology, Pasadena, CA 91125, USA}
\affiliation{Center for Data Driven Discovery, California Institute of Technology, Pasadena, CA 91125, USA}

\author[0000-0002-8532-9395]{Frank J. Masci}
%\author{Frank J. Masci}
\affiliation{IPAC, California Institute of Technology, 1200 E. California Blvd, Pasadena, CA 91125, USA}

\author[0000-0002-7226-0659]{Michael S. Medford}
%\author{Michael S. Medford}
\affiliation{University of California, Berkeley, Department of Astronomy, Berkeley, CA 94720}
\affiliation{Lawrence Berkeley National Laboratory, 1 Cyclotron Rd., Berkeley, CA 94720}

\author[0000-0002-0387-370X]{Reed Riddle}
%\author{Reed Riddle}
\affiliation{Caltech Optical Observatories, California Institute of Technology, Pasadena, CA 91125, USA} 

\author[0000-0001-7648-4142]{Ben Rusholme}
%\author{Ben Rusholme}
\affiliation{IPAC, California Institute of Technology, 1200 E. California Blvd, Pasadena, CA 91125, USA}

\begin{abstract}

  In this work, we aimed to derive the $gri$-band period-luminosity (PL) and period-luminosity-color (PLC) relations for late-type contact binaries, for the first time, located in the globular clusters, using the homogeneous light curves collected by the Zwicky Transient Factory (ZTF). We started with 79 contact binaries in 15 globular clusters, and retained 30 contact binaries in 10 globular clusters that have adequate number of data points in the ZTF light curves and unaffected by blending. Magnitudes at mean and maximum light of these contact binaries were determined using a fourth-order Fourier expansion, while extinction corrections were done using the {\tt Bayerstar2019} 3D reddening map together with adopting the homogeneous distances to their host globular clusters. After removing early-type and ``anomaly'' contact binaries, our derived $gri$-band PL and period-Wesenheit (PW) relations exhibit a much larger dispersion with large errors on the fitted coefficients. Nevertheless, the $gr$-band PL and PW relations based on this small sample of contact binaries in globular clusters were consistent with those based on a larger sample of nearby contact binaries. Good agreements of the PL and PW relations suggested both samples of contact binaries in the local Solar neighborhood and in the distant globular clusters can be combined and used to derive and calibrate the PL, PW and PLC relations. The final derived $gr$-band PL, PW and PLC relations were much improved than those based on the limited sample of contact binaries in the globular clusters. 

\end{abstract}

%\keywords{ {\bf TBD}}

\section{Introduction}

Contact binaries (CB) are binary systems with two stars overflow their Roche lobes, majority of them have orbital period less than a day. CB include more massive early-type systems and less massive late-type systems -- commonly known as W UMa (W Ursae Majoris) type binaries. In this work, we used the term CB to refer for both types, although majority of the CB mentioned in the cited literature are belong to W UMa system. For reviews on CB, see (for example) \citet{smith1984}, \citet{webbink2003} and \citet{pacz2006}. In analog to classical pulsating stars (e.g., Cepheids and RR Lyrae), CB also exhibit a period-luminosity (PL) relation or a period-luminosity-color (PLC) relation, albeit the physics behind their PL and PLC relations is different than the classical pulsating stars. Brief theoretical framework on the PL and PLC relations for CB can be found, for example, in \citet{rucinski1994,rucinski1996,rucinski2004} and \citet{chen2016}, and will not be repeated here. \citet{chen2016} and more recent work from \citet{chen2018} demonstrated the PL relations for CB could be used as distance indicators at a level that is competitive to the classical standard candles such as Cepheids.

Some earlier work on deriving the PL or PLC relations for CB can be found in \citet{rucinski1974} and \citet[][equation 22 in the paper]{mochnacki1981}. Since then, a number of papers have been published by Rucinski and co-workers to calibrate the PL and PLC relations for CB in the $BVI$-bands using parallax measurements, CB in open clusters and a few CB with visual companions \citep[see][]{rucinski1994,rucinski97a,rucinski1997,rucinski1997b,rucinski2006,mateo2017}. Independently, \citet{eker2009} improved the calibration of optical PLC relation using 31 systems with most accurate {\it Hipparcos} parallaxes by applying the Lutz-Kelker corrections \citep{lk1973} and extending the PLC calibration to the near infrared $JHK$ bands. The $K$-band PL relation was also presented in \citet{muraveva2014}. Later, \citet{chen2016} derived the $JHK$-band PL relations using 66 CB in open clusters (42 systems), moving groups (4 systems) and those with accurate parallaxes from {\it Hipparcos} (20 systems), as well as the $V$-band PL relation based on a much larger sample. In \citet{chen2018}, twelve bands PL relations spanning from optical to infrared were derived using a sample of $\sim 183$ CB with parallax measurements based on {\it Tycho-Gaia} astrometric solution (as part of the Gaia Data Release 1). The calibrated PL or PLC relations presented in these work were mainly based on the late-type W UMa systems. In case of the early-type CB, the $J$-band PL relation  was derived in \citet{chen2016}, while the optical PLC relation was presented in \citet{pawlak2016} based on 64 early-type CB located in the Large Magellanic Cloud. Using a much larger sample of CB, \citet{jay2020} showed that early-type and late-type CB follow a different PL relation. Finally, \citet{sun2020} divided CB into three sub-classes and derived the mid-infrared $W1$-band PL relation for each of them.

Besides open clusters and parallax measurements, globular clusters provide a viable and yet independent route to calibrate their PL and PLC relations. CB in globular clusters shared the same distances at which independent and reliable distances of globular clusters from various methods can be found in literature. Majority of the globular clusters are located in high Galactic latitudes, therefore the effect of extinction is tended to be minimal. The CB in globular clusters also provides a consistency check of the PL and PLC relations extending to the old populations and low metallicity environment \citep[e.g. see][]{rucinski2000}. Since majority of the CB observed by LAMOST have sub-Solar metallicity \citep{qian2017}, the route of using globular clusters to calibrate the PL and PLC relations might be more appropriate. Even though not numerous, CB indeed could be found in a number of globular clusters. For instance, \citet{li2010} provided a list of globular clusters containing CB, the number of the CB in the 17 listed globular clusters ranged from 1 to 18 for a total of 230 systems. Other compilations of CB in globular clusters, in the format of $N_{GC}/N_{CB}$ (where $N_{GC}$ and $N_{CB}$ is the total number of globular clusters and CB, respectively), include \citet[][5/15]{hut1992}, \citet[][3/10]{rucinski1995}, \citet[][7/24]{mateo1996}, \citet[][14/68]{rucinski2000}\footnote{The number of CB should be 68 (as listed in Table 2 of the paper) instead of 86. Also, a fraction of the CB (about 1/3 to half) were foreground or background stars.}, and \citet[][5/11]{li2017}.

The main goal of our work is to derive the $gri$-band PL and PLC relations using CB found in the globular clusters. The optical PL or PLC relations derived in the literature mentioned earlier do not include the $gri$-band, this motivated us to derive such relations using the homogeneous photometric data obtained from the Zwicky Transient Facility \citep[ZTF,][]{bellm2017,bel19,gra19}. Our derived $gri$-band PL and/or PLC relations can then be applied to the large number of CB observed with the same or similar filters set, for example, those CB observed with the Dark Energy Camera \citep[DECam,][]{flaugher2015}, discovered in ZTF Data Release 2 \citep{chen2020}, and the faint CB to be discovered by the Vera C. Rubin Observatory's Legacy Survey of Space and Time \citep[LSST,][]{ivezic2019}. In Section \ref{sec2}, we described the selection of globular clusters and the CB within them, as well as the ZTF light curves data for the selected CB. Period refinements for these selected CB were given in Section \ref{sec3}. We then performed light curves fitting using Fourier expansion to derive their magnitudes at mean and maximum light in Section \ref{sec4}. Fitting of the PL and PLC relations will be done in Section \ref{sec5} and \ref{sec6}, respectively. An example on the application of the derived PL relation will be presented in Section \ref{sec7}, followed by conclusions in Section \ref{sec8}.

\section{Sample and Data}\label{sec2}

\subsection{Selection of Contact Binaries in Globular Clusters}

The samples of CB in globular clusters were obtained from \citet[][hereafter the Clement's Catalog]{clement2017}, an updated version of \citet{clement2001}. We first selected globular clusters that are observable with ZTF ($\delta_{J2000}>-30^\circ$), for those selected globular clusters we searched for variable stars classified as ``EW'' or ``EC'' in the Clement Catalog. Those EW or EC binaries marked as foreground or suspected foreground stars in the Clement Catalog were further removed. After applying the mentioned selection criteria, the preliminary list of CB included 45 EW and 25 EC in 15 globular clusters.

\begin{deluxetable*}{lcccccrrl}
  %\movetableright=-1in
  \tabletypesize{\scriptsize}
  \tablecaption{Basic properties of the globular clusters adopted in this work.\label{tab_gc}}
  \tablewidth{0pt}
  \tablehead{
    \colhead{Name} &
    \colhead{$\alpha_{J2000}$} &
    \colhead{$\delta_{J2000}$} &
    \colhead{$D$\tablenotemark{a}} &
    \colhead{$E(B-V)$\tablenotemark{b}} &
    \colhead{$[F\mathrm{e}/H]$\tablenotemark{b}} &
    \colhead{$N_{CB}$\tablenotemark{c}} &
    \colhead{$N_{CB}^{ZTF}$\tablenotemark{d}} &
    \colhead{Reference\tablenotemark{e}}
  }
  \startdata
  M4     & 16:23:35.22	& $-$26:31:32.7 & $1.97\pm0.04$ & 0.35 & $-1.16$ & 20 & 12 & 1,2,3,4,5 \\
  M9     & 17:19:11.26	& $-$18:30:57.4 & $8.40$        & 0.38 & $-1.77$ & 3  & 3  & 6,7 \\	
  M12    & 16:47:14.18	& $-$01:56:54.7 & $4.67\pm0.21$ & 0.19 & $-1.37$ & 3  & 3  & 8,9 \\
  M13    & 16:41:41.24	& +36:27:35.5   & $6.77\pm0.28$ & 0.02 & $-1.53$ & 2  & 1  & 10,11 \\	
  M15    & 21:29:58.33	& +12:10:01.2 & $10.22\pm0.13$& 0.10 & $-2.37$ & 2  & 2  & 12 \\
  M22    & 18:36:23.94	& $-$23:54:17.1 & $3.23\pm0.08$ & 0.34 & $-1.70$ & 17 & 9  & 13,14,15 \\
  M30    & 21:40:22.12	& $-$23:10:47.5 & $8.00\pm0.57$ & 0.03 & $-2.27$ & 4  & 2  & 16 \\ 
  M71    & 19:53:46.49	& +18:46:45.1 & $3.99\pm0.20$ & 0.25 & $-0.78$ & 9  & 9  & 17 \\
  NGC4147& 12:10:06.30	& +18:32:33.5 & $18.20$       & 0.02 & $-1.80$ & 5  & 5  & 18 \\
  NGC5466& 14:05:27.29	& +28:32:04.0 & $16.90$       & 0.00 & $-1.98$ & 2  & 1  & 19 \\	
  NGC6401& 17:38:36.60	& $-$23:54:34.2 & $7.70$        & 0.72 & $-1.02$ & 6  & 4  & 20 \\	
  NGC6712& 18:55:06.04	& $-$22:42:05.3 & $6.95\pm0.39$ & 0.22 & $-1.26$ & 2  & 2  & 21 \\
  \enddata
  \tablenotetext{a}{Distance $D$ (in kpc) adopted from \citet{baumgardt2019}.}
  \tablenotetext{b}{These values were adopted from the Milky Way Globular Clusters Catalog \citep[December 2010 version;][]{harris1996,harris2010}.}
  \tablenotetext{c}{The number of CB (see text for details).}
  \tablenotetext{d}{The number of CB that contain at least one ZTF data point in any filters.}
  \tablenotetext{e}{The references on CB are: (1) \citet{kaluzny1997}; (2) \citet{kaluzny2013}; (3) \citet{nascimbeni2014}; (4) \citet{stetson2014}; (5) \citet{safonova2016}; (6) \citet{clement1996}; (7) \citet{af2013}; (8) \citet{vb2002}; (9) \citet{kaluzny2015}; (10) \citet{pk2004}; (11) \citet{deras2019}; (12) \citet{jeon2001}; (13) \citet{kaluzny2001}; (14) \citet{pk2003}; (15) \citet{roz2017}; (16) \citet{pk2004}; (17) \citet{park2000}; (18) \citet{lata2019}; (19) \citet{mateo1990}; (20) \citet{soszynski2016}; (21) \citet{deras2020}.}
\end{deluxetable*}

This preliminary list of CB was updated based on new discoveries published in recent years. New CB that were confirmed to be members of globular clusters were added for M4 \citep[][1 new CB added]{safonova2016}, M10 \citep[][1 new CB added]{roz2018}, M12 \citep[][3 new CB added]{kaluzny2015}, M13 \citep[][1 new CB added]{deras2019}, M22 \citep[][6 new CB added]{roz2017}, NGC4147 \citep[][5 new CB added]{lata2019} and NGC6712 \citep[][2 new CB added]{deras2020}. Positions and periods for CB in M10, M22 and NGC6712 were also updated, whenever available, from \citet{af2020}, \citet{roz2017} and \citet{deras2020}, respectively. KT-42 in M22 was reclassified as EW variable \citep{roz2017} hence it was added to the list. Similarly, CB that turned out not a member of the globular clusters or being re-classified as other types of variable stars were removed. These include: V2 in M12 \citep{kaluzny2015}; KT-01, KT-03, KT-15, KT-39, KT-40 \& KT-41 in M22 \citep{roz2017}; V27, V28 \& V29 in NGC6712 \citep{deras2020}; V97 in M14 \citep{cp2018}; and V10 in NGC288 \citep{lee2016}. Therefore, our final list consists of 79 CB in 15 globular clusters.

\subsection{Light Curves from ZTF}

ZTF carried out a multi-band time-domain optical surveys on northern sky using the Samuel Oschin 48-inch Schmidt telescope, located at the Palomar Observatory. For more details on ZTF, see \citet{bel19} and \citet{gra19}, and will not be repeated here. Observing time, hence the data right, of ZTF were divided into partner surveys, public surveys and those reserved for the California Institute of Technology (Caltech). The ZTF imaging data were reduced and calibrated with a dedicated reduction pipeline, as detailed in \citet{mas19}. ZTF light curves for the CB in our list were extracted, using an 1~arc-second match radius, from the PSF (point spread function) catalogs generated from the dedicated ZTF pipeline. These PSF catalogs include the public and Caltech data from the ZTF Public Data Release 3\footnote{\url{https://www.ztf.caltech.edu/page/dr3}} and the ZTF private date spanning from commissioning (October 2017) to 30 September 2020. However, only 53 CB in 12 globular clusters have ZTF light curves data, as summarized in Table \ref{tab_gc}, and no ZTF light curves data were extracted for CB in M10, M80 and NGC2419. The extracted ZTF light curves data for these 53 CB were provided in Table \ref{tab_data}, with 1 to $\sim650$ data points per ZTF light curves (in either one of the ZTF's $gri$ filters). We further removed 10 CB (69\_K50, 70\_K51 \& 100\_N13 in M4; V3 in M12; KT-08 \& KT-33 in M22; P8 \& P12 in M71; V43 \& V54 in NGC6401) with sum of the data points per light curves in all $gri$-band that was less than 20, leaving 43 CB in our sample. All of these 43 CB have ZTF light curves data in $gr$-band, and some (21) of them included additional $i$-band light curves. For the remaining light curves, we excluded outliers that were deviated by more than $\pm0.75$~mag from the median values of the light curves, because the light curve amplitudes for CB in general would not exceed $1.2$~mag \citep[for example, see][]{rucinski1998,rucinski2001,dey2015,chen2018}. Therefore, we adopted $\pm0.75$~mag as a conservative cut to remove the outliers.

\begin{deluxetable}{llcccc}
  %\movetableright=-1in
  \tabletypesize{\scriptsize}
  \tablecaption{ZTF light curves data for CB in the globular clusters.\label{tab_data}}
  \tablewidth{0pt}
  \tablehead{
    \colhead{G.C.} &
    \colhead{CB Name} &
    \colhead{ZTF Filter} &
    \colhead{MJD} &
    \colhead{Mag.} &
    \colhead{Mag\_Error} 
  }
  \startdata
  M12  &   V3   &   g & 58617.4159838 & 19.169 & 0.250 \\
  M12  &   V3   &   g & 58672.1844329 & 19.027 & 0.334 \\
  M12  &   V3   &   g & 58715.1481944 & 19.583 & 0.341 \\
  M12  &   V3   &   r & 58300.3152894 & 18.136 & 0.185 \\
  M12  &   V3   &   r & 58331.1678009 & 18.270 & 0.346 \\
  $\cdots$& $\cdots$& $\cdots$& $\cdots$& $\cdots$& $\cdots$ \\
  \enddata
  \tablecomments{Table \ref{tab_data} is published in its entirety in the machine-readable format. A portion is shown here for guidance regarding its form and content.}
\end{deluxetable}

\section{Periods Refinement}\label{sec3}

Since it is well known that CB will undergo period change \citep[e.g., see][]{vv1972,wolf2000,rl2006,pilecki2007,lohr2013}, we therefore decided to obtain the orbital periods ($P$) for our sample of CB from ZTF light curves, and not adopting the published periods reported in the literature.

The Lomb-Scargle \citep{lomb1976,scargle1982} periodogram is a widely used method to search for periodicity of variable stars, which is based on fitting the unevenly sampled light curve data with a sinusoidal model \citep[for a review on Lomb-Scargle periodogram, see][]{vdp2018}. Because the shape of the CB light curve is near sinusoidal, it is common to apply Lomb-Scargle periodogram to search for their orbital periods. Majority of the codes that implement the Lomb-Scargle periodogram algorithm, however, rely on the light curve data taken in the same filter. An exception is {\tt gatspy} \citep{vdp2015,vdp2016} which can be run on either the single-band or the multi-band light curve data. Given the fact that shapes and amplitudes of the light curve for CB are almost independent of filters and ZTF multi-band observations were not in simultaneous (or near simultaneous), we applied three period search methods on the multi-band light curves for the 43 CB in the sample. The first two methods used the Lomb-Scargle periodogram in a different way, at which the first method constructed an effectively ``single-band'' light curve by combining the multi-band light curve data, while the second method utilized the multi-band version of the Lomb-Scargle periodogram. The third method is independent of Lomb-Scargle periodogram, at which a non-parametric smoother called the SuperSmoother \citep{reimann1994} was used to search for periodicity. The SuperSmoother algorithm does not rely on the assumption of sinusoidal-like light curve. The three period search methods applied on a given CB are: 

\begin{deluxetable*}{lllccc}
  %\movetableright=-1in
  \tabletypesize{\scriptsize}
  \tablecaption{Comparison of periods found with Method I, II and III to the published periods.\label{tab_period}}
  \tablewidth{0pt}
  \tablehead{
    \colhead{G.C.} &
    \colhead{Var. Name} &
    \colhead{Published Period (days)} &
    \colhead{Method I (days)} &
    \colhead{Method II (days)} &
    \colhead{Method III (days)} 
  }
  \startdata
M12	&	V6	& 0.256196	& {\bf 0.256193} & 0.278736 & 0.230562 \\
M12	&	V8	& 0.435135 	& {\bf 0.435132} & 0.435145 & 0.435164 \\
M13	&	V57	& 0.285416 	& 0.285408 & {\bf 0.285406} & 0.285408 \\
M15	&	V157	& 0.23306  	& {\bf 0.246187} & 0.246194 & 0.246191 \\
M15	&	V158	& 0.23576 	& 0.236162 & {\bf 0.236164} & 0.236167 \\
M22	&	KT-07	& 0.329797	& {\bf 0.329787} & 0.666617 & 0.331186 \\
M22	&	KT-23	& 0.298523 	& {\bf 0.298527} & 0.206080 & 0.299401 \\
M22	&	KT-43	& 0.22052 	& {\bf 0.224590} & 0.209436 & 0.198765 \\
M22	&	V118	& 0.280272 	& {\bf 0.280376} & 1.965056 & 0.280477 \\
M4	&	63\_K44	& 0.26358410 	& {\bf 0.263586} & 0.244123 & 0.238292 \\
M4	&	66\_K47	& 0.26987521 	& {\bf 0.269882} & 0.278171 & 0.244124 \\
M4	&	67\_K48	& 0.28269412 	& {\bf 0.294306} & 0.396847 & 0.262592 \\
M4	&	72\_K53	& 0.30844869 	& {\bf 0.308442} & 0.207441 & 0.316414 \\
M4	&	74\_K55	& 0.31070264 	& {\bf 0.310697} & 0.323251 & 0.291010 \\
M4	&	110\_C6	& 0.26297749 	& {\bf 0.262979} & 0.244133 & 0.281642 \\
M4	&	111\_C7	& 0.27161155 	& {\bf 0.271614} & 0.331127 & 0.248663 \\
M4	&	NV4	& 0.34444 	& {\bf 0.344350} & 0.278170 & 0.366307 \\
M71	&	P1	& 0.348905 	& 0.348905 & 1.997643 & {\bf 0.348903} \\
M71	&	P2	& 0.367188 	& {\bf 0.367177} & 0.367177 & 0.399362 \\
M71	&	P3	& 0.37386 	& {\bf 0.399229} & 0.327954 & 0.378340 \\
M71	&	P5	& 0.404380 	& 0.404384 & {\bf 0.404388} & 0.404395 \\
M71	&	P11	& 0.244461 	& 0.245115 & 1.996124 & {\bf 0.245114} \\
M71	&	P21	& 0.353 	& 0.357881 & 1.032355 & {\bf 0.357381} \\
M9	&	V21	& 0.7204518 	& 0.720477 & 1.945550 & {\bf 0.720453} \\
M9	&	V24	& 0.366784 	& {\bf 0.366788} & 0.449998 & 0.354914 \\
M9	&	V32	& 0.34460 	& {\bf 0.415818} & 0.984914 & 0.334858 \\
NGC5466	&	V28	& 0.342144 	& 0.342145 & {\bf 0.342147} & 0.342140 \\
NGC6401	&	V41\tablenotemark{a}	& 0.5873744 	& {\bf 0.587386} & 0.665054 & 0.598863 \\
NGC6712	&	V30\tablenotemark{a}	& 0.504748 	& {\bf 0.676754} & 0.667863 & 0.503702 \\
NGC6712	&	V31\tablenotemark{b}	& 0.418762 	& {\bf 0.418769} & 0.428192 & 0.442552 \\
  \enddata
  \tablenotetext{a}{Range of the searched periods is set to $[0.05,~0.40]$~days.}
  \tablenotetext{b}{Range of the searched periods is set to $[0.05,~0.25]$~days.}
  \tablecomments{For each CB, the adopted periods from Method I, II and III are marked in boldface.}
\end{deluxetable*}

\begin{figure*}
  \epsscale{2}
  \gridline{
    \fig{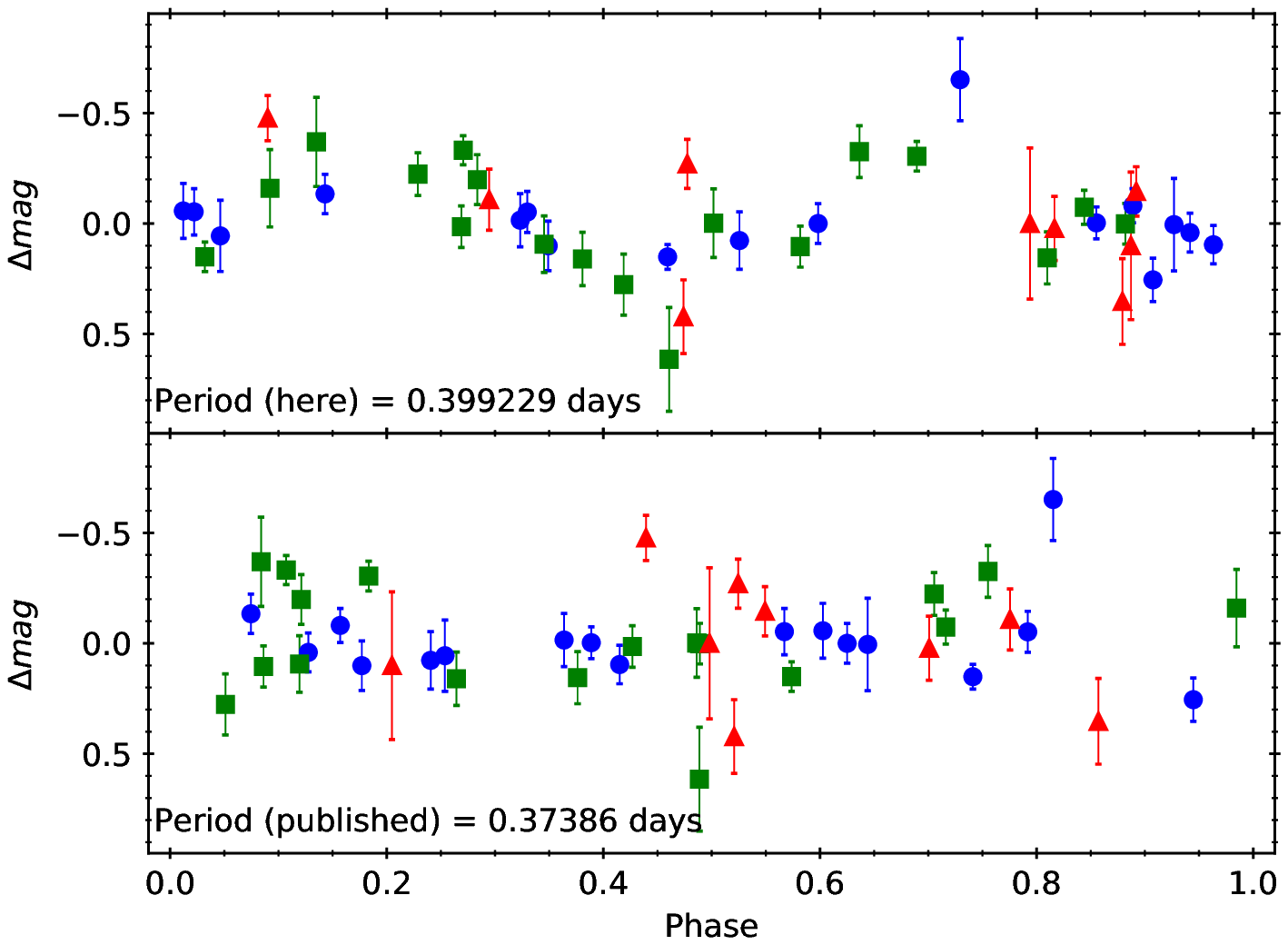}{0.34\textwidth}{(a) M71 P3, $\Delta P = 36.53$~minutes}
    \fig{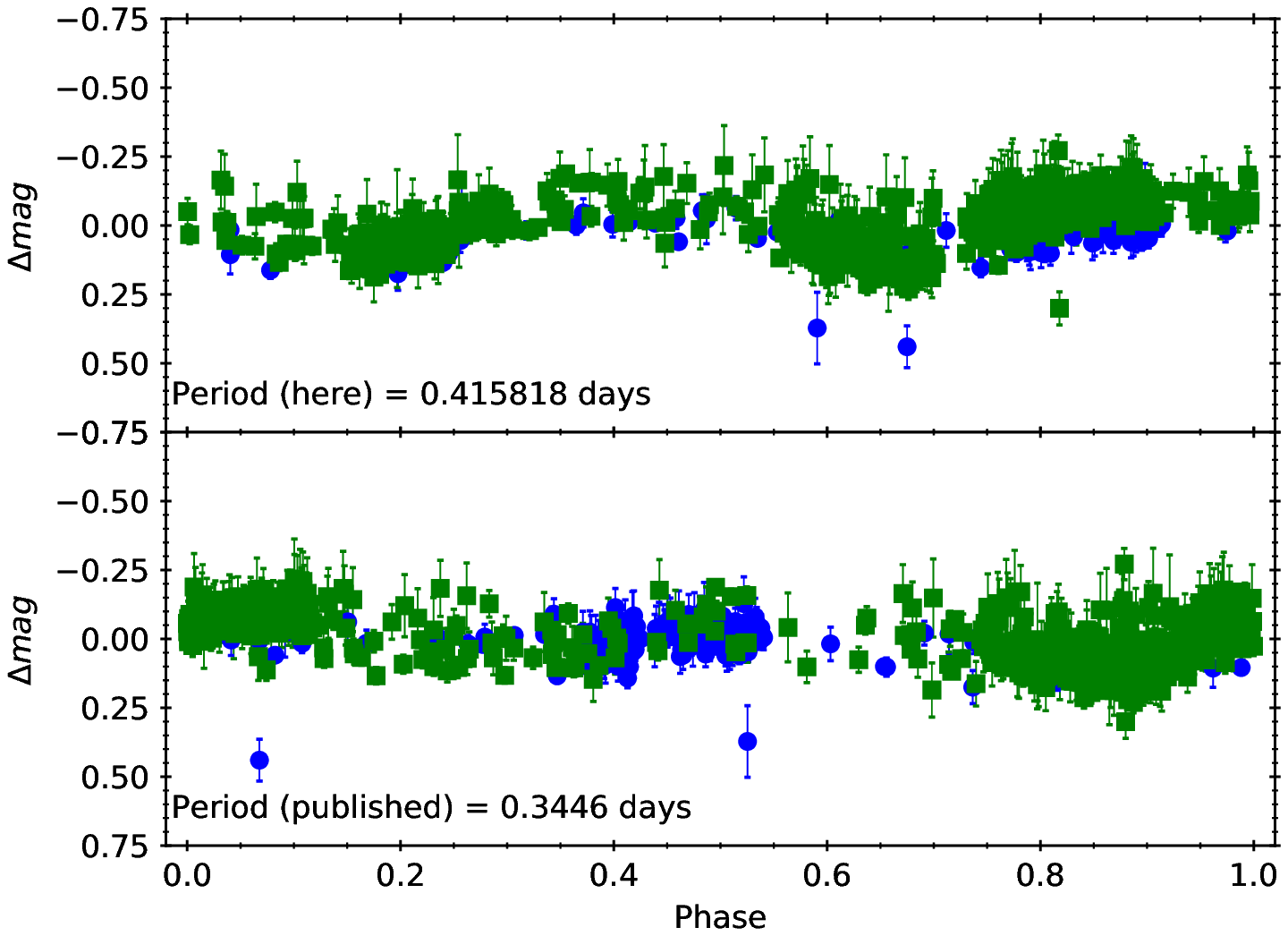}{0.34\textwidth}{(b) M9 V32, $\Delta P = 1.71$~hours}
    \fig{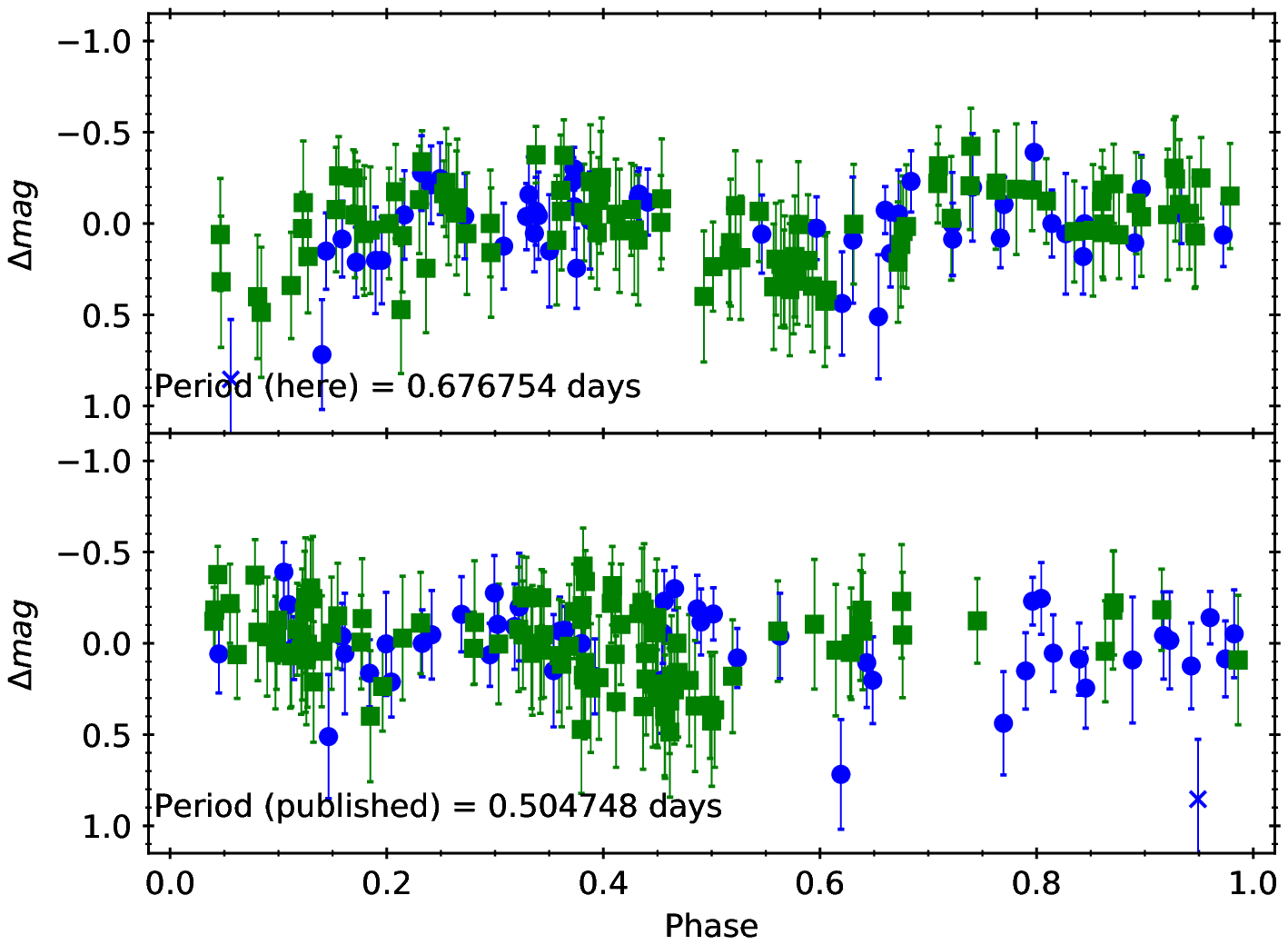}{0.34\textwidth}{(c) NGC6712 V30, $\Delta P = 4.13$~hours}
  }
  \caption{Comparison of the phased light curves folded with periods found here (top panel in each sub-figures) and the published periods given in the literature (bottom panel in each sub-figures) for the 3 CB with the most discrepant periods. $\Delta P$ is the difference between the period found here and the published period. In each light curves, the blue circles, green squares and red triangles are for the $g$-, $r$- and $i$-band data, respectively. Crosses represent the rejected outliers of the light curves. For a given CB, median values of each light curves in the $gr$-band (and $i$-band if available) were subtracted from the respected light curves before they were combined together. Similar plots for other CB were presented in the Appendix.}\label{fig_compareP}
\end{figure*}

\begin{itemize}
\item Method I: The $gr(i)$-band light curves were combined after subtracting the median values of each light curves, hereafter referred as the combined light curves. The Lomb-Scargle periodogram, implemented either in {\tt astropy} \citep{astropy2013,astropy2018} or {\tt gatspy}, was used to search for periods in the range of $0.1$ to $2.0$~days on the combined light curve, and twice of the best period found was adopted.
\item Method II: The multi-band Lomb-Scargle periodogram, {\tt LombScargleMultiband} implemented in {\tt gatspy}, was run on the $gr(i)$-band light curves with period range specified in between $0.1$ to $2.0$~days, and twice of the best period found was adopted.
\item Method III: The multi-band SuperSmoother periodogram, {\tt SuperSmootherMultiband} implemented in {\tt gatspy}, was run on the $gr(i)$-band light curves with period range restricted within 10\% of the published period.
\end{itemize}

Results of the periods found with these three methods are listed in Table \ref{tab_period}, together with final adopted periods and the published periods. In several CB, we used a smaller period range when applying Method I \& II (as indicated in Table \ref{tab_period}). The returned periods from these three methods were then used to fold the ZTF light curves and visually inspected. Period that can fold the $gr(i)$-band light curves closely resembling the CB-like light curve was adopted and marked as boldface in Table \ref{tab_period}. As can be seen from Table \ref{tab_period}, Method I generally gives the best-fit periods, nevertheless in several cases Method II or III provides a better period to fold the light curves. While visually inspecting these folded light curves, we found that light curves for 13 CB (including all of the CB in M30 and NGC4147, respectively) were seriously affected by blending (such as the light curves were flat or too bright when compared to the published $V$- or $I$-band data), therefore they were removed from the sample. For the remaining 30 CB (listed in Table \ref{tab_cb}), periods found in this work agreed within 1~second or less with the published periods for 19 CB, and another 4 CB the agreements were within 1~minute. The rest of the 7 CB the difference of the periods varied from $\sim6$~minutes to $\sim4$~hours. Figure \ref{fig_compareP} presents the phased light curves folded with periods used in this work and the published periods for the top {\bf 3} CB that have the largest deviations, clearly the periods found in this work can fold the light curves better. Comparisons of the folded light curves with published periods and the periods adopted in this work are presented in the Appendix.

\section{Light Curves Fitting with Fourier Expansion}\label{sec4}

%\movetabledown=2cm
%\begin{rotatetable*}
\begin{deluxetable*}{llcccccccccccccc}

  %\movetableright=-6in
  \tabletypesize{\scriptsize}
  \tablecaption{Properties of the contact binaries in globular clusters.\label{tab_cb}}
  \tablewidth{0pt}
  \tablehead{
    \colhead{G.C.} &
    \colhead{Var. Name} &
    \colhead{$P$} &
    \colhead{$N_g$\tablenotemark{a}} &
    \colhead{$N_r$\tablenotemark{a}} &
    \colhead{$N_i$\tablenotemark{a}} &
    \colhead{$\langle g\rangle$} &
    \colhead{$\langle r\rangle$} &
    \colhead{$\langle i\rangle$} &
    \colhead{$g_X$} &
    \colhead{$r_X$} &
    \colhead{$i_X$} &
    \colhead{$\sigma_g$} &
    \colhead{$\sigma_r$} &
    \colhead{$\sigma_i$} &
    \colhead{$E$\tablenotemark{b}} \\
    \colhead{} &
    \colhead{} &
    \colhead{(days)} &
    \colhead{} &
    \colhead{} &
    \colhead{} &
    \colhead{(mag)} &
    \colhead{(mag)} &
    \colhead{(mag)} &
    \colhead{(mag)} &
    \colhead{(mag)} &
    \colhead{(mag)} &
    \colhead{(mag)} &
    \colhead{(mag)} &
    \colhead{(mag)} &
    \colhead{(mag)}   
  }
  \startdata
M12	& V6		& 0.256193	& 45	& 98	& 4	& $19.56\pm0.02$	& $18.94\pm0.02$	& $18.81\pm0.10$	& 19.38	& 18.76	& 18.63		& 0.15	& 0.21	& 0.06		& $0.110\pm0.020$ \\
M12	& V8		& 0.435132	& 66	& 122	& 4	& $16.57\pm0.00$	& $16.51\pm0.00$	& $16.50\pm0.02$	& 16.53	& 16.47	& 16.46		& 0.02	& 0.04	& 0.02		& $0.184\pm0.002$ \\
M13	& V57		& 0.285406	& 647	& 656	& 143	& $18.59\pm0.00$	& $18.36\pm0.00$	& $18.31\pm0.01$	& 18.41	& 18.19	& 18.14		& 0.10	& 0.12	& 0.08		& $0.000\pm0.000$ \\
M15	& V157		& 0.246187	& 69	& 160	& 59	& $20.76\pm0.02$	& $19.92\pm0.01$	& $19.57\pm0.02$	& 20.55	& 19.72	& 19.36		& 0.15	& 0.13	& 0.13		& $0.124\pm0.002$ \\
M15	& V158		& 0.236164	& 122	& 181	& 73	& $20.09\pm0.01$	& $19.77\pm0.01$	& $19.61\pm0.02$	& 19.92	& 19.59	& 19.44		& 0.14	& 0.13	& 0.15		& $0.060\pm0.001$ \\
M22	& KT-07		& 0.329787	& 31	& 543	& 0	& $18.13\pm0.01$	& $17.29\pm0.01$	& $\cdots$		& 17.96	& 17.12	& $\cdots$	& 0.10	& 0.08	& $\cdots$	& $0.366\pm0.002$ \\
M22	& KT-23		& 0.298527	& 32	& 543	& 0	& $16.84\pm0.01$	& $16.53\pm0.00$	& $\cdots$		& 16.72	& 16.41	& $\cdots$	& 0.11	& 0.07	& $\cdots$	& $0.338\pm0.002$ \\
M22	& KT-43		& 0.224590	& 11	& 13	& 0	& $17.69\pm0.11$	& $16.94\pm0.11$	& $\cdots$		& 17.54	& 16.80	& $\cdots$	& 0.07	& 0.18	& $\cdots$	& $0.419\pm0.006$ \\
M22	& V118		& 0.280376	& 23	& 534	& 0	& $19.09\pm0.02$	& $18.33\pm0.01$	& $\cdots$		& 19.00	& 18.24	& $\cdots$	& 0.11	& 0.10	& $\cdots$	& $0.294\pm0.004$ \\
M4	& 63\_K44	& 0.263586	& 21	& 29	& 0	& $18.26\pm0.01$	& $17.42\pm0.01$	& $\cdots$		& 18.14	& 17.30	& $\cdots$	& 0.07	& 0.04	& $\cdots$	& $0.448\pm0.004$ \\
M4	& 66\_K47	& 0.269882	& 22	& 29	& 0	& $17.37\pm0.01$	& $16.64\pm0.01$	& $\cdots$		& 17.23	& 16.50	& $\cdots$	& 0.04	& 0.03	& $\cdots$	& $0.395\pm0.009$ \\
M4	& 67\_K48	& 0.294306	& 15	& 25	& 0	& $17.15\pm0.03$	& $16.34\pm0.03$	& $\cdots$		& 16.97	& 16.17	& $\cdots$	& 0.11	& 0.14	& $\cdots$	& $0.446\pm0.002$ \\
M4	& 72\_K53	& 0.308442	& 18	& 12	& 0	& $15.88\pm0.03$	& $15.27\pm0.03$	& $\cdots$		& 15.75	& 15.15	& $\cdots$	& 0.02	& 0.08	& $\cdots$	& $0.446\pm0.002$ \\
M4	& 74\_K55	& 0.310697	& 20	& 29	& 0	& $17.25\pm0.01$	& $16.52\pm0.01$	& $\cdots$		& 17.11	& 16.39	& $\cdots$	& 0.05	& 0.05	& $\cdots$	& $0.446\pm0.002$ \\
M4	& 110\_C6	& 0.262979	& 21	& 29	& 0	& $18.58\pm0.02$	& $17.67\pm0.01$	& $\cdots$		& 18.45	& 17.54	& $\cdots$	& 0.07	& 0.03	& $\cdots$	& $0.280\pm0.005$ \\
M4	& 111\_C7	& 0.271614	& 17	& 28	& 0	& $19.68\pm0.03$	& $18.50\pm0.02$	& $\cdots$		& 19.46	& 18.28	& $\cdots$	& 0.13	& 0.05	& $\cdots$	& $0.320\pm0.016$ \\
M4	& NV4		& 0.344350	& 23	& 29	& 0	& $15.99\pm0.00$	& $15.16\pm0.00$	& $\cdots$		& 15.84	& 15.01	& $\cdots$	& 0.06	& 0.03	& $\cdots$	& $0.376\pm0.005$ \\
M71	& P1		& 0.348903	& 297	& 565	& 171	& $18.59\pm0.00$	& $17.83\pm0.00$	& $17.54\pm0.00$	& 18.40	& 17.63	& 17.34		& 0.08	& 0.06	& 0.04		& $0.193\pm0.006$ \\
M71	& P2		& 0.367177	& 297	& 562	& 170	& $18.17\pm0.00$	& $17.47\pm0.00$	& $17.20\pm0.01$	& 17.97	& 17.27	& 17.00		& 0.07	& 0.09	& 0.16		& $0.266\pm0.004$ \\
M71	& P3		& 0.399229	& 17	& 18	& 9	& $18.39\pm0.03$	& $17.82\pm0.03$	& $17.26\pm0.05$	& 18.11	& 17.54	& 16.99		& 0.12	& 0.16	& 0.29		& $0.266\pm0.004$ \\
M71	& P5		& 0.404388	& 305	& 569	& 171	& $18.17\pm0.00$	& $17.52\pm0.00$	& $17.29\pm0.00$	& 18.01	& 17.36	& 17.13		& 0.06	& 0.05	& 0.05		& $0.216\pm0.007$ \\
M71	& P11		& 0.245114	& 118	& 498	& 164	& $20.91\pm0.02$	& $19.69\pm0.01$	& $19.18\pm0.01$	& 20.66	& 19.44	& 18.93		& 0.20	& 0.12	& 0.12		& $0.157\pm0.004$ \\
M71	& P21		& 0.357881	& 136	& 370	& 79	& $16.59\pm0.00$	& $15.80\pm0.00$	& $15.46\pm0.00$	& 16.45	& 15.66	& 15.33		& 0.03	& 0.03	& 0.03		& $0.304\pm0.006$ \\
M9	& V21		& 0.720453	& 164	& 607	& 0	& $17.23\pm0.00$	& $16.51\pm0.00$	& $\cdots$		& 17.14	& 16.42	& $\cdots$	& 0.05	& 0.03	& $\cdots$	& $0.356\pm0.005$ \\
M9	& V24		& 0.366788	& 151	& 595	& 0	& $17.77\pm0.00$	& $17.11\pm0.00$	& $\cdots$		& 17.58	& 16.91	& $\cdots$	& 0.06	& 0.07	& $\cdots$	& $0.376\pm0.002$ \\
M9	& V32		& 0.415818	& 165	& 610	& 0	& $16.24\pm0.00$	& $15.51\pm0.00$	& $\cdots$		& 16.16	& 15.43	& $\cdots$	& 0.06	& 0.07	& $\cdots$	& $0.366\pm0.002$ \\
NGC5466	& V28		& 0.342147	& 145	& 240	& 98	& $18.45\pm0.01$	& $18.51\pm0.01$	& $18.67\pm0.01$	& 18.33	& 18.39	& 18.55		& 0.08	& 0.08	& 0.08		& $0.000\pm0.000$ \\
NGC6401	& V41		& 0.587386	& 6	& 167	& 76	& $20.60\pm0.10$	& $19.19\pm0.01$	& $18.33\pm0.02$	& 20.43	& 19.02	& 18.16		& 0.26	& 0.12	& 0.11		& $1.108\pm0.004$ \\
NGC6712	& V30		& 0.676754	& 49	& 106	& 0	& $19.62\pm0.04$	& $19.24\pm0.04$	& $\cdots$		& 19.47	& 19.09	& $\cdots$	& 0.18	& 0.16	& $\cdots$	& $0.414\pm0.006$ \\
NGC6712	& V31		& 0.418769	& 90	& 464	& 27	& $19.05\pm0.01$	& $18.64\pm0.01$	& $18.25\pm0.03$	& 18.91	& 18.51	& 18.12		& 0.16	& 0.28	& 0.36		& $0.411\pm0.002$ \\
  \enddata
  \tablenotetext{a}{Number of data points per light curve in a given $gri$ filter.}
  \tablenotetext{b}{$E$ is the mean reddening value obtained from the {\tt Bayerstar2019} 3D reddening map \citep{green2019} with the adopted distances listed in Table \ref{tab_gc}.}
  
\end{deluxetable*}
%\end{rotatetable*}

The near sinusoidal shape of the CB's light curves indicates these light curves can be well represented by a truncated Fourier expansion \citep{rucinski1973,rucinski1993,coker2013}. Indeed, low-order Fourier expansion has been used to separate out CB from other types of eclipsing binaries \citep[mainly the Algol and Beta Lyrae types; for example in][]{rucinski1997b,rucinski1997c,selam2004,derekas2007,lee2014,muraveva2014,ned2015,yang2020} or to classify CB against different types of variable stars in time-series surveys \citep[for example, see][]{pojmanski2002,eyer2005,debosscher2007,hoffman2009,kim2014,masci2014,jay2019}. In this work, a $4^{\mathrm{th}}$-order Fourier expansion \citep{derekas2007,deb2011,dey2015,chen2018} in the following form was adopted to fit the folded ZTF light curves for the CB listed in Table \ref{tab_cb}:

\begin{eqnarray}
  m(\phi) & = & m_0 + \sum_{j=1}^4 [a_j \cos (2\pi j \phi) + b_j \sin (2\pi j \phi)],
\end{eqnarray}

\noindent where $m(\phi)$ are magnitudes, and $\phi$ from 0 to 1 represent the (orbital) phases of the light curves after folded with adopted periods in Section \ref{sec3}. $m_0$, $a_j$ and $b_j$ are the Fourier coefficients to be fitted.

\begin{figure*}
  \epsscale{2}
  \gridline{
    \fig{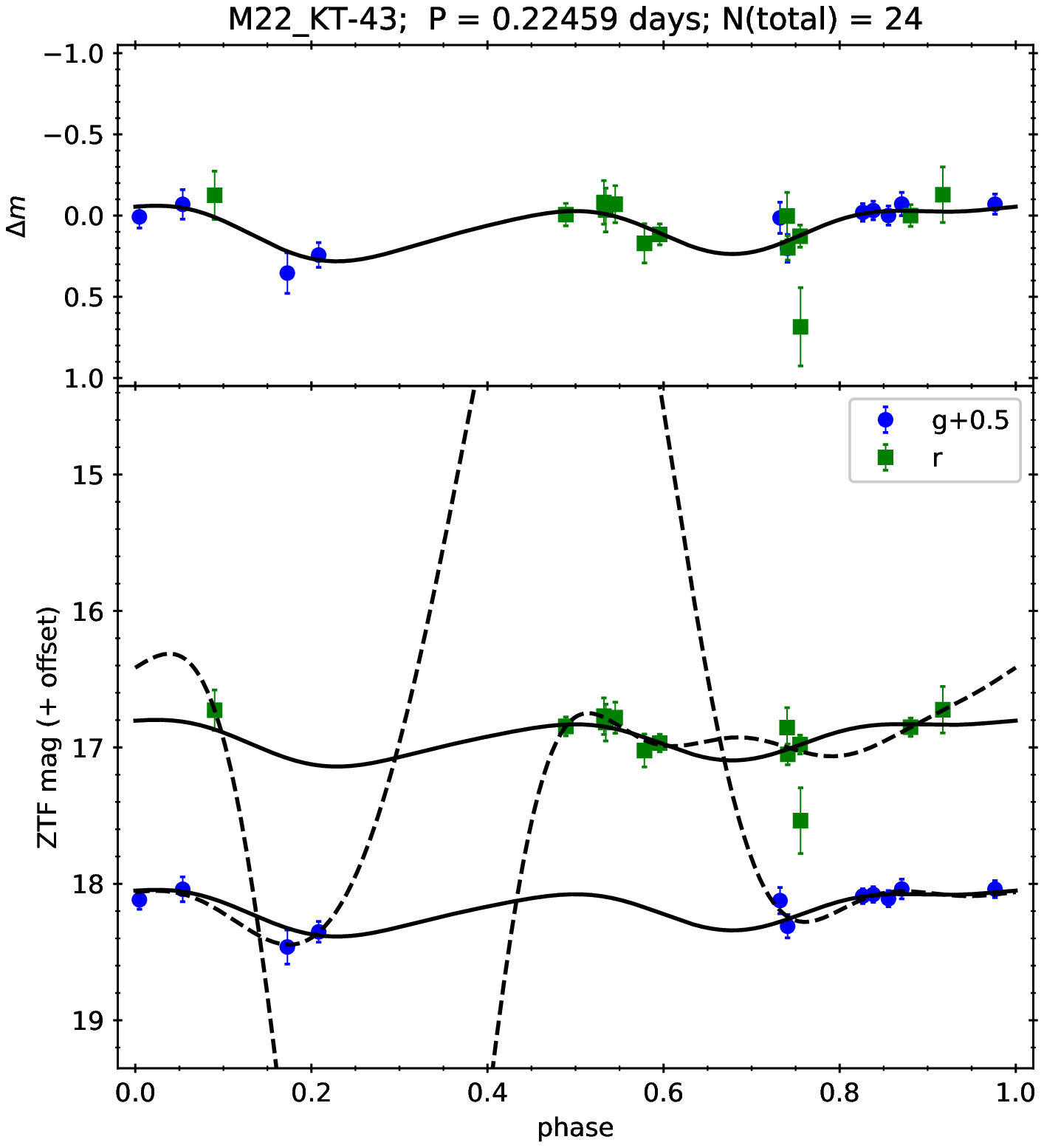}{0.34\textwidth}{(a) CB with small number of data points ($N<50$).}
    \fig{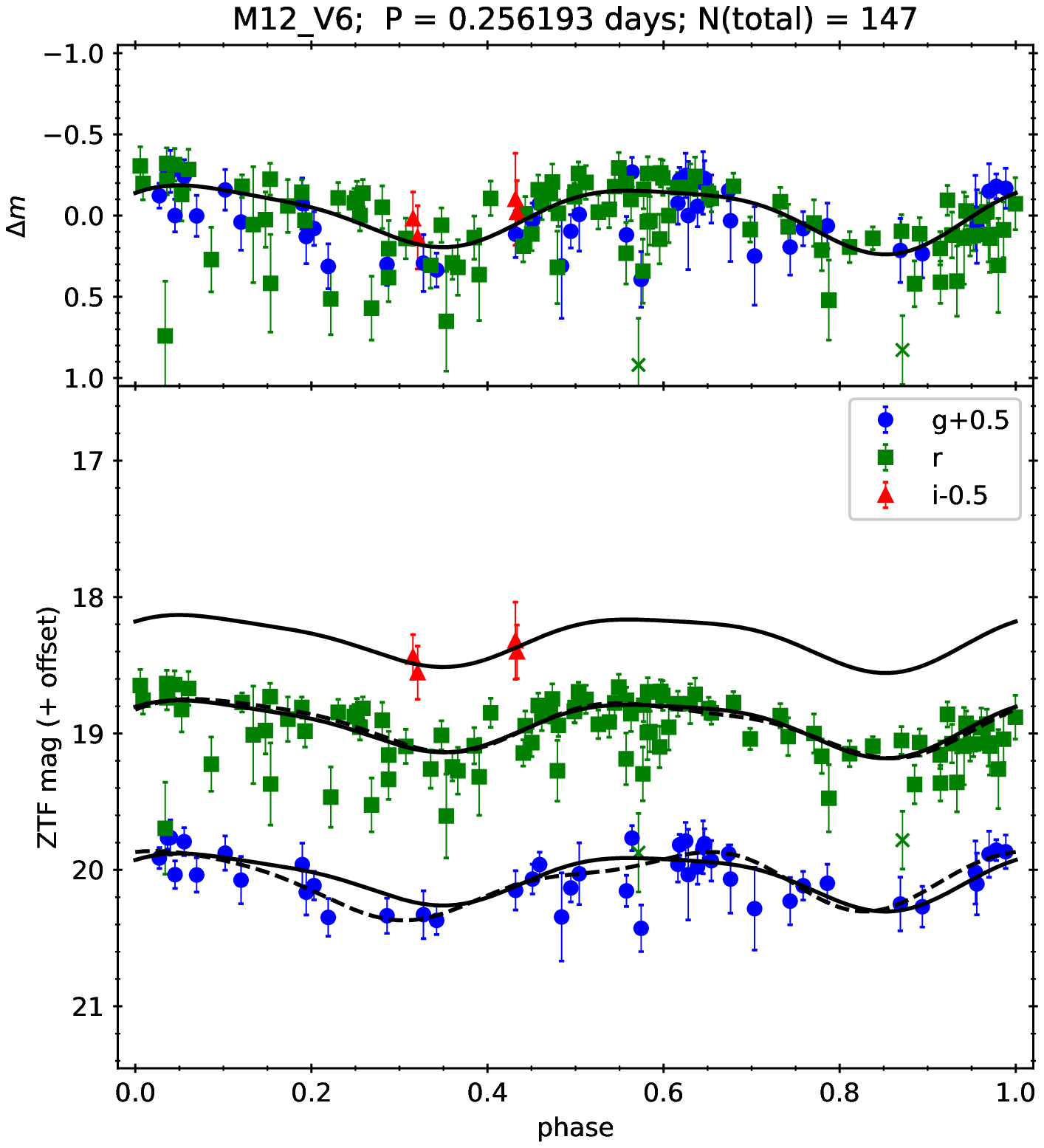}{0.34\textwidth}{(b) CB with moderate number of data points ($50>N>200$).}
    \fig{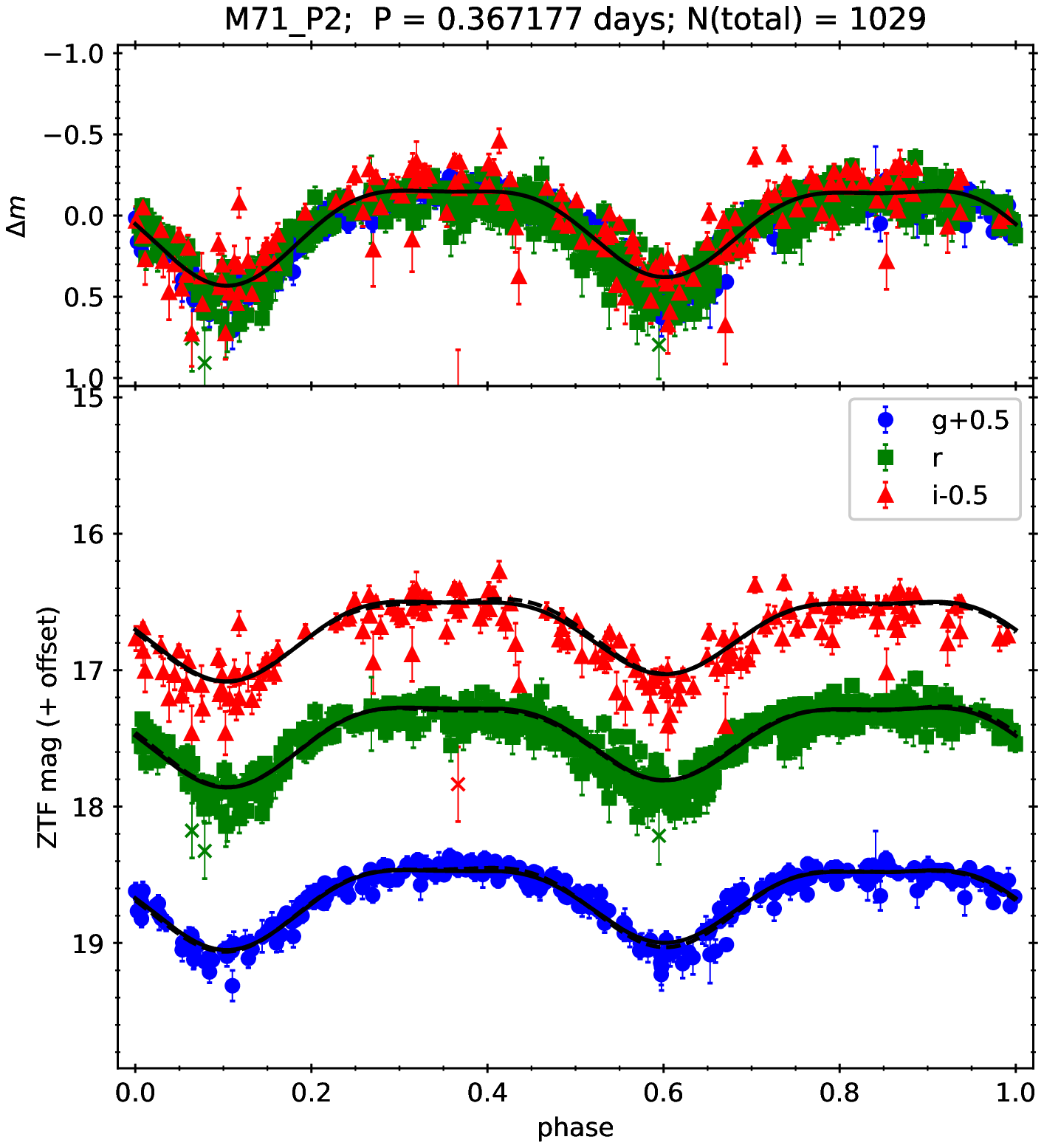}{0.34\textwidth}{(c) CB with large number of data points ($N>200$).}
  }
  \caption{Examples of the CB light curves fitted with a $4^{\mathrm{th}}$-order Fourier expansion, where $N(\mathrm{total})=N_g+N_r+N_i$ is the total number of data points in the ZTF $gr(i)$-band light curves. In the lower panels, the dashed curves are the fitted Fourier expansion to individual $gr(i)$-band light curves, while the solid curves are the template light curves based on the combined light curves, as presented in the top panels, shifted vertically to fit the $gr(i)$-band light curves. For each CB, the template light curves were constructed by fitting the $4^{\mathrm{th}}$-order Fourier expansion to the combined light curves. The blue circles, green squares and red triangles represent the $g$-, $r$- and $i$-band data, respectively. Crosses are the rejected outliers of the light curves. Similar plots for other CB were presented in the Appendix.}\label{fig_lc}
\end{figure*}

Since equation (1) contains 9 Fourier coefficients, the ZTF $gr(i)$ band light curves with more than 9 data points were fitted with the $4^{\mathrm{th}}$-order Fourier expansion. However, the fittings would be influenced by how even the data points were distributed on the folded light curves. Figure \ref{fig_lc}(a) shows an example of a CB with small number of data points in the $g$- and $r$-band light curves, at which the fittings of the $4^{\mathrm{th}}$-order Fourier expansion were unacceptable due to the presence of a large gap in the folded light curves (as demonstrated with the dashed curves in the plot).

To remedy this problem, for each CB we constructed a template light curve by fitting the $4^{\mathrm{th}}$-order Fourier expansion to the combined light curve, because the combined light curve has a much larger number of data points per light curve to constrain the fitted Fourier coefficients better. Examples of the the constructed template light curves were shown in the top panels of Figure \ref{fig_lc} as solid curves. The template light curve was then shifted vertically with an amount of $\Delta m$ to fit the individual $gr(i)$-band light curves for a given CB by minimizing their error sum of squares (SSE). Lower panel of Figure \ref{fig_lc}(a) (for M22 KT-43) clearly demonstrated the improvement of using the template light curve to fit the sparsely sampled ZTF light curves (similar improvement can also be seen in the $g$-band light curve of M12 V6). Another advantage of using the template light curve is it can be use to fit light curves with number of data points that was less than 9, as in the case of $i$-band light curve for M12 V6, see Figure \ref{fig_lc}(b). When the individual $gr(i)$-band light curves were well sampled (with a large number of data points per light curves, such as 200 or more), both of the fitted template light curves and the fitted $4^{\mathrm{th}}$-order Fourier expansion on the individual $gr(i)$-band light curves were (almost) indistinguishable, as shown in the lower panel of Figure \ref{fig_lc}(c) (for M71 P2; and the $r$-band light curve of M12 V6). 

\begin{figure}
  %\epsscale{1.2}
  \plotone{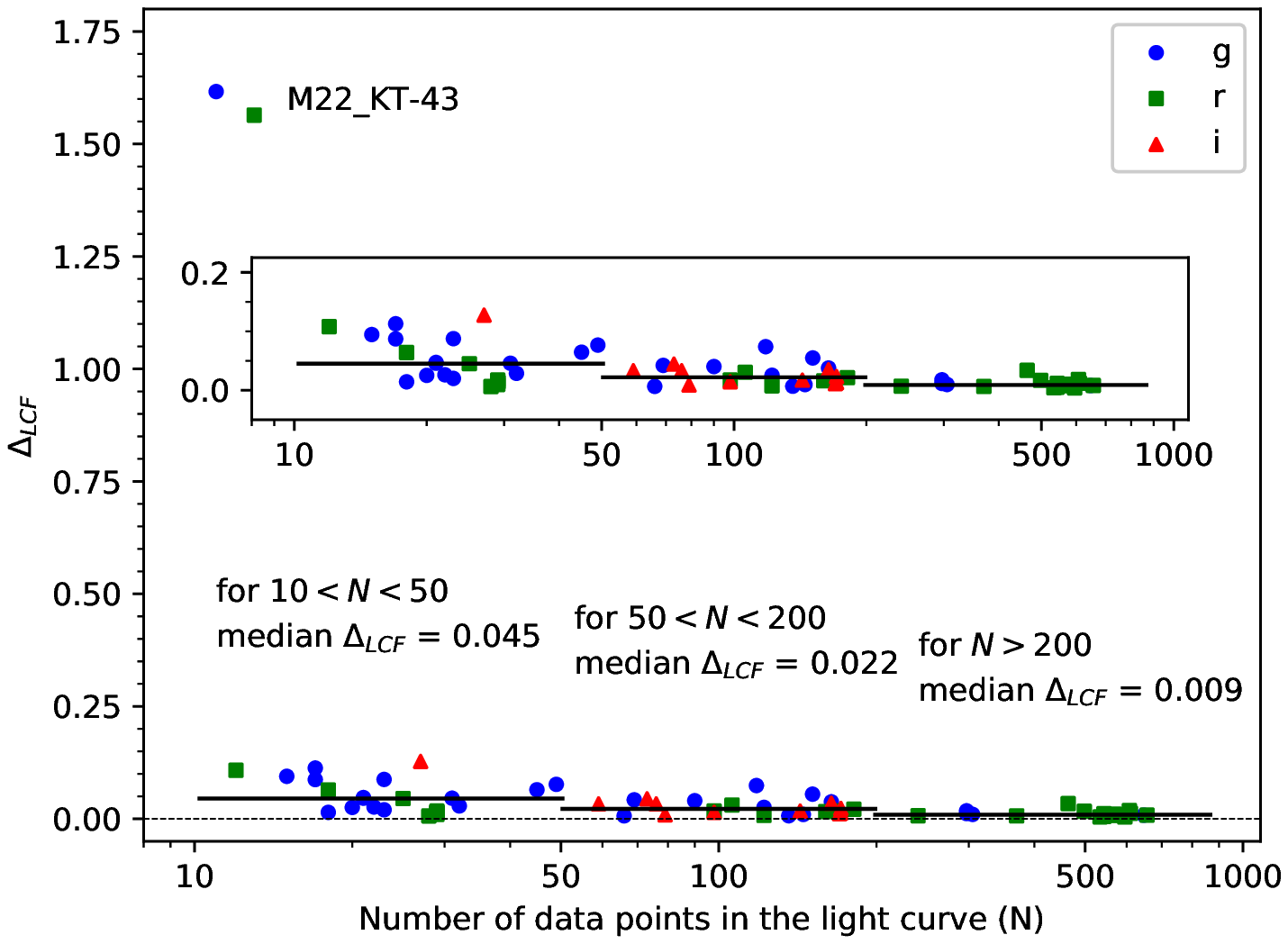}
  \caption{The averaged absolute difference between the template light curve and the individual $gr(i)$-band light curve, $\Delta_{LCF}$, as a function of $N$, the number of data points in the individual $gr(i)$-band light curves, where $N$ is divided into three bins as indicated. Median values of $\Delta_{LCF}$ in each bins are represented as solid horizontal lines. The dashed horizontal line represents $\Delta_{LCF}=0$. The inset Figure is an enlarged version of $\Delta_{LCF}$ vs. $N$, after excluding the two outlier points for M22 KT-43.}\label{fig_deltalc}
\end{figure}

We compare the fitted light curves between the template light curves and the individual $gr(i)$-band light curves (i.e. the solid and dashed curves in the bottom panels of Figure \ref{fig_lc}), all fitted with a $4^{\mathrm{th}}$-order Fourier expansion, by calculating the averaged absolute difference between them. The averaged absolute difference between the template light curve, $m(\phi)_{\mathrm{TEMPLATE}}$, and the individual $gr(i)$-band light curve, $m(\phi)_{\mathrm{SINGLE}}$, is defined as $\Delta_{LCF} = \langle ~|m(\phi)_{\mathrm{TEMPLATE}} - m(\phi)_{\mathrm{SINGLE}}|~ \rangle$, where $m(\phi)$ represents both of the fitted template and single-band light curves based on equation (1) for 1000 phase points ($\phi$) evenly distributed between 0 and 1. Figure \ref{fig_deltalc} presents the resulted $\Delta_{LCF}$ as a function of the number of data points ($N$) on individual $gr(i)$-band light curves. Excluding the data points from M22 KT-43 (see Figure \ref{fig_lc}(a)), we bin the values of $\Delta_{LCF}$ into three bins of $10<N<50$, $50<N<200$, and $N>200$. The median values of $\Delta_{LCF}$ decrease from $0.045$~mag to $0.009$~mag from the first bin to the last bin, indicating the fitted Fourier expansion on individual $gr(i)$-band light curve will be closer to the template light curve when $N$ increases.

Mean magnitudes of the individual $gr(i)$-band light curves were equivalent to the sum of the $m_0$ term from the fitting of template light curves and the vertical shift $\Delta m$, they were listed in Table \ref{tab_cb} as $\langle m\rangle$, where $m=\{g,r,i\}$. The minimum locus of the fitted template light curves were adopted as the magnitudes at maximum light of each CB, $m_X$, in Table \ref{tab_cb}. The dispersion $\sigma_m$ of fitting the template light curves to individual $gr(i)$-band light curves were also given in Table \ref{tab_cb}.

\section{The Period-Luminosity Relations}\label{sec5}

\begin{figure*}
  %\epsscale{2}
  \plottwo{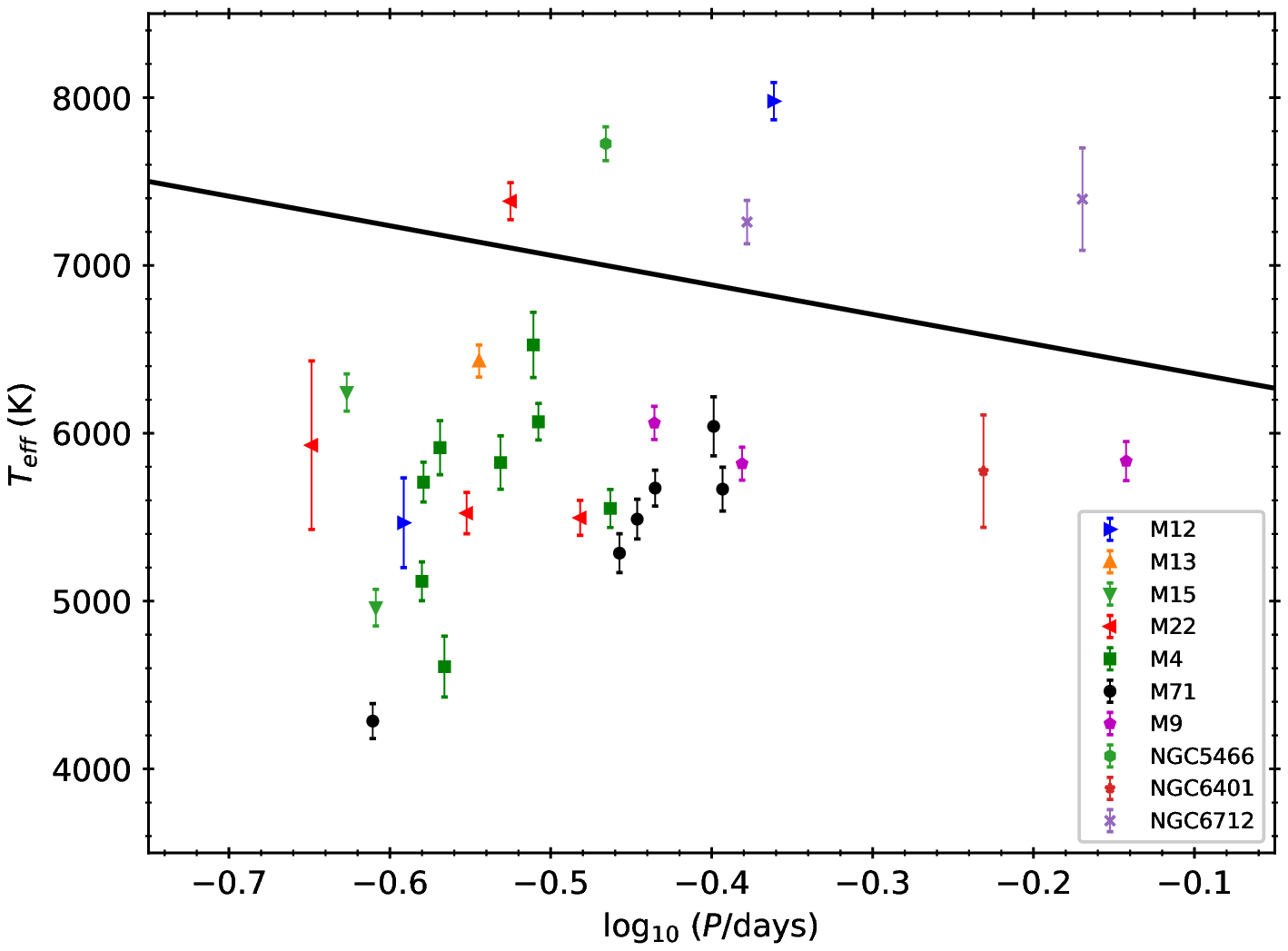}{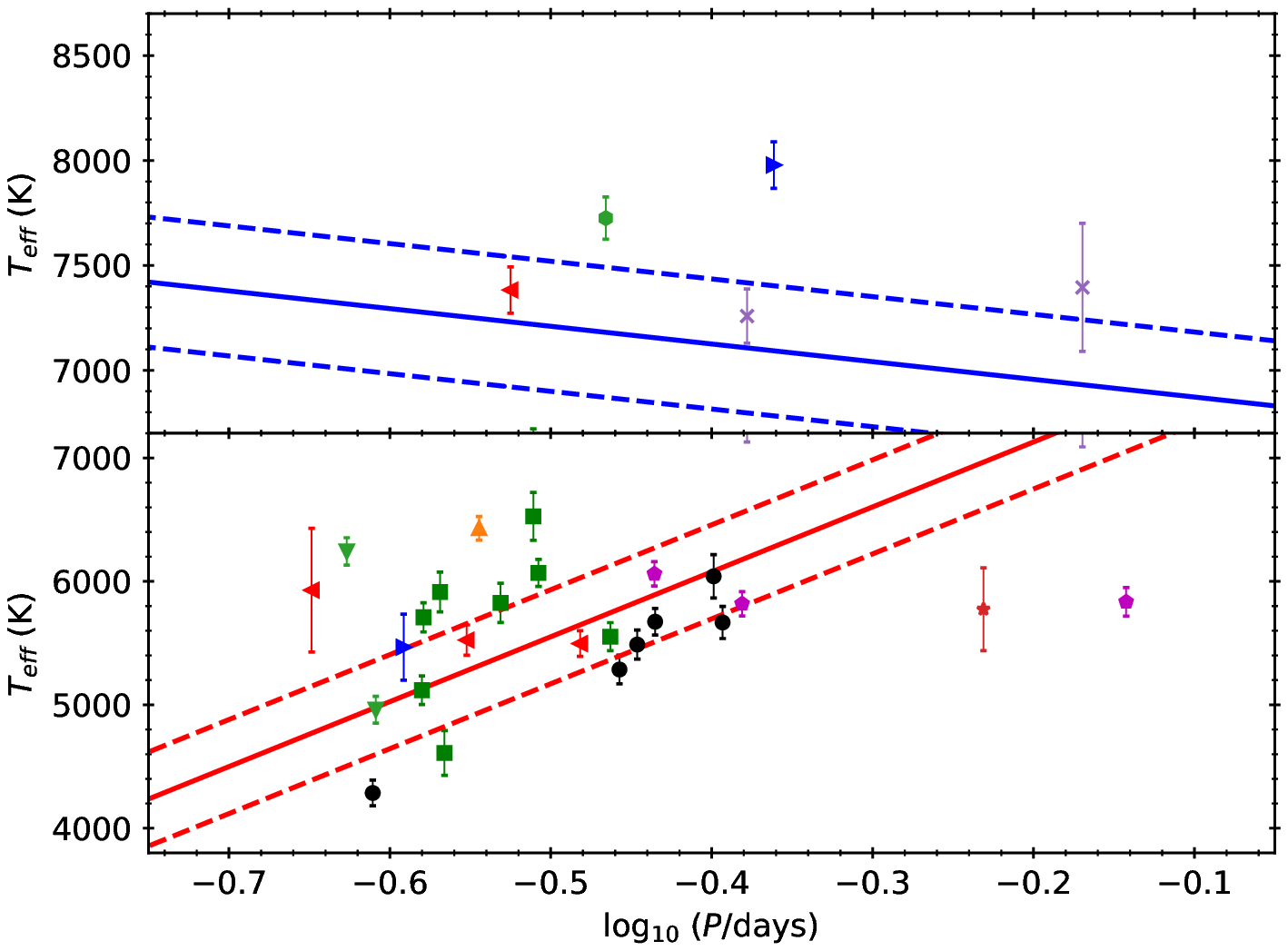}
  \caption{Effective temperatures, {\bf $T_{\mathrm{eff}}$,} of the CB in our sample as a function of orbital periods, where the effective temperatures were converted from the extinction corrected $(g-r)$ colors using the converting relation given in \citet{fukugita2011}. The error bars represent the quadrature sum of the errors converted from the colors and $\sigma_T$, where $\sigma_T=93$K is the dispersion of the converting relation \citep{fukugita2011}. The black solid line in the left panel is the period-temperature relation, proposed in \citet{jay2020} to separate out the early-type and late-type CB. The blue and red solid lines in the right panel are the period-temperature relations for early-type and late-type CB, respectively, adopted from \citet{jay2020}, and the dashed lines represent the $\pm1\sigma$ dispersion of these relations.} \label{fig_teff}
\end{figure*}

Using the periods as well as the mean magnitudes and the magnitudes at maximum light obtained from previous sections, together with extinction corrections and published distances to individual globular clusters, the PL relations can be derived. Values of reddening $E$ toward each of the CB was obtained using the {\tt Bayerstar2019} 3D reddening map\footnote{\url{http://argonaut.skymaps.info/}} presented in \citet{green2019}, and listed in the last column of Table \ref{tab_cb}. Absolute magnitudes at mean or maximum light of the CB were then calculated via $M_{\{g,r,i\}} = m_{\{g,r,i\}} - R_{\{g,r,i\}}E - 5\log D + 5$, where distance $D$ is in parsec as given in Table \ref{tab_gc}. Since ZTF photometry is calibrated to the Pan-STARRS1 system \citep{mas19}, the values of extinction coefficient are $R_{\{g,r,i\}} = \{3.518,\ 2.617,\ 1.971\}$ \citep{green2019}. Errors from $m$, $E$ and $D$ were propagated to the total errors on $M$. In case of magnitudes at maximum light, the dispersion of the fitted light curves $\sigma$, listed in Table \ref{tab_cb}, were adopted as the errors for $m$. For those distances without errors, a 10\% error was assumed \citep{baumgardt2019}.

\begin{figure}
  %\epsscale{1.2}
  \plotone{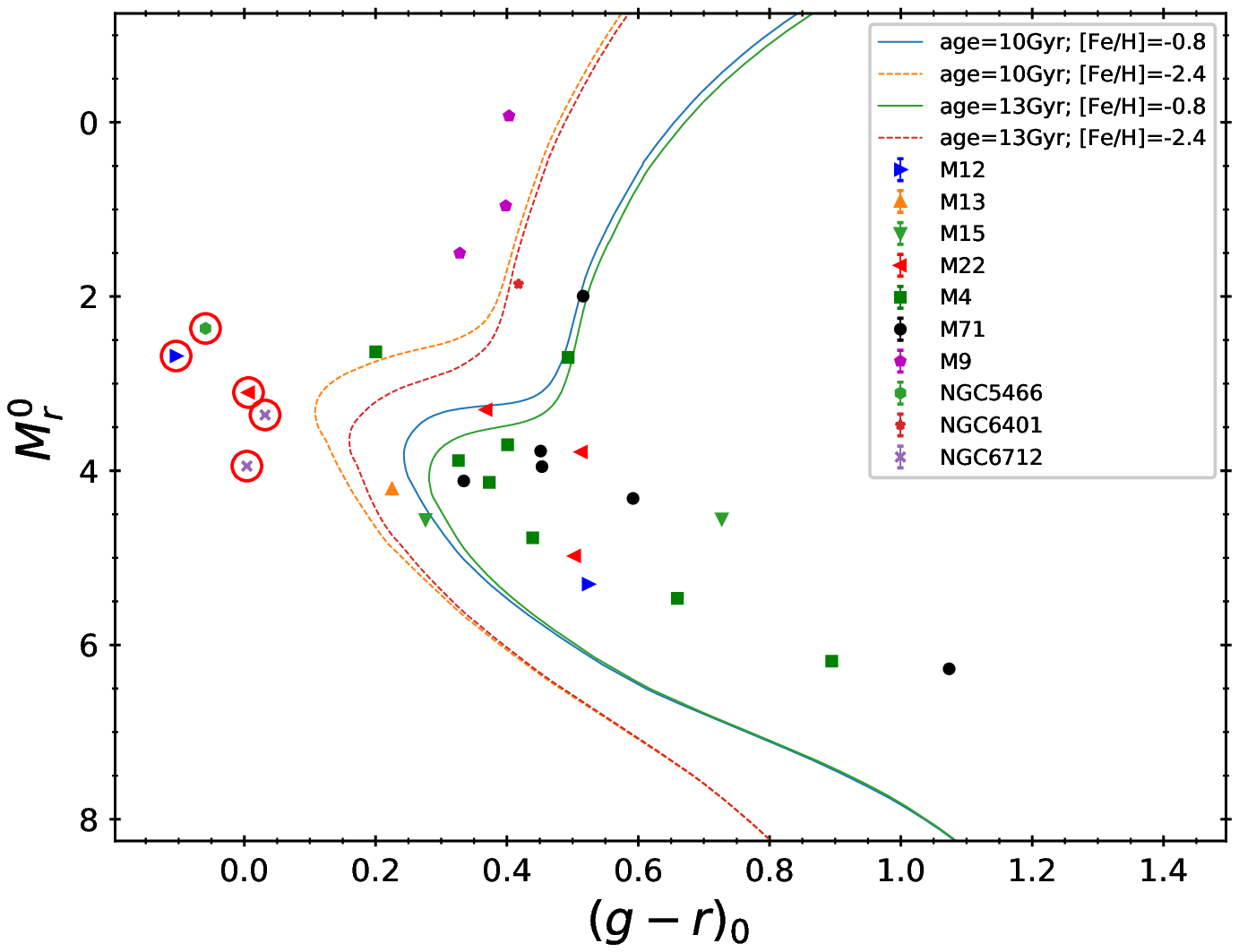}
  \caption{Color-magnitude diagram for the CB in our sample. $M_r^0$ and $(g-r)_0$ denote the extinction corrected absolute magnitude and color, respectively. The red circles are the early-type CB identified in Figure \ref{fig_teff} using the period-temperature relation. The model isochrones were taken from the Darmouth Stellar Evolution Database \citep{dotter2007,dotter2008}. These isochrones were chosen to roughly bracket the metallicity listed in Table \ref{tab_gc}.}\label{fig_cmd}
\end{figure}

\begin{figure*}
  \epsscale{2}
  \gridline{
    \fig{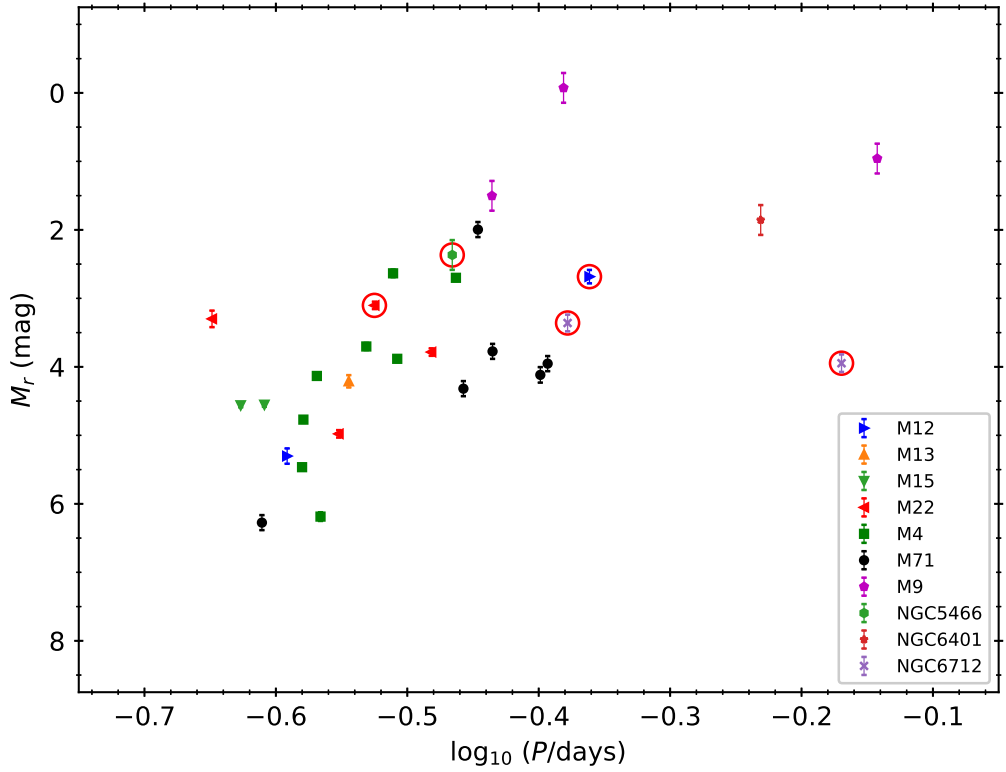}{0.875\textwidth}{(a) The $r$-band PL relation at mean light.}
  }
  \gridline{
    \fig{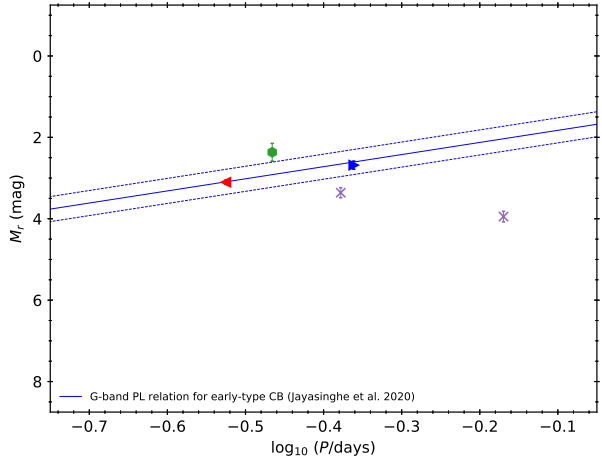}{0.52\textwidth}{(b) Same as (a) but for early-type CB.}
    \fig{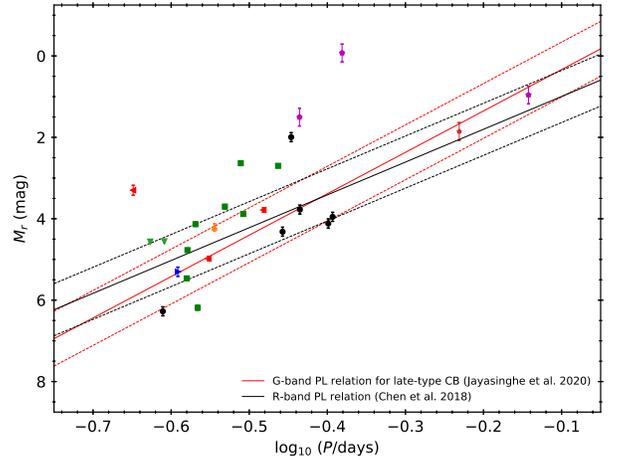}{0.52\textwidth}{(c) Same as (a) but for late-type CB.}
  }
  \caption{Panel (a): The $r$-band PL relation for 30 CB in our sample. The red circles indicate the early-type CB selected based on the period-temperature relation (see Figure \ref{fig_teff}). Panel (b): Same as panel (a) but for the early-type CB, the solid line is the $G$-band PL relation taken from \citet{jay2020}.  Panel (c): Same as panel (a) but for the late-type CB, the solid red and black lines are the $G$-band and $R$-band PL relation taken from \citet{jay2020} and \citet{chen2018}, respectively. In both panel (b) and (c), the dashed lines are the corresponding $\pm2\sigma$ dispersion of the adopted PL relations.} \label{fig_iniPL}
\end{figure*}

\begin{figure*}
  \epsscale{1.15}
  \plottwo{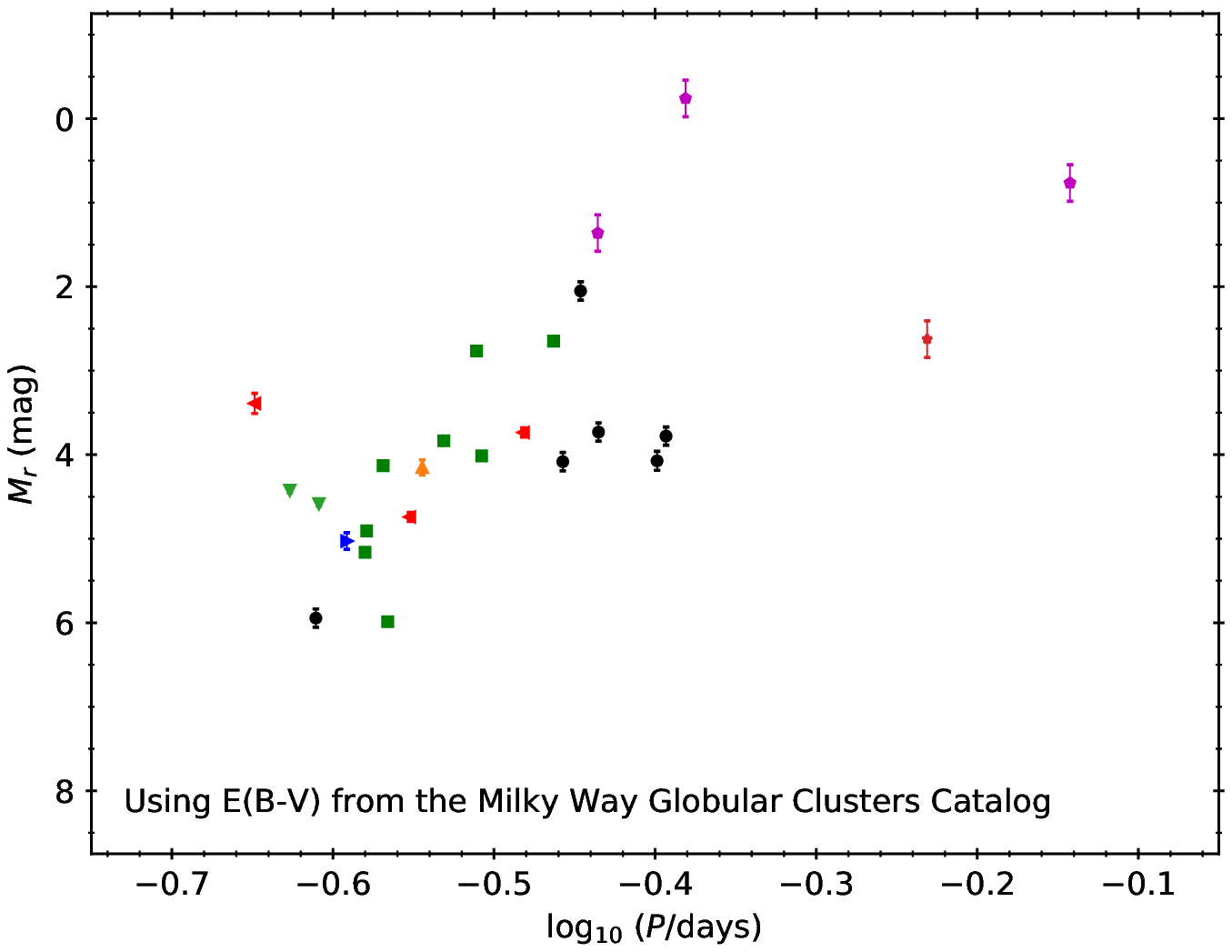}{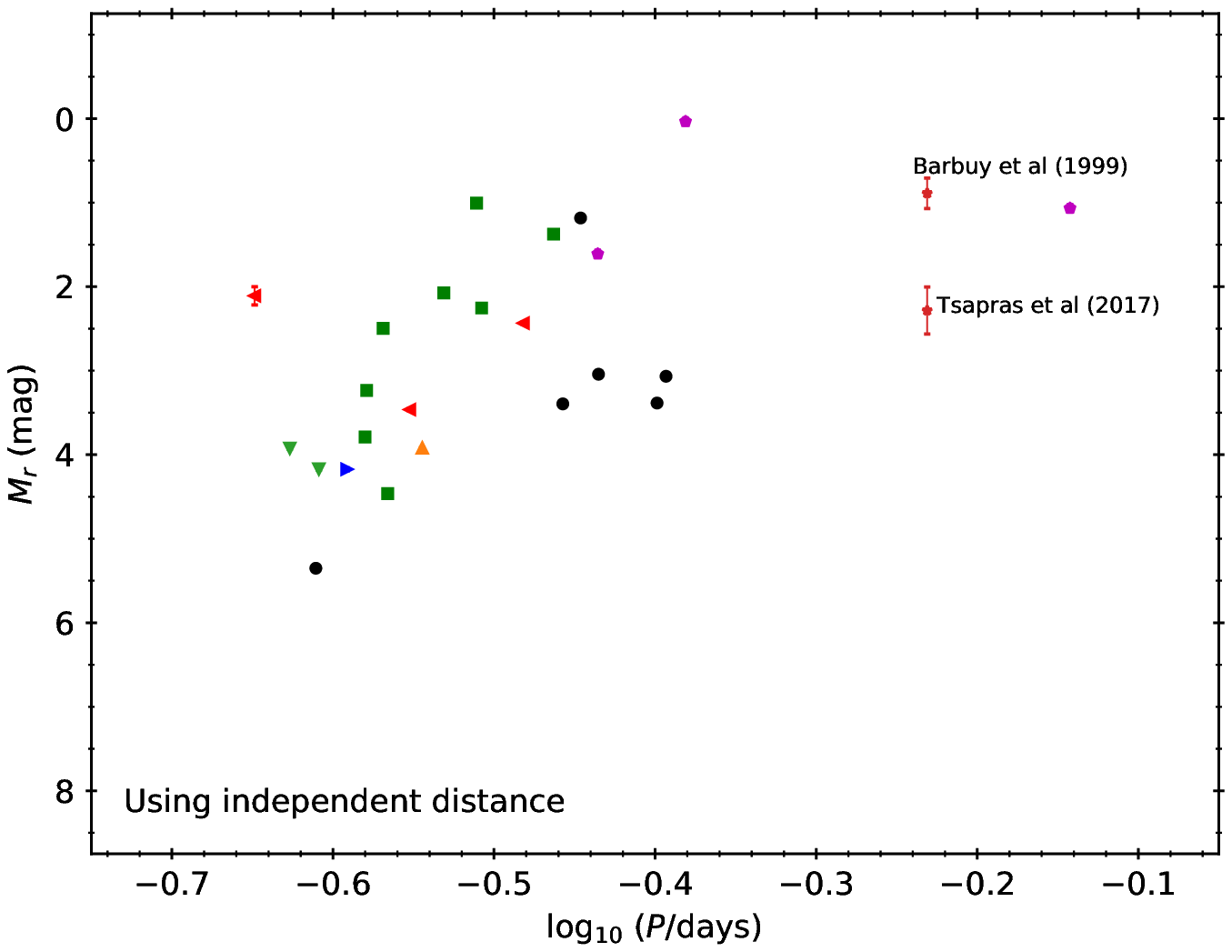}
  \plottwo{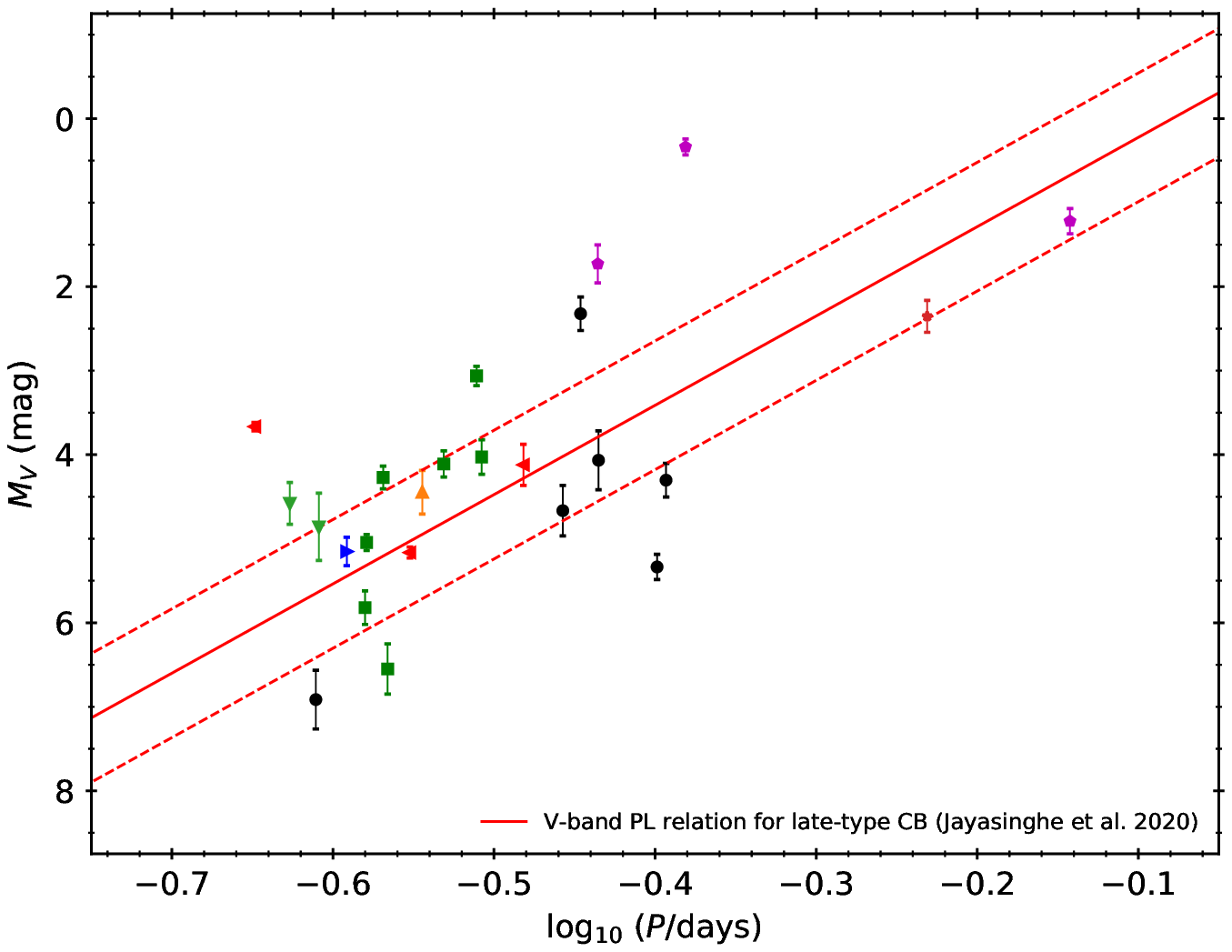}{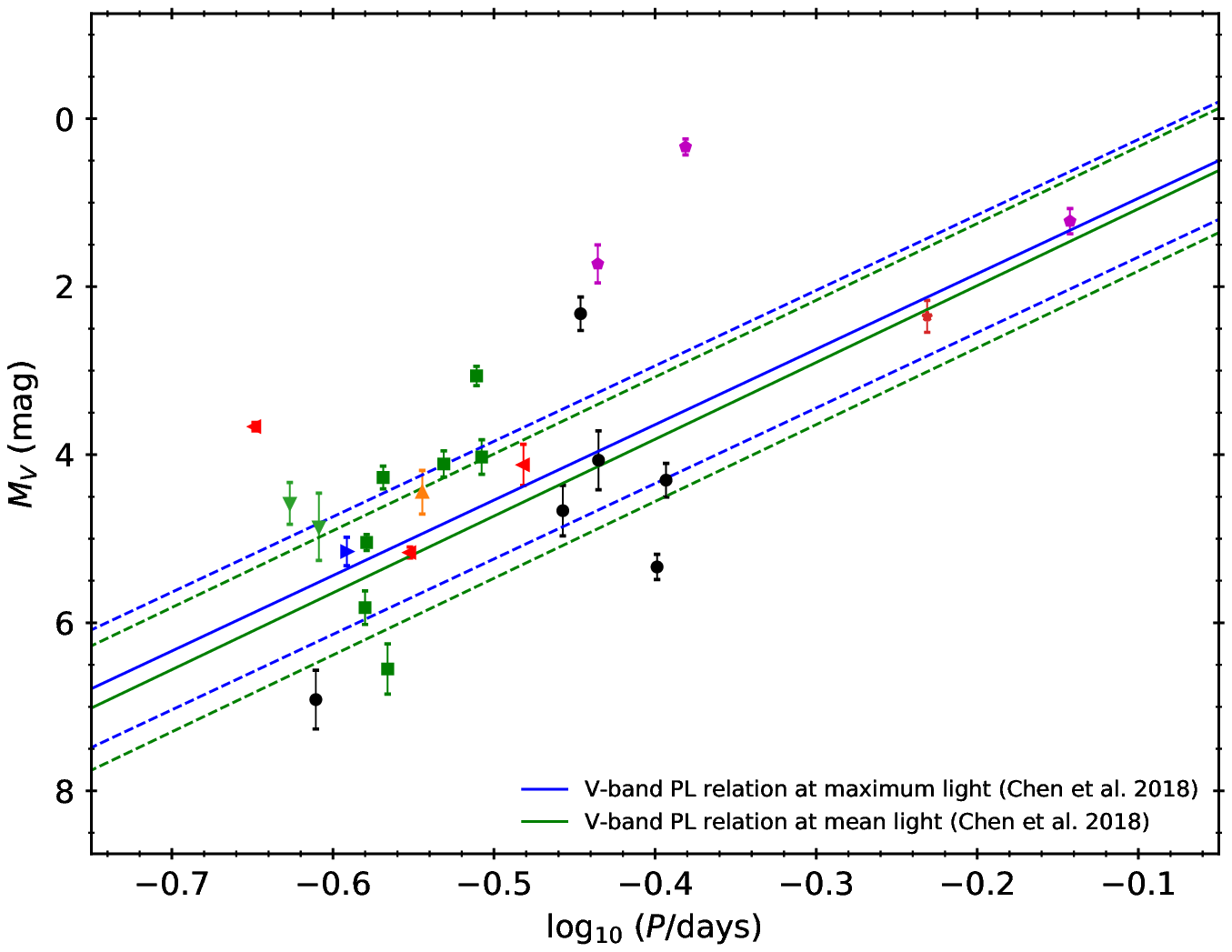}
  \caption{Test of the PL relation by replacing extinctions (upper-left panel), distance (upper-right panel) and apparent magnitudes (lower panels) -- see text for more details. The solid lines in the lower panels are the published $V$-band PL relations taken from \citet[][in the lower-left panel]{jay2020} and \citet[][in the lower-right panel]{chen2018}; the dashed lines are the corresponding $\pm2\sigma$ dispersion.} \label{fig_testpl}
\end{figure*}

\subsection{Identifying Early-Type CB}\label{sec51}

\citet{chen2016} and \citet{chen2018} applied a period cut at $\log P_{\mathrm{cut}} = -0.25$ (or $P_{\mathrm{cut}}=0.562$~days) to remove the early-type CB that have $\log P > -0.25$ \citep[also, see][]{rucinski2006}. In Table \ref{tab_cb}, there are three CB with periods longer than $P_{\mathrm{cut}}$ (V21 in M9, V41 in NGC6401 \& V30 in NGC6712). However, \citet{jay2020} proposed to use a period-temperature relation to separate out the early-type and late-type CB. Left panel of Figure \ref{fig_teff} presents the effective temperatures, {\bf $T_{\mathrm{eff}}$,} of the CB in our sample as a function of orbital periods, where the effective temperatures were converted from the extinction corrected $(g-r)$ colors, $T_{\mathrm{eff}}/10^4K = 1.09/[(g-r)+1.47]$ \citep{fukugita2011}, and the solid line is the period-temperature relation, $T_{eff}=6710K - 1760K \log(P/0.5~\mathrm{day})$, given in \citet{jay2020}. Five CB (V8 in M12, KT-23 in M22, V28 in NGC5466, V30 \& V31 in NGC6712) clearly located above this relation, indicating they are early-type CB. On the color-magnitude diagram (CMD, see Figure \ref{fig_cmd}), these early-type CB appeared to occupy the blue stragglers region. Nevertheless, investigation of the connection between CB and blue stragglers is out of the scope of this work. Theoretical and empirical of such investigations can be found in, for example, \citet{stepien2015} and \citet{ferraro2009}, respectively.

\subsection{The $r$-Band PL Relation at Mean Light}\label{sec52}

Figure \ref{fig_iniPL}(a) presents the $r$-band PL relation at mean light for the CB listed in Table \ref{tab_cb} as a representative PL relation. Plots for the PL relation at $r$-band maximum light, as well as in the $g$- and $i$-band, were similar to Figure \ref{fig_iniPL}(a). The five identified early-type CB were marked as open circles in Figure \ref{fig_iniPL}(a), and their PL relation was presented separated in Figure \ref{fig_iniPL}(b). Since the effective wavelength of $r$-band is close to the Gaia's $G$-band,\footnote{The effective wavelengths in $G$- and $r$- and $R$-band are 5857.56\AA, 6156.36\AA~and 6695.58\AA, respectively. These values are adopted from the SVO Filter Profile Service \citep{rod2012,rod2020}.\label{fnote4}} the $G$-band PL relation adopted from \citet{jay2020} was overlaid on Figure \ref{fig_iniPL}(b) as a guidance. Except the longest period early-type CB (V30 in NGC6712), other four early-type CB seems following the $G$-band PL relation for the early-type CB that is different from the late-type CB. Hence, these early-type CB were excluded in the subsequent analysis, as the number of them is too small to fit a meaningful PL relation.   

The $r$-band PL relation at mean light for the remaining late-type CB, shown in Figure \ref{fig_iniPL}(c), exhibits a large scatter. The large scatter of the PL relation can also be seen from globular clusters that hosting 3 or more late-type CB (M4, M9, M22 \& M71). Similar to Figure \ref{fig_iniPL}(b), the $G$- and $R$-band (again, due the effective wavelength of $R$-band is close to the $r$-band)\footnote{See footnote \ref{fnote4}.} PL relations taken from \citet{jay2020} and \citet{chen2018}, respectively, were overlaid on Figure \ref{fig_iniPL}(c). Late-type CB with fainter absolute magnitudes at a given (orbital) period seems to follow the published $G$- or $R$-band PL relation. Since the absolute magnitudes were determined with only three parameters ($m$, $D$ and $E$\footnote{More precisely, using the same {\tt Bayerstar2019} 3D reddening map, as $E$ in general is a function of distance $D$.}), we examined each of them by using alternate values for one of them while keeping the remaining two parameters fixed, in order to determine if any of these parameters cause the observed large scatter of the PL relation. The results were presented in Figure \ref{fig_testpl}.

\begin{figure*}
  \epsscale{1.15}
  \plottwo{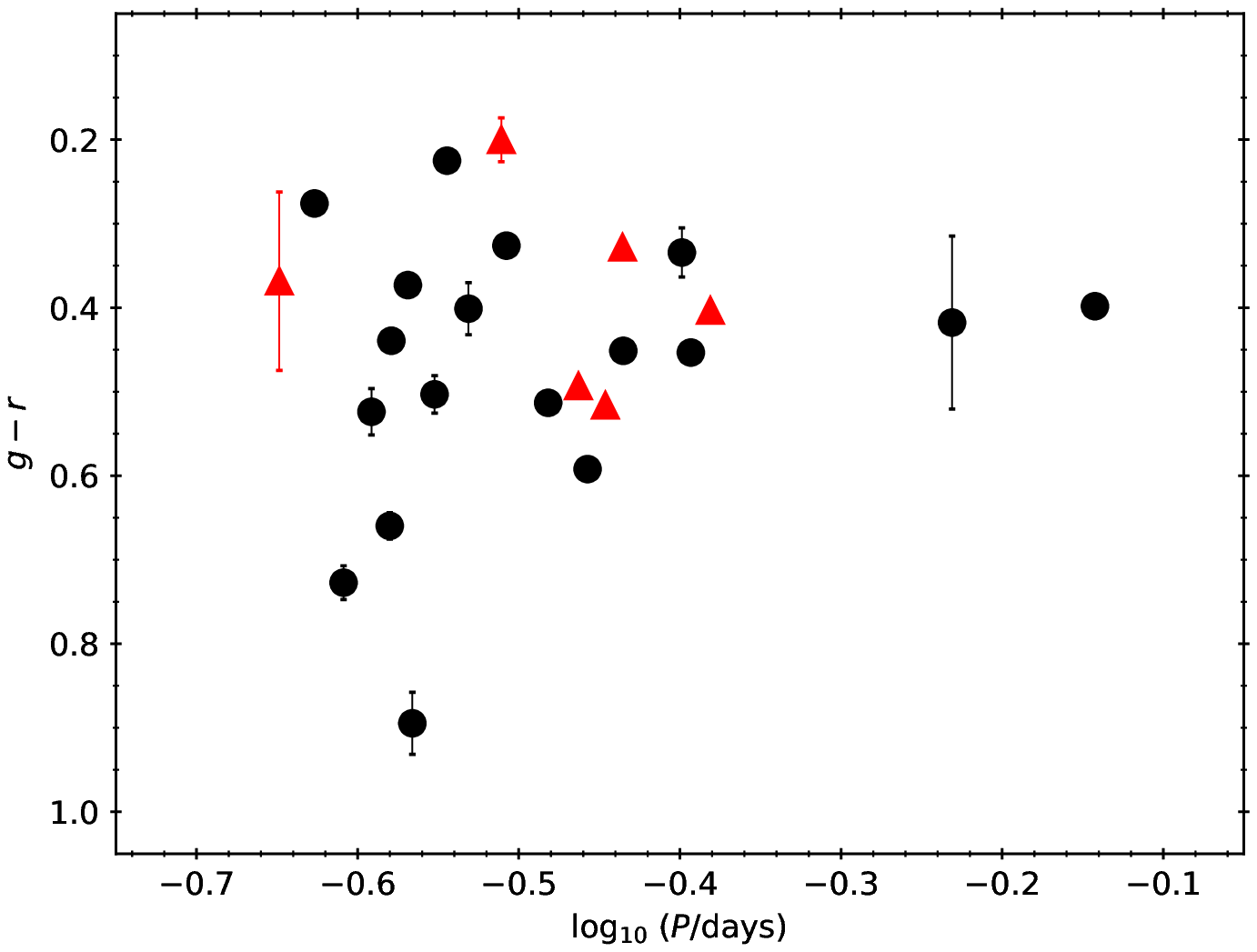}{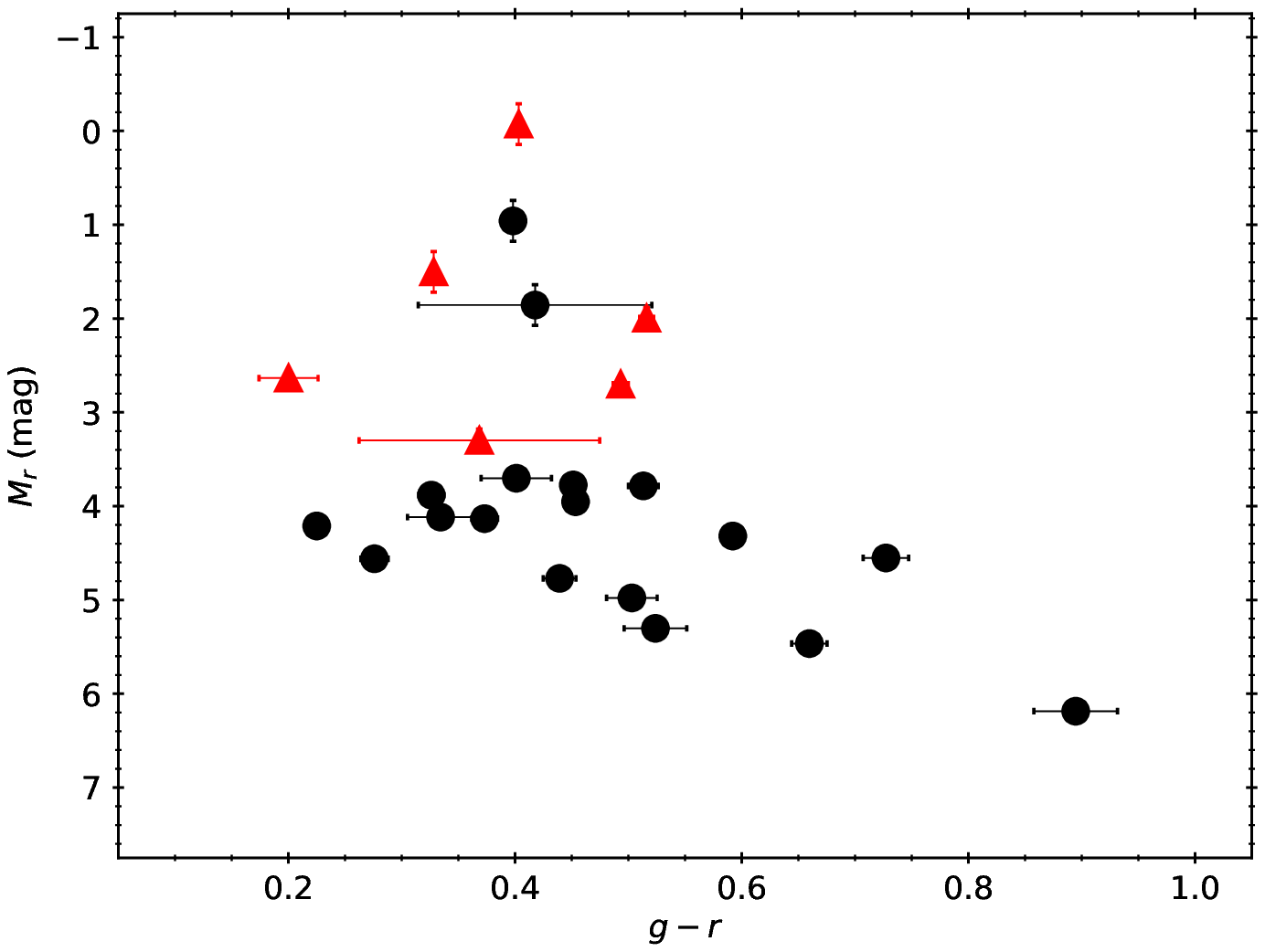}
  \caption{Locations of the 6 rejected CB (red triangles) while fitting the PL relations on the extinction corrected period-color relation (left panel) and the CMD (right panel), against other nominal CB (black circles).}\label{fig_check}
\end{figure*}

{\it Changing the extinction:} Upper-left panel of Figure \ref{fig_testpl} is similar to Figure \ref{fig_iniPL}(c), except the $E(B-V)$ values listed in Table \ref{tab_gc} were used to correct for extinction.\footnote{The $E(B-V)$ values were converted to $E$ using the relation of $E(B-V)=0.884E$ as given in \url{http://argonaut.skymaps.info/usage}.} 

{\it Changing the distance:} Independent distance measurements were available from \citet{wk2017} for six globular clusters (M4, M12, M13, M15, M22 \& M71). In case of M9, a weighted mean of $7.99\pm0.16$~kpc, measured from using the ab- and c-type RR Lyrae \citep{af2013}, was adopted. For the bulge globular cluster NGC6401, the published distances covered a wide range including $\approx 6.35\pm0.81$~kpc \citep[][by using RR Lyrae]{tsapras2017} and $\approx 12.0\pm1.0$~kpc \citep[][by using horizontal branch on $VI$-band CMD]{barbuy1999}. Therefore, these two NGC6410 distances were adopted to test their impact on the PL relation. The resulted PL relation was shown in upper-right panel of Figure \ref{fig_testpl}. Note that V41 in NGC6401 was plotted twice using the two mentioned distances.

{\it Changing the apparent magnitudes:} The Clement's Catalog or the references listed in Table \ref{tab_gc} included the $V$-band magnitudes for the CB in our sample. However, these $V$-band magnitudes were mixed with magnitudes at mean light and at maximum light, and no associated errors were reported. The $V$-band PL relation, either at mean light or at maximum light, was displayed in lower panels of Figure \ref{fig_testpl}, where the ``error bars'' in these plots represent the half amplitude of these CB. The $V$-band PL relation from \citet{jay2020} was overlaid on the lower-left panel, while the $V$-band PL relations from \citet{chen2018} at both mean and maximum light were included in lower-right panel of Figure \ref{fig_testpl} as a guidance.

\begin{figure*}
  \epsscale{1.15}
  \plottwo{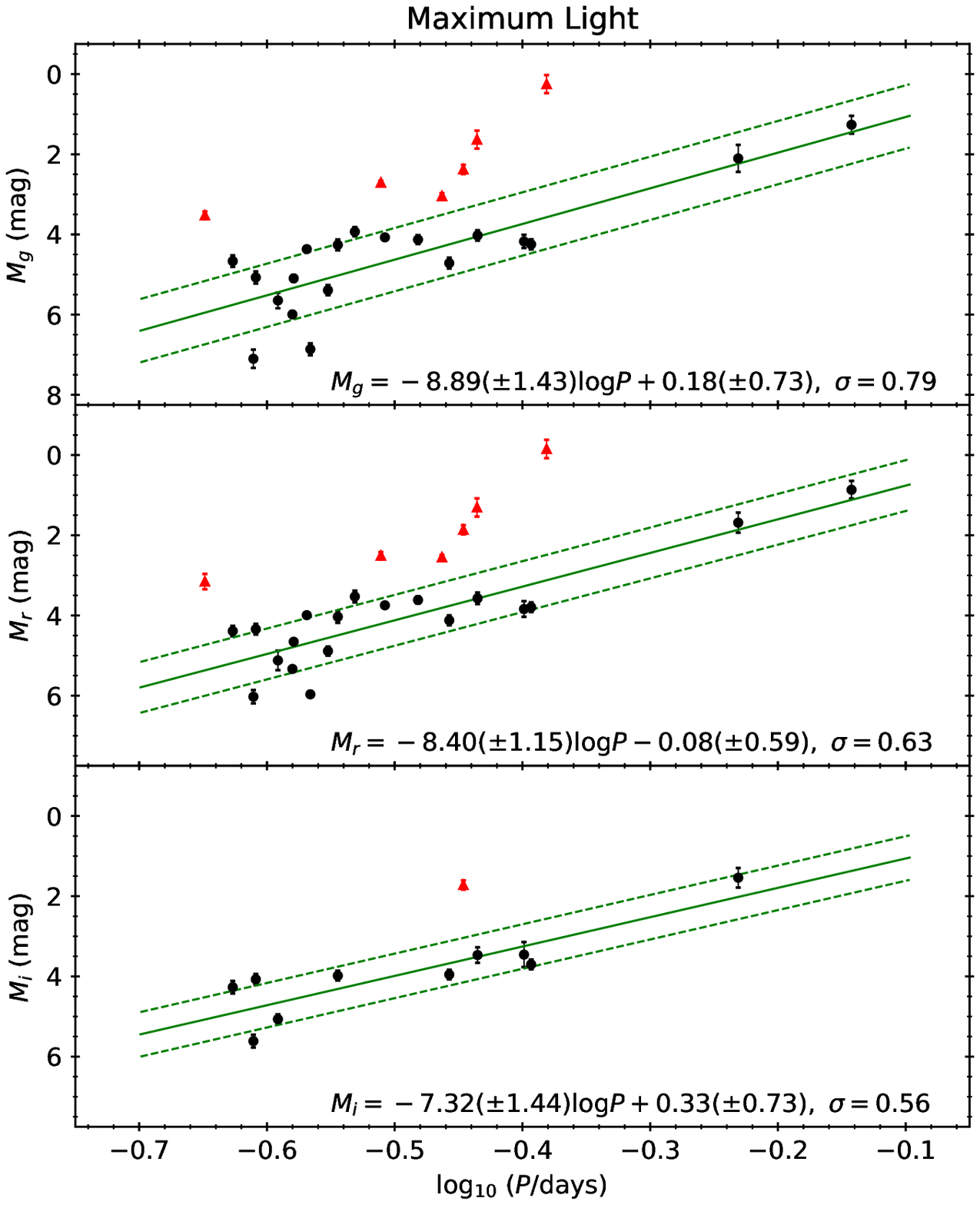}{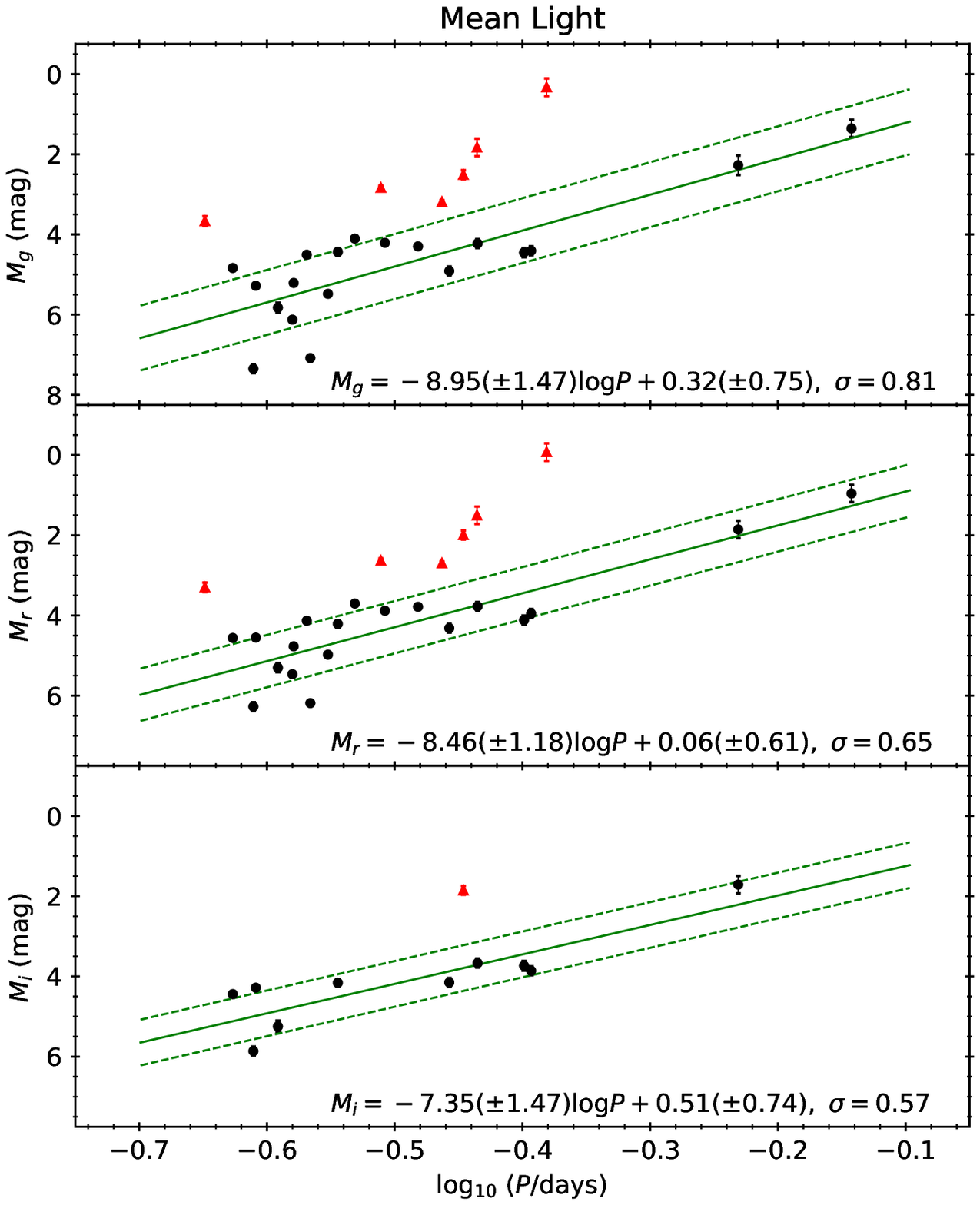}
  \caption{Extinction corrected PL relations of the late-type CB located in globular clusters. The solid lines were fitted $gri$-band PL relations at maximum (left panels) and mean (right panels) light to the final 19 CB in the sample (filled black circles). These PL relations were fitted with a linear regression in the form of $M=a\log P + b$; the fitting results were given in the lower-right corners of each panels, where $\sigma$ is the dispersion of the fitted PL relations (the dashed lines represent the $\pm 1\sigma$ of the PL relations). The filled red triangles were those CB being excluded in the fitting of the PL relations.}\label{fig_plfit}
\end{figure*}

All of the PL relations displayed in Figure \ref{fig_testpl} were similar to Figure \ref{fig_iniPL}(c), which also displayed a large scatter, suggesting this could be intrinsic. There were 6 outliers CB in Figure \ref{fig_iniPL}(c) that are brighter than $\sim 2\sigma$ boundaries of either the \citet{chen2018} or \citet{jay2020} PL relations, and seem to follow a different PL relation. The colors, and hence their temperatures, of these 6 CB do not appear to be anomalous when compared to other nominal CB, as shown in the left panel of Figure \ref{fig_check}. On the CMD, these 6 CB are brighter than majority of the remaining CB and located in the subgiant region (see Figure \ref{fig_cmd} and right panel of Figure \ref{fig_check}), suggesting one of the component in these systems could be a subgiant. A detailed investigation of their nature is beyond the scope of this work. Nevertheless, after removing these 6 outliers CB a PL relation could be fit to the remaining 19 late-type CB.

\subsection{The Derived $gri$-Band Period-Luminosity and Period-Wesenheit Relations}\label{sec53}

The $gri$-band PL relations at both mean and maximum light were fitted to the final sample of 19 CB, as shown in Figure \ref{fig_plfit}. These $gri$-band PL relations at maximum light are:

\begin{eqnarray}
  M_g & = & -8.89(\pm1.43)\log P + 0.18(\pm0.73),\ \sigma=0.79, \\
  M_r & = & -8.40(\pm1.15)\log P - 0.08(\pm0.59),\ \sigma=0.63, \\
  M_i & = & -7.32(\pm1.44)\log P + 0.33(\pm0.73),\ \sigma=0.56,
\end{eqnarray}

\noindent where $\sigma$ is the dispersion of the fitted PL relation. The corresponding PL relations at mean light are:

\begin{eqnarray}
  M_g & = & -8.95(\pm1.47)\log P + 0.32(\pm0.75),\ \sigma=0.81, \\
  M_r & = & -8.46(\pm1.18)\log P + 0.06(\pm0.61),\ \sigma=0.65, \\
  M_i & = & -7.35(\pm1.47)\log P + 0.51(\pm0.74),\ \sigma=0.57.
\end{eqnarray}

\noindent Due to small sample size and possible intrinsic large scatters, errors on the coefficients of the fitted PL relations, and the associated dispersions, were much larger than those presented in \citet{chen2018} and \citet{jay2020}. The two longest period late-type CB could be eliminated based on period cut of $\log P = -0.25$ (see Section \ref{sec51}), or based on their locations on the CMD (i.e. occupying the same subgiant regions as the 6 outlier CB, see Section \ref{sec52}). If they were removed from the sample, the slopes of the fitted PL relations would become shallower, at the same time doubling the errors on both of the fitted slopes and intercepts, even thought the dispersions of the PL relations remained nearly the same. Therefore, we decided to keep these two CB in our sample because they provide important constraints on the fitted PL relations.

\begin{figure*}
  \epsscale{1.15}
  \plottwo{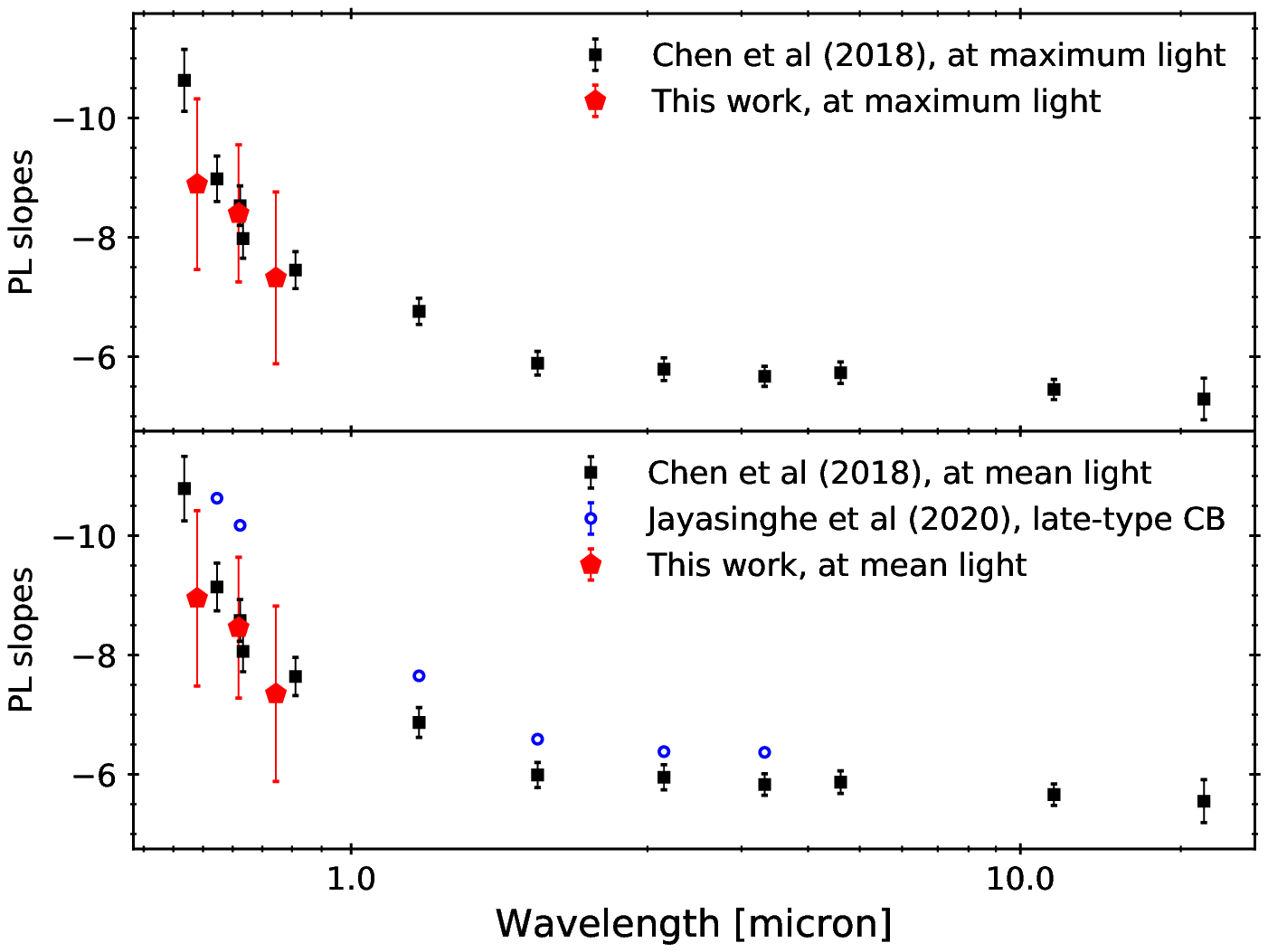}{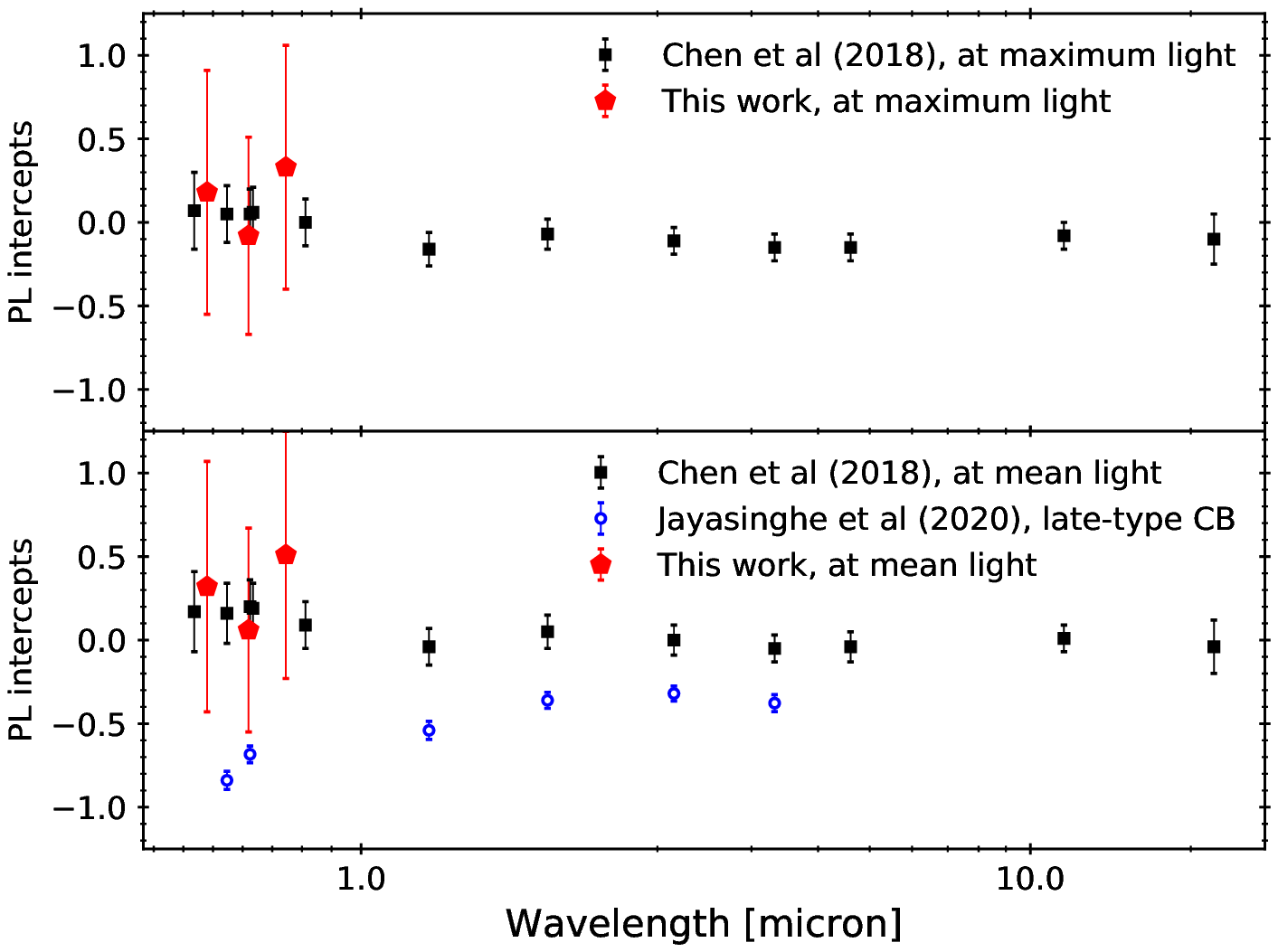}
  \caption{Comparisons of the slopes (left panel) and intercepts (right panel) of the multi-band PL relations from \citet{chen2018} and \citet{jay2020} with the $gri$-band PL relations derived in this work (Figure \ref{fig_plfit}). The intercepts of the PL relations from \citet{jay2020} have been adjusted to the form of $M=A\log P + B$ \citep[instead of $M=A\log P/0.5 + B$ in][]{jay2020}.}\label{fig_plcompare}
\end{figure*}

\citet{chen2018} showed that the slopes of the PL relations exhibit a trend as a function of wavelengths. Albeit their large errors, slopes of the fitted $gri$-band PL relations presented in Figure \ref{fig_plfit} were fully consistent with such trend as demonstrated in the left panel of Figure \ref{fig_plcompare}. Similarly, right panel of Figure \ref{fig_plcompare} showed that the intercepts of the $gri$-band PL relations were in agreement with the multi-band PL relations from \citet{chen2018}. Interestingly, both slopes and intercepts of the PL relations derived in \citet{jay2020} do not agree with those from \citet{chen2018} at any given filters. We also noted the dispersions of the fitted $gri$-band PL relations, given in equation (2) to (7), decrease from the longer wavelength $g$-band filter to the shorter wavelength $i$-band filter, a similar trend is also seen in the PL dispersions reported in \citet{chen2018} and \citet{jay2020}.

\begin{figure*}
  \epsscale{1.15}
  \plottwo{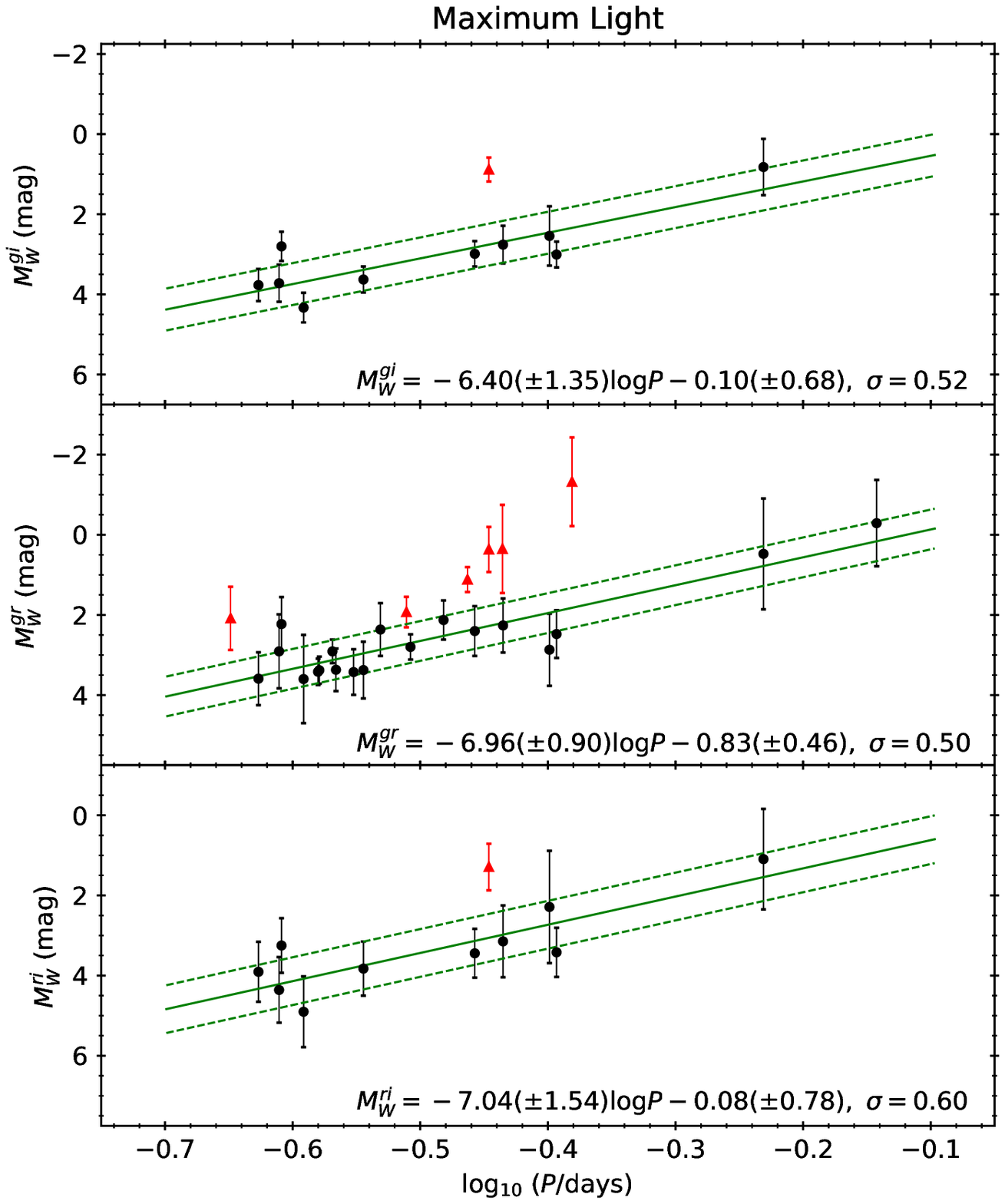}{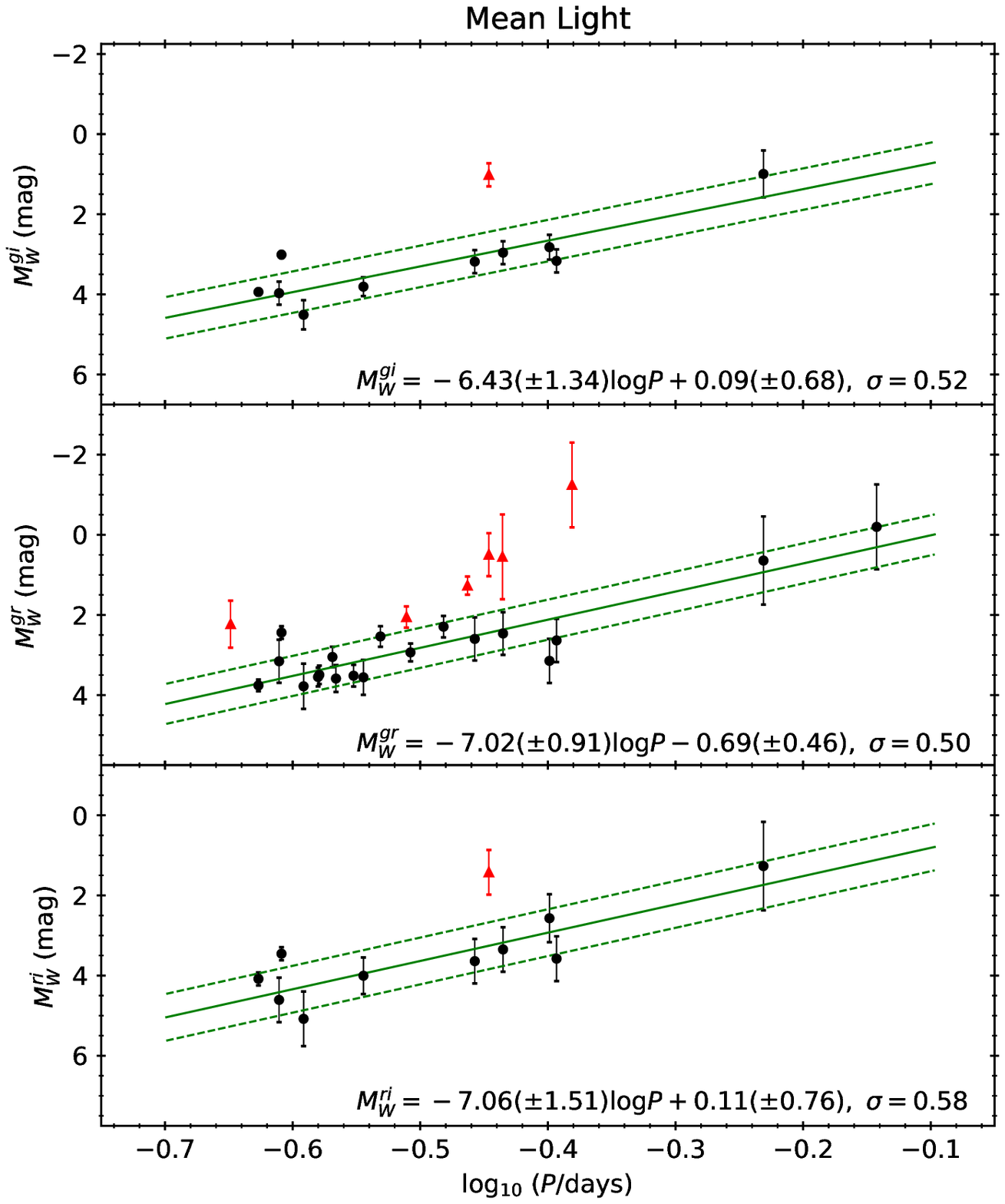}
  \caption{Similar to Figure \ref{fig_plfit}, but for the period-Wesenheit relations at maximum (left panels) and mean (right panels) light. The absolute magnitudes of Wesenheit indexes were calculated as $M_W^c = W^c - 5\log D + 5$, where $c=gi$, $gr$ or $ri$. The filled black circles are the final 19 CB in the sample, and the filled red triangles were those CB being excluded in the fitting of the PW relations.}\label{fig_pwfit}
\end{figure*}

In addition to the $gri$-band PL relations, we have also fitted the period-Wesenheit (PW) relations for the late-type CB in our sample. The extinction-free Wesenheit indexes \citep{madore1991} were defined as:

\begin{eqnarray}
  W^{gi} & = & g - 2.274 (g-i), \\
  W^{gr} & = & r - 2.905 (g-r), \\
  W^{ri} & = & r - 4.051 (r-i). 
\end{eqnarray}

\noindent The resulted PW relations and their linear fittings were presented in Figure \ref{fig_pwfit} and listed below:

\begin{eqnarray}
  M_W^{gi} & = & -6.40(\pm1.35)\log P - 0.10(\pm0.68),\ \sigma=0.52, \\
  M_W^{gr} & = & -6.96(\pm0.90)\log P - 0.83(\pm0.46),\ \sigma=0.50, \\
  M_W^{ri} & = & -7.04(\pm1.54)\log P - 0.08(\pm0.78),\ \sigma=0.60,
\end{eqnarray}

\noindent for the PW relations at maximum light. Similarly, at mean light the PW relations are:

\begin{eqnarray}
  M_W^{gi} & = & -6.43(\pm1.34)\log P + 0.09(\pm0.68),\ \sigma=0.52, \\
  M_W^{gr} & = & -7.02(\pm0.91)\log P - 0.69(\pm0.46),\ \sigma=0.50, \\
  M_W^{ri} & = & -7.06(\pm1.51)\log P + 0.11(\pm0.76),\ \sigma=0.58.
\end{eqnarray}

\noindent In analogue to the $W_{JK}$ PW relation presented in \citet{jay2020}, dispersions of these PW relations were reduced to values between $\sim 0.5$~mag and $\sim0.6$~mag when compared to the $gri$-band PL relations. For example, dispersion of the $W^{gr}$ PW relation was reduced by $\sim38\%$ and $\sim 22\%$ when compared to the dispersions of the $g$- and $r$-band PL relations, respectively.

\begin{figure}
  \epsscale{1.15}
  \plottwo{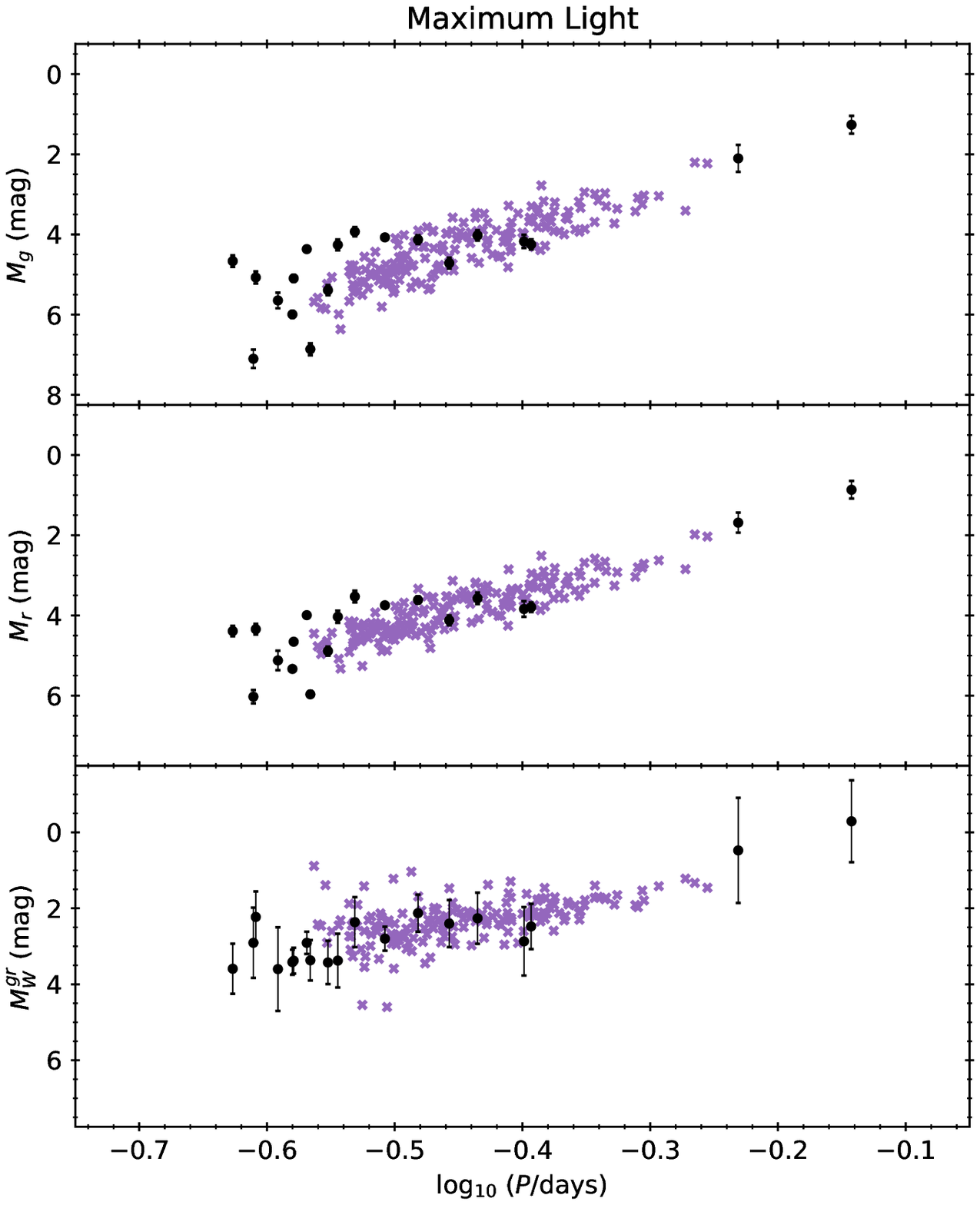}{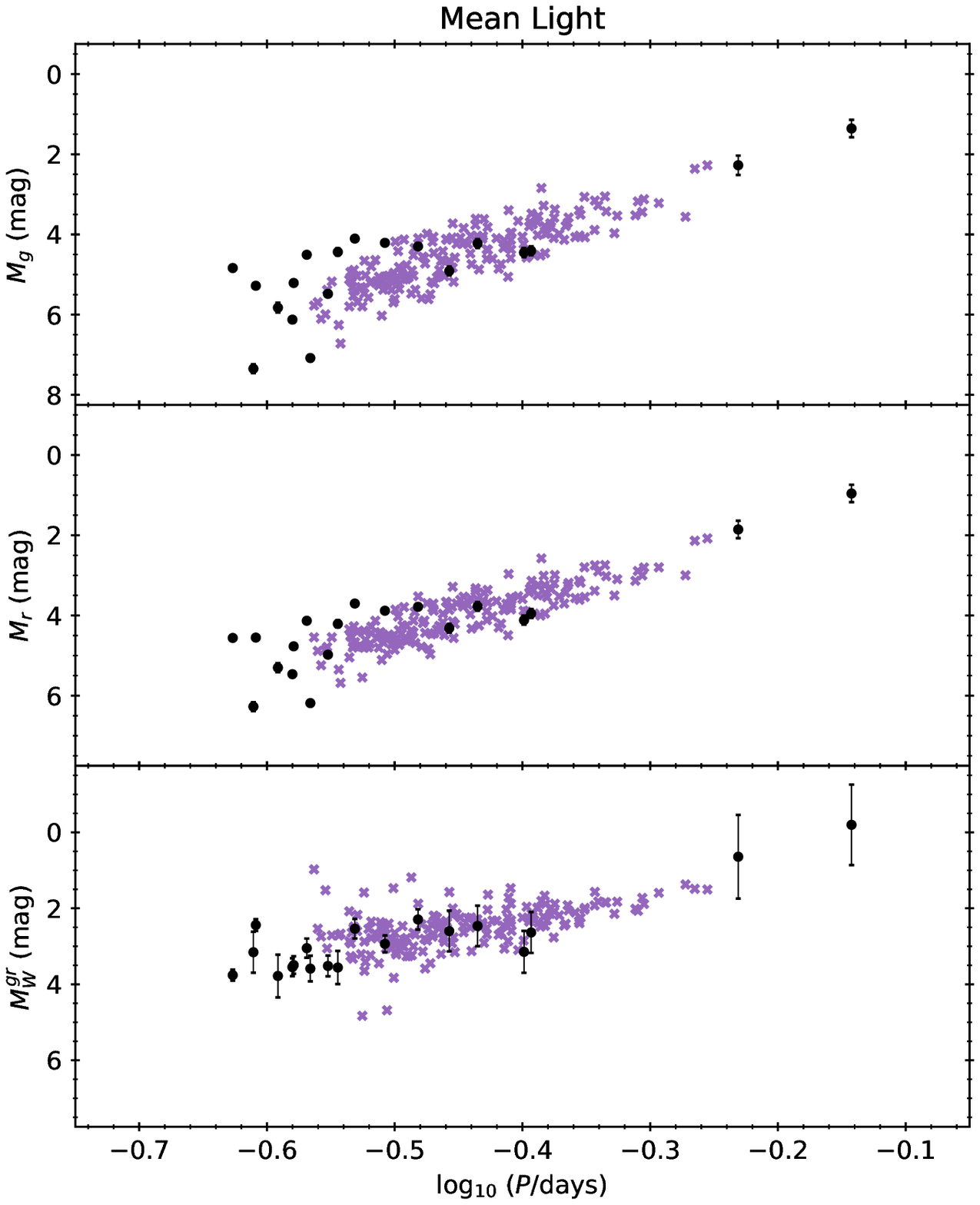}
  \caption{Comparisons of the PL and PW relations at maximum light (left panels) and at mean light (right panels) for the 19 CB in globular clusters (black circles) and the 178 nearby CB (purple crosses) given in \citet{chen2018} after transforming the $BV$-band photometry to the $gr$-band using the transformation provided in \citet{tonry2012}.}\label{fig_plchen}
\end{figure}

Even though about 100 CB in \citet{chen2018} were located within the footprint of ZTF, these nearby sample of CB were too bright for ZTF as observations of ZTF saturate around $\sim 14$~mag. Instead, the $BV$-band photometry of the 183 CB in \citet{chen2018} were transformed to $gr$-band using the quadratic transformation relation given in \citet{tonry2012}. We did not transform the Johnson $I$-band photometric data in \citet{chen2018} to $i$-band because the transformation given in \citet{tonry2012} was in Cousins system, and we tended to avoid ``double transformations''. After transformed the $BV$-band photometry to $gr$-band, we checked the effective temperatures of these CB, following the same procedures done in Section \ref{sec51}. Five CB in the \citet{chen2018} sample turned out to be early-type CB, and hence they were removed from the sample. Top and middle panels of Figure \ref{fig_plchen} compared the $gr$-band PL relations at maximum light (left panels) and at mean light (right panels) between our sample of CB in the globular clusters and the transformed $gr$-band photometry for the rest of the nearby CB sample, at which a good agreement is clearly seen. Combining these two independent sets of CB samples, referred as the combined CB sample, we derive the following PL relations at maximum light:

\begin{eqnarray}
  M_g  & = & -9.32(\pm0.41) \log P + 0.15 (\pm0.19),\ \sigma=0.45, \\
  M_r  & = & -8.20(\pm0.33) \log P + 0.14 (\pm0.15),\ \sigma=0.35. 
\end{eqnarray}

\noindent While the corresponding PL relations at mean light are:

\begin{eqnarray}
  M_g  & = & -9.45(\pm0.43) \log P + 0.26 (\pm0.20),\  \sigma=0.47, \\
  M_r  & = & -8.34(\pm0.34) \log P + 0.25 (\pm0.16),\  \sigma=0.37. 
\end{eqnarray}

\noindent Errors on the slopes and intercepts on these PL relations, derived from the combined CB sample, are much reduced when compared to their counterparts based on the 19 CB in globular clusters. The PL dispersions derived from the combined CB sample were also reduced by $\sim42\%$ to $\sim44\%$ in comparison to the PL dispersions using only the 19 CB in globular clusters.

In contrast, data points based on the \citet{chen2018} sample exhibit a larger scatter on the PW relations when comparing to the PL relations (especially around $\log P \sim -0.5$), as demonstrated in the bottom panels of Figure \ref{fig_plchen}. After applying an iterative $2.5\sigma$-clipping algorithm, we derived the following PW relation at maximum light:

\begin{equation}
  M_W^{gr} =  -5.16(\pm0.33) \log P + 0.06 (\pm0.15),\sigma=0.32, 
\end{equation}

\noindent Similarly, the PW relation at mean light is:

\begin{equation}
  M_W^{gr} =  -5.26(\pm0.34) \log P + 0.17 (\pm0.15),\sigma=0.33. 
\end{equation}

\noindent These $M_W^{gr}$ PW relations show a substantial improvement over the PW relations only based on the 19 CB in globular clusters as given in equation (12) and (15). The dispersions of $0.32$~mag and $0.33$~mag for these $M_W^{gr}$ PW relations are also the smallest when compared to the $gr$-band PL relations derived from the combined CB sample, and comparable to the dispersions of the $G$- and $R$-band PL relations given in \citet{chen2018} and the $G$-band PL relation from \citet{jay2020}.

\section{The Period-Luminosity-Color Relations}\label{sec6}

\begin{figure*}
  \epsscale{1.15}
  \plottwo{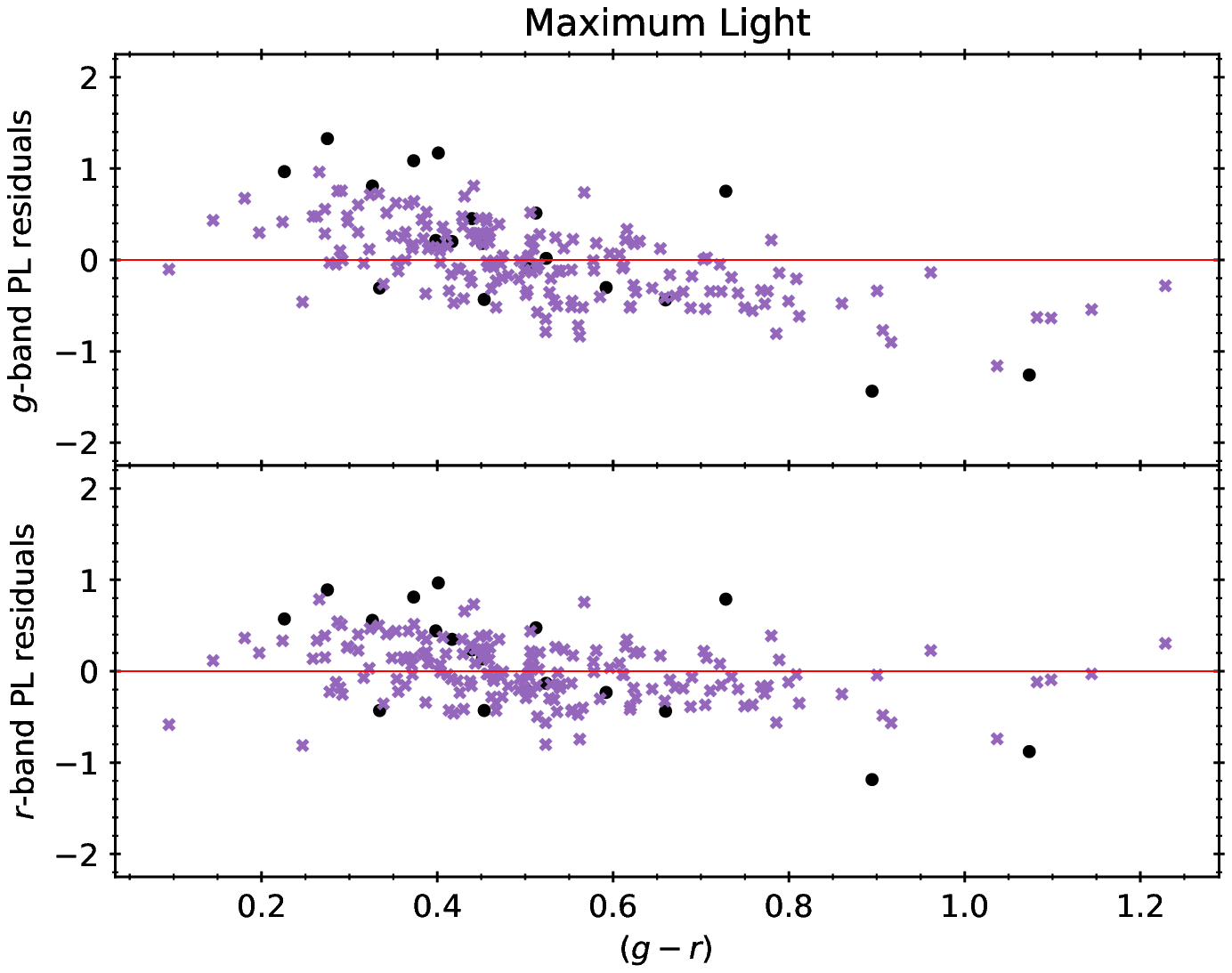}{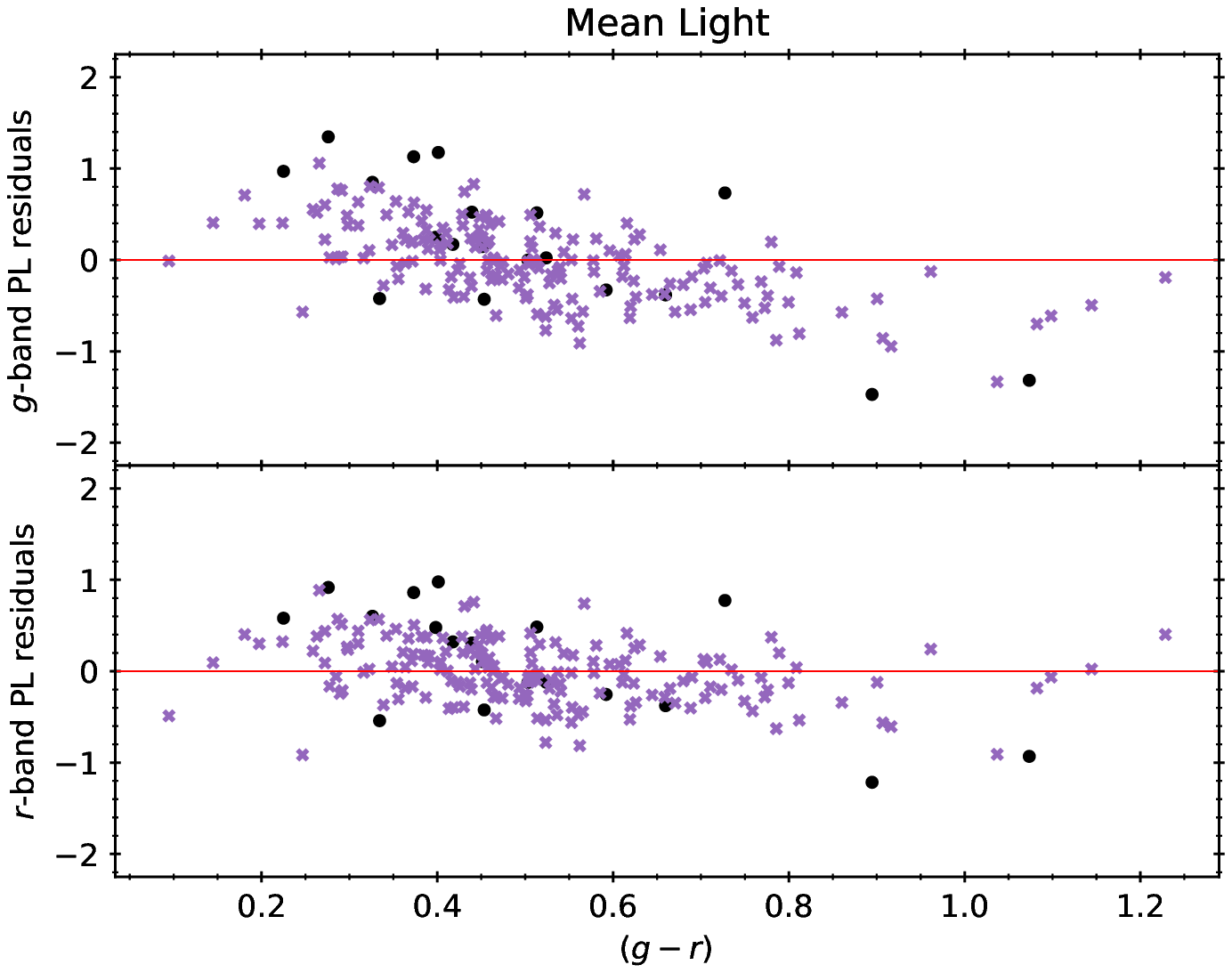}
  \caption{Residuals of the PL relations at maximum light (left panels) and at mean light (right panels) as a function of extinction corrected $(g-r)$ colors. The (red) horizontal lines represent zero residuals. In all panels, the 19 CB in the globular clusters and the 178 nearby CB were marked as black circles and purple crosses, respectively.}\label{fig_res}
\end{figure*}

\begin{figure*}
  \epsscale{1.15}
  \plottwo{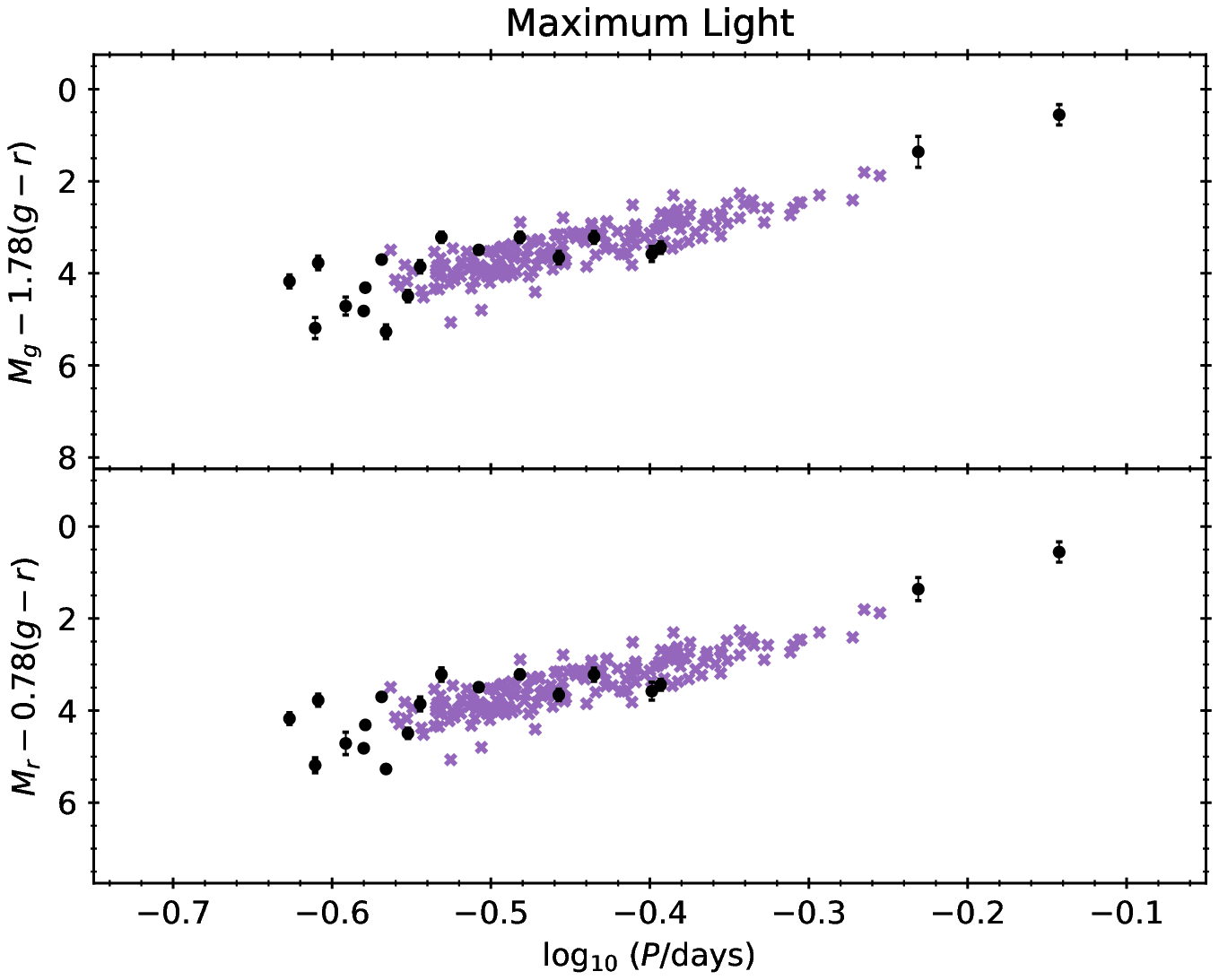}{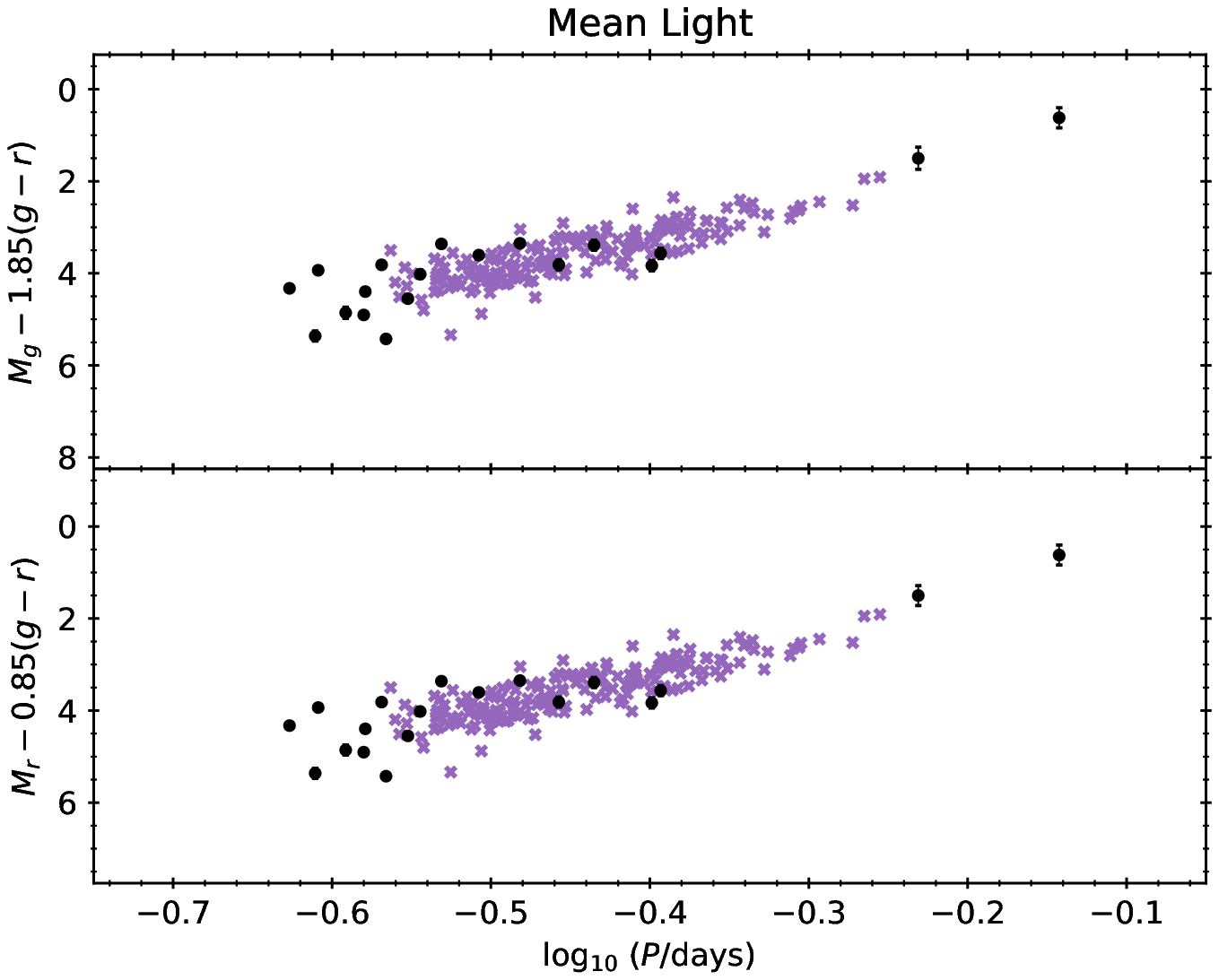}
  \caption{Color-corrected PL relations at maximum light (left panels) and at mean light (right panels). In all panels, the 19 CB in the globular clusters and the 178 nearby CB were marked as black circles and purple crosses, respectively.}\label{fig_plc}
\end{figure*}

Earlier works mentioned in the Introduction focused on the derivation of PLC relations. This is because from theoretical consideration the effective temperature of the CB system, and hence the color term, is expected to be presence in the correlation between the orbital period and absolute magnitudes for the CB \citep[for example, see][]{rucinski1994,rucinski1997}, although other works \citep{mateo2017} indicated the color term might not need to be included. Indeed, residuals of the PL relations derived in Section \ref{sec53} were well (anti-)correlated with the $(g-r)$ colors, as shown in Figure \ref{fig_res}. The slopes were found to be $-1.47(\pm0.14)$~mag and $-0.67(\pm0.13)$~mag for the $g$- and $r$-band mean light PL relations, respectively. Similarly, slopes of the residuals for the $g$- and $r$-band maximum light PL relations are $-1.42(\pm0.13)$~mag and $-0.63(\pm0.12)$~mag, respectively. Including the $(g-r)$ color term in the linear regression, we obtained the following extinction corrected PLC relation at maximum light with the combined CB sample:

\begin{equation}
  M_{(g,r)}  =  0.13 (\pm0.14) - 7.32(\pm0.34)\log P + \alpha_{(g,r)}(g-r),
\end{equation}

\noindent where $\alpha_g = 1.78(\pm0.14)$ and $\alpha_r = 0.78(\pm0.14)$. Both of the derived PLC relation carries a same dispersion of $0.33$~mag, which is similar to the PW relations. The corresponding PLC relation at mean light is: 

\begin{equation}
  M_{(g,r)}  =  0.24 (\pm0.14) - 7.39(\pm0.35)\log P + \alpha_{(g,r)}(g-r),
\end{equation}

\noindent where $\alpha_g = 1.85(\pm0.14)$ and $\alpha_r = 0.85(\pm0.14)$, with the same dispersion of $0.34$~mag. It is worth to point out that the PLC relations at maximum and mean light, presented in equation (23) and (24) respectively, are consistent to each others, and both slopes and color terms ($\alpha$) are the same within the errors. Furthermore, dispersons of these PLC relations are similar to the dispersions of the $M_W^{gr}$ PW relations derived in Section \ref{sec53}. Figure \ref{fig_plc} presents the PL relations after removing the color term, at which the reduction of the dispersion on the $g$-band PL relation can be clearly seen.

\section{An Example of Application}\label{sec7}

\begin{figure*}
  \epsscale{1.15}
  \plottwo{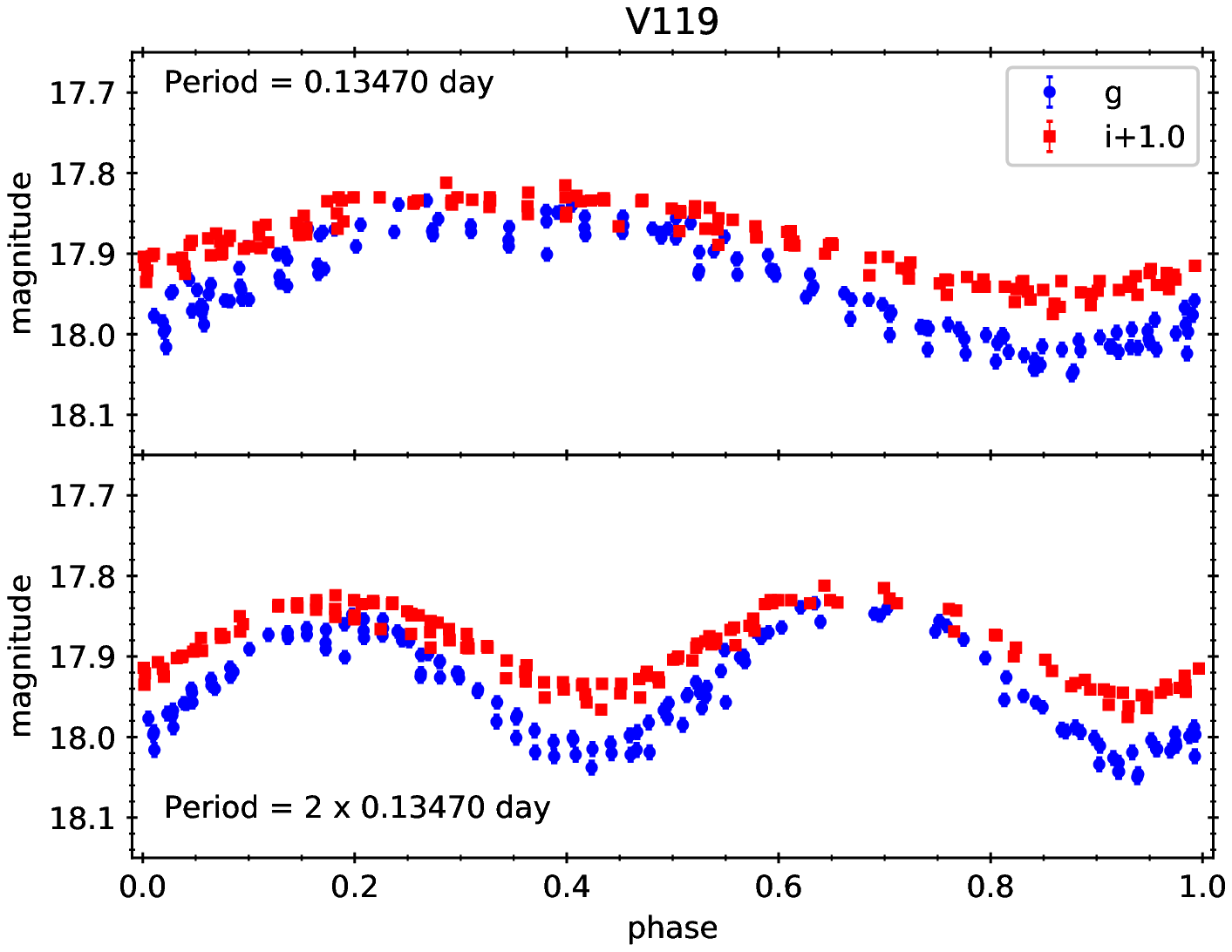}{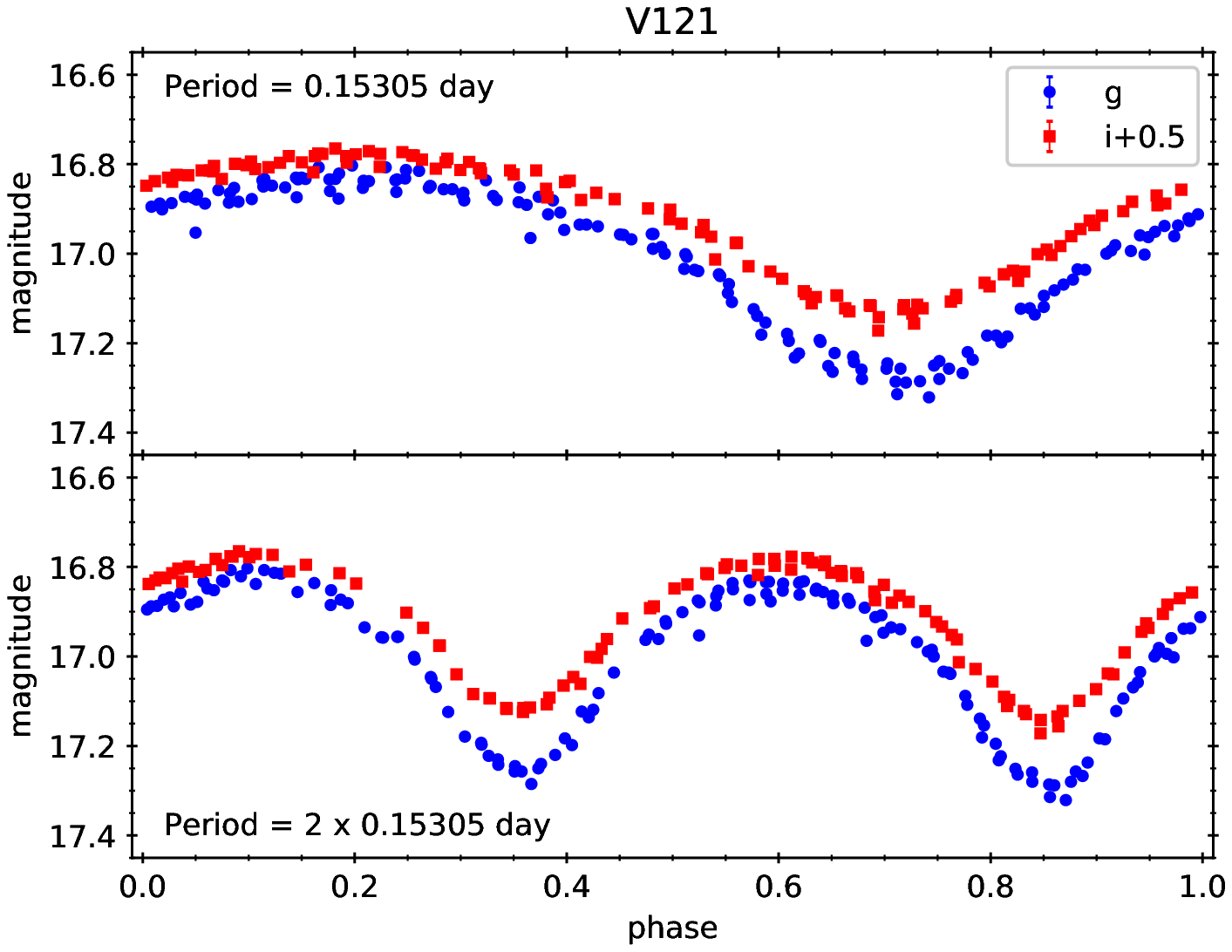}
  \plottwo{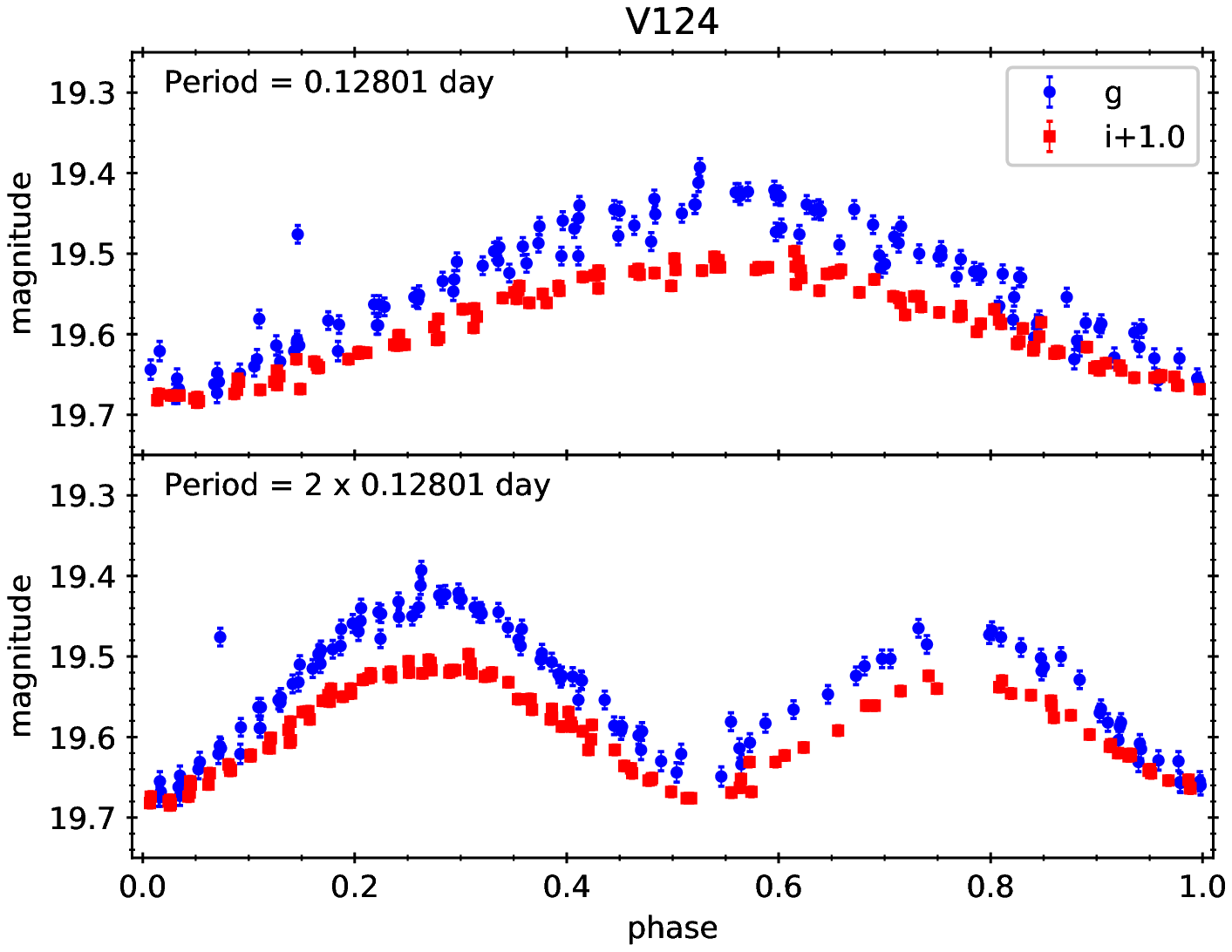}{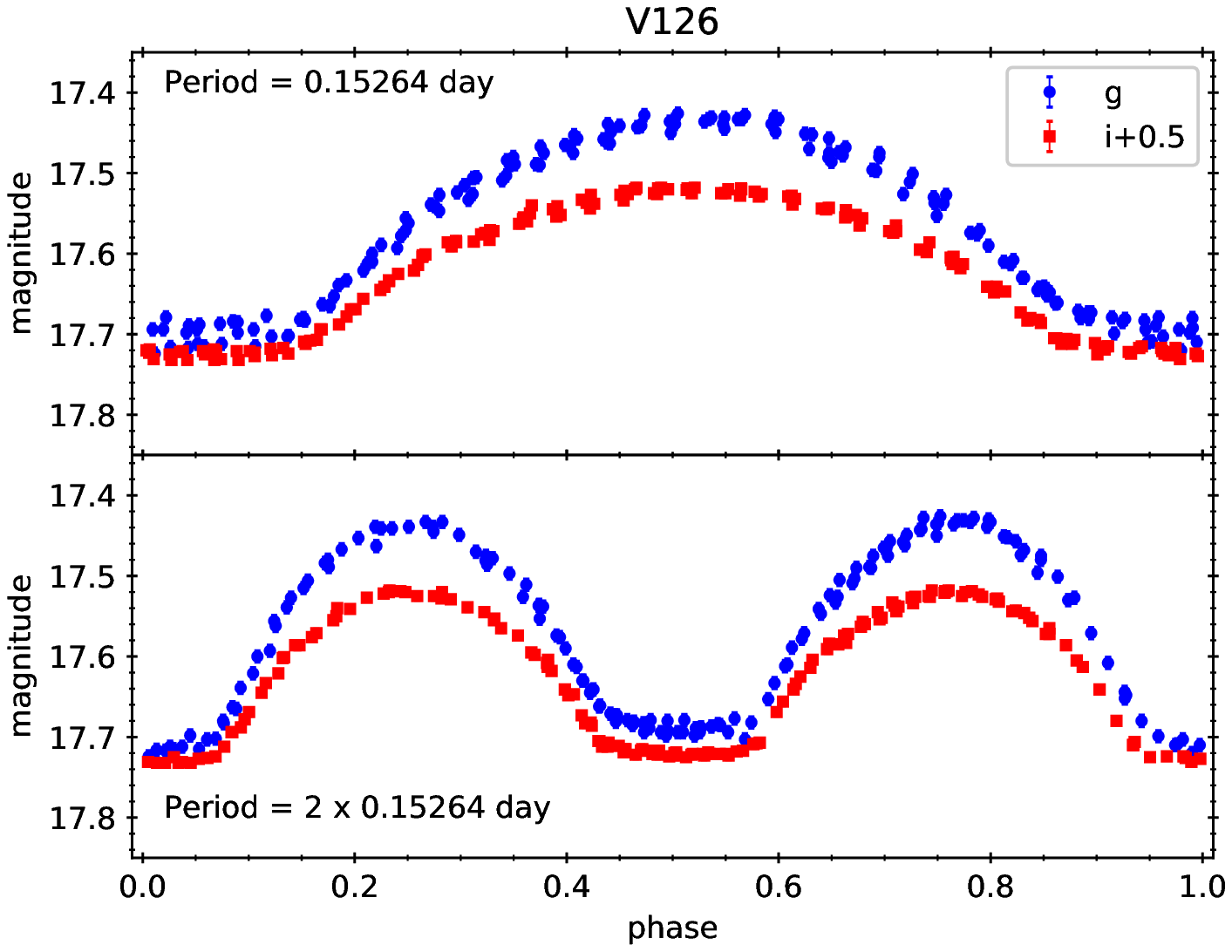}
  \caption{Light curves for the four CB reported in \citet{vivas2020} at which their published periods need to be doubled. The top panels of each sub-figures are light curves folded with periods given in \citet{vivas2020}, while the bottom panels are the folded light curves with twice of the published periods. }\label{fig_vivas}
\end{figure*}

Recently, \citet{vivas2020} conducted time-series observations in $gi$-band on a dwarf galaxy Crater II using the DECam installed on the 4m Blanco Telescope. These authors found 14 eclipsing binaries within the survey area. Based on their locations on the CMD, the authors concluded these eclipsing binaries are foreground objects. The $g$-band PL relation given in equation (19) can be used to confirm foreground nature of them. Visual inspecting the light curves of these eclipsing binaries revealed 6 of them (V118, V119, V120, V121, V124 \& V126) have CB-like light curves. As illustrated in Figure \ref{fig_vivas}, the reported periods in \citet{vivas2020} need to be doubled for V119, V121, V124 \& V126, because CB light curves are expected to exhibit two minima with nearly equal depth. Since the photometry of these DECam observations were calibrated to the SDSS (Sloan Digital Sky Survey) system \citep{vivas2020}, we converted the mean $g$-band magnitudes for these six CB to the Pan-STARRS1 (PS1) system via the transformation equations provided in \citet{abbott2021}. We did not convert the corresponding mean $i$-band magnitudes because \citet{vivas2020} observations did not include the $r$-band, hence no $(r-i)$ colors available to perform such transformation. Assuming these eclipsing binaries are late-type CB, using the extinction provided by \citet{vivas2020} and the $g$-band PL relation given in equation (19) we found that the distance moduli of these 6 CB ranged from 11.28~mag to 13.48~mag (with errors of $\sim 0.45$~mag, dominated by the dispersion of the $g$-band PL relation), which are much closer than the distance modulus of Crater II \citep[$20.33\pm0.07$~mag,][]{vivas2020}. Therefore, the foreground nature of these eclipsing binaries were confirmed.

\section{Conclusion}\label{sec8}

In this work, we present, for the first time, the $gr(i)$-band PL, PW and PLC relations for the late-type CB. We started a sample of 79 CB in 15 globular clusters visible from the Palomar Observatory, however about three quarters of them were eliminated due to either extrinsic (CB with no or lack of ZTF data, and those affected by blending) or intrinsic (CB belongs to early-type, and the 6 outliers CB) reasons. The remaining 19 CB (in 8 globular clusters), at which the number is same or similar to some early work \citep{rucinski1994,rucinski97a}, were used to fit the $gri$-band PL relations and the PW relations. Even though the errors on the coefficients (slopes and intercepts) and the dispersions of the PL (and PW) relations for this sample of CB were several times larger than the published PL relations at similar filters, these slopes and intercepts of the $gri$-band PL relations were consistent to the local sample of CB as derived in \citet{chen2018}, suggesting CB in globular clusters is a viable and yet independent route to calibrate and/or cross-check the PL relations. By transforming the $BV$ photometry of the local CB samples in \citet{chen2018} to $gr$-band, the PL and PW relations for these two independent CB samples were shown to be in good agreements, and hence both samples of CB can be combined to improve the derivation of the PL and PW relations in the $gr$-band, as the dispersions of these relations were greatly reduced when using the combined CB sample. We have also derived the PLC relations using the combined CB sample as the residuals of the PL relations were well (anti-)correlated to the $(g-r)$ colors. Finally, we applied the $g$-band PL relation to derive the distance moduli of 6 CB found in \citet{vivas2020}, proving that they were all foreground objects and unrelated to the dwarf galaxy Crater II. We envision these relations, especially the $gr$-band PW and PLC relations that exhibit the smallest dispersion, will be useful in future distance scale work to trace distances using the late-type CB, for example, located in the Galactic halo or dwarf galaxies. We expect a large number of late-type CB to be discovered in time-domain surveys that include the $gr(i)$ filters, such as the SkyMapper Southern Survey \citep{onken2019}, the Dark Energy Survey \citep{des2016}, and the Vera C. Rubin Observatory's LSST \citep{ivezic2019}, in the near future. The globular clusters based $gri$-band PL, PW and PLC relations could be potentially improved by increasing the number of (late-type) CB in globular clusters from future ZTF data releases, or dedicated observations with larger aperture telescopes.

\acknowledgments

We thank the useful discussions and comments from an anonymous referee to improve the manuscript. We thank the funding from Ministry of Science and Technology (Taiwan) under the contract 107-2119-M-008-014-MY2, 107-2119-M-008 012, 108-2628-M-007-005-RSP and 109-2112-M-008-014-MY3.

Based on observations obtained with the Samuel Oschin Telescope 48-inch Telescope at the Palomar Observatory as part of the Zwicky Transient Facility project. Major funding has been provided by the U.S. National Science Foundation under Grant No. AST-1440341 and by the ZTF partner institutions: the California Institute of Technology, the Oskar Klein Centre, the Weizmann Institute of Science, the University of Maryland, the University of Washington, Deutsches Elektronen-Synchrotron, the University of Wisconsin-Milwaukee, and the TANGO Program of the University System of Taiwan. 

This research has made use of the SIMBAD database and the VizieR catalogue access tool, operated at CDS, Strasbourg, France. This research made use of Astropy,\footnote{\url{http://www.astropy.org}} a community-developed core Python package for Astronomy \citep{astropy2013, astropy2018}.

This research has made use of the SVO Filter Profile Service (\url{http://svo2.cab.inta-csic.es/theory/fps/}) supported from the Spanish MINECO through grant AYA2017-84089

\facility{PO:1.2m}

\software{{\tt astropy} \citep{astropy2013,astropy2018}, {\tt gatspy} \citep{vdp2015}, {\tt Matplotlib} \citep{hunter2007},  {\tt NumPy} \citep{harris2020}, {\tt SciPy} \citep{virtanen2020}}

\appendix

\section*{ZTF Light Curves for the 30 Contact Binaries}

\begin{figure*}
  %\epsscale{2}
  \begin{tabular}{ccc}
    \includegraphics[scale=0.34]{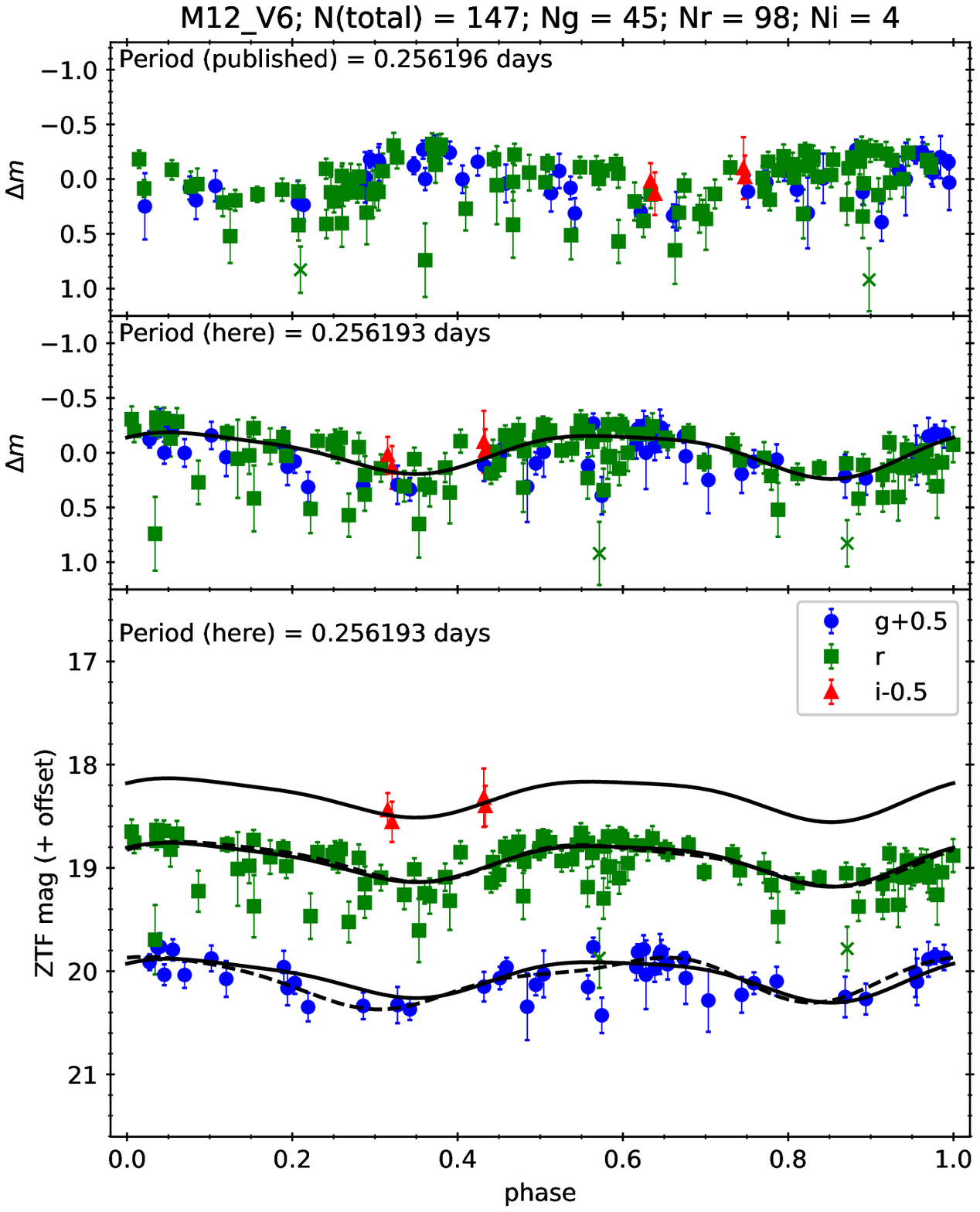} &  \includegraphics[scale=0.34]{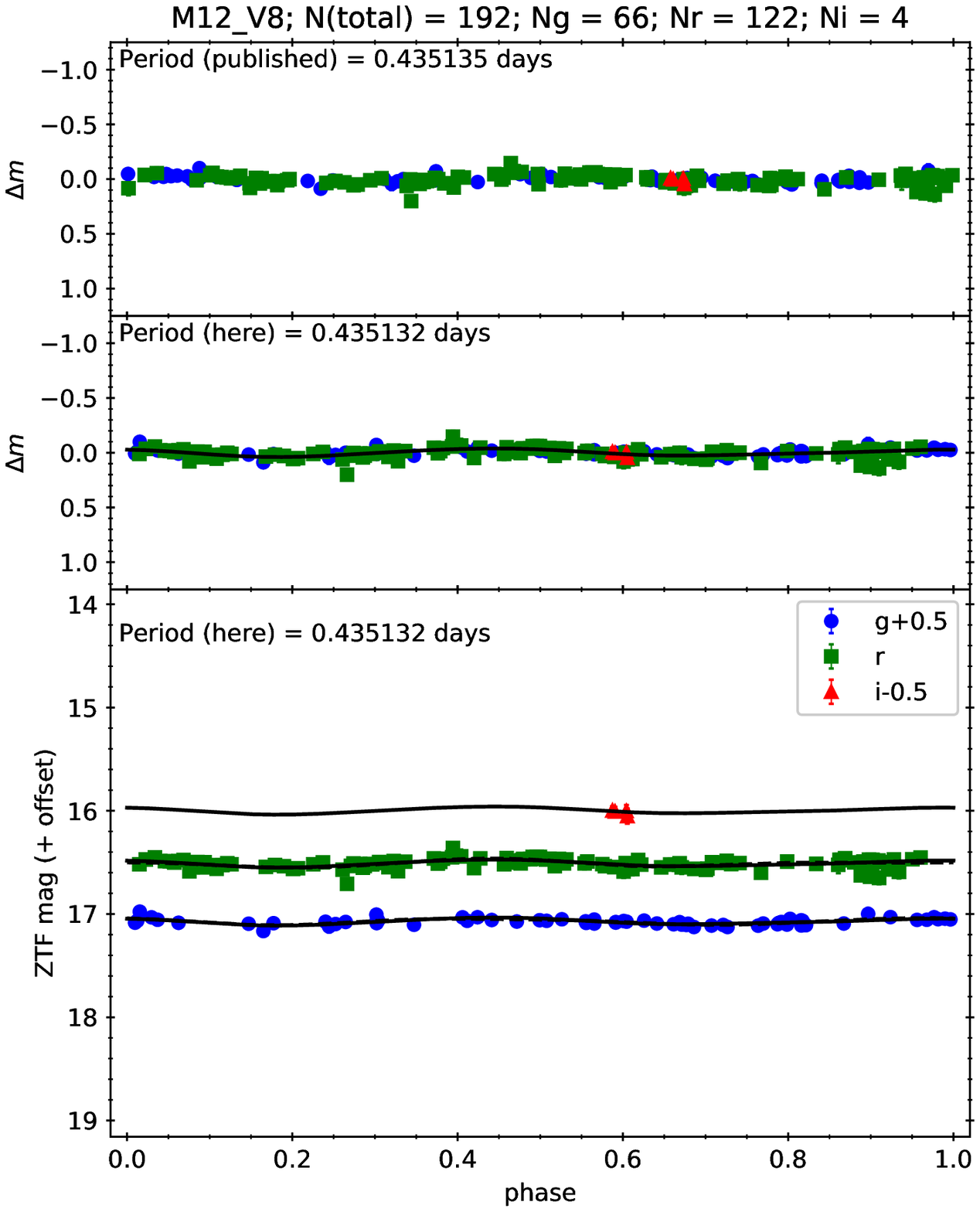} &  \includegraphics[scale=0.34]{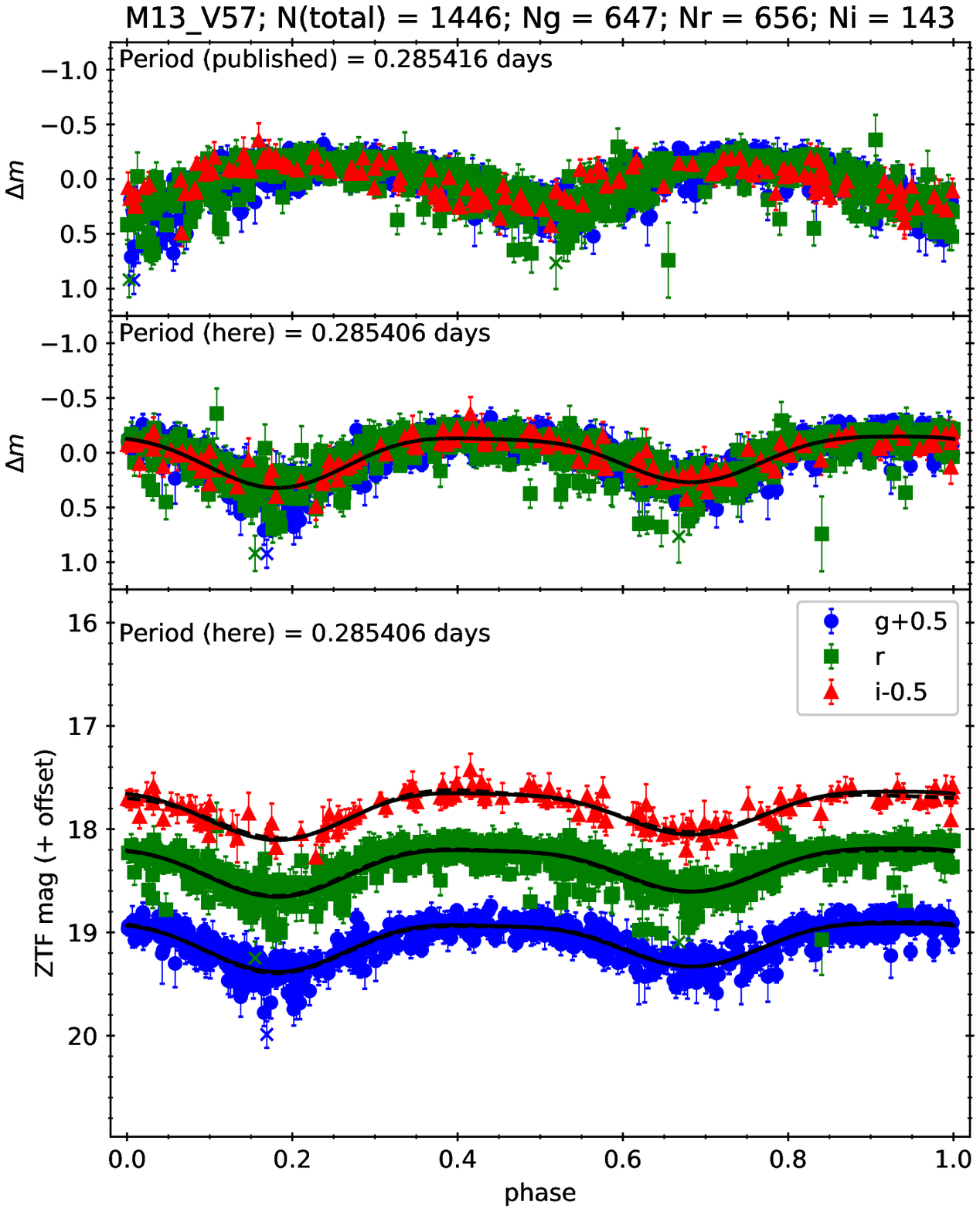} \\
    \includegraphics[scale=0.34]{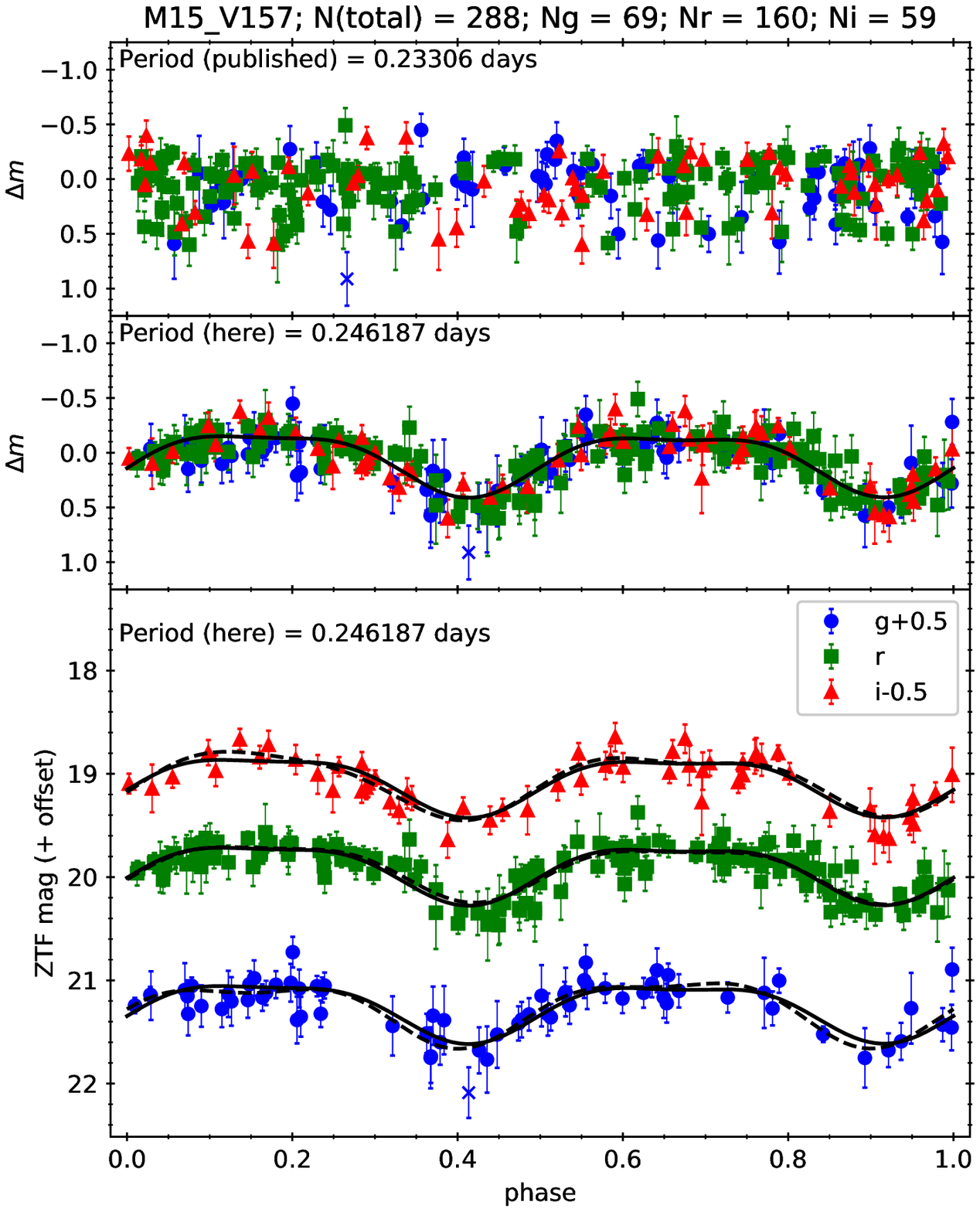} &  \includegraphics[scale=0.34]{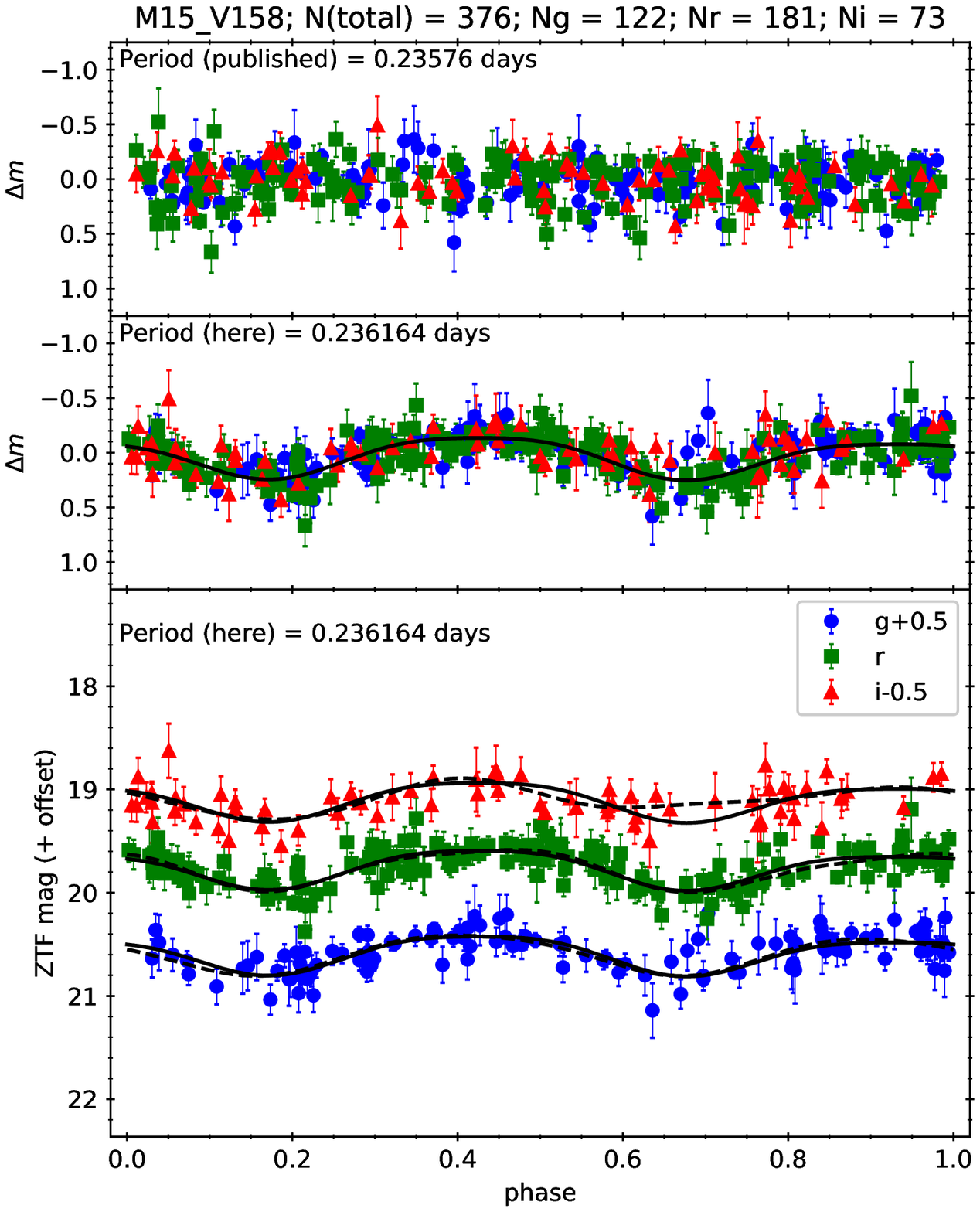} &  \includegraphics[scale=0.34]{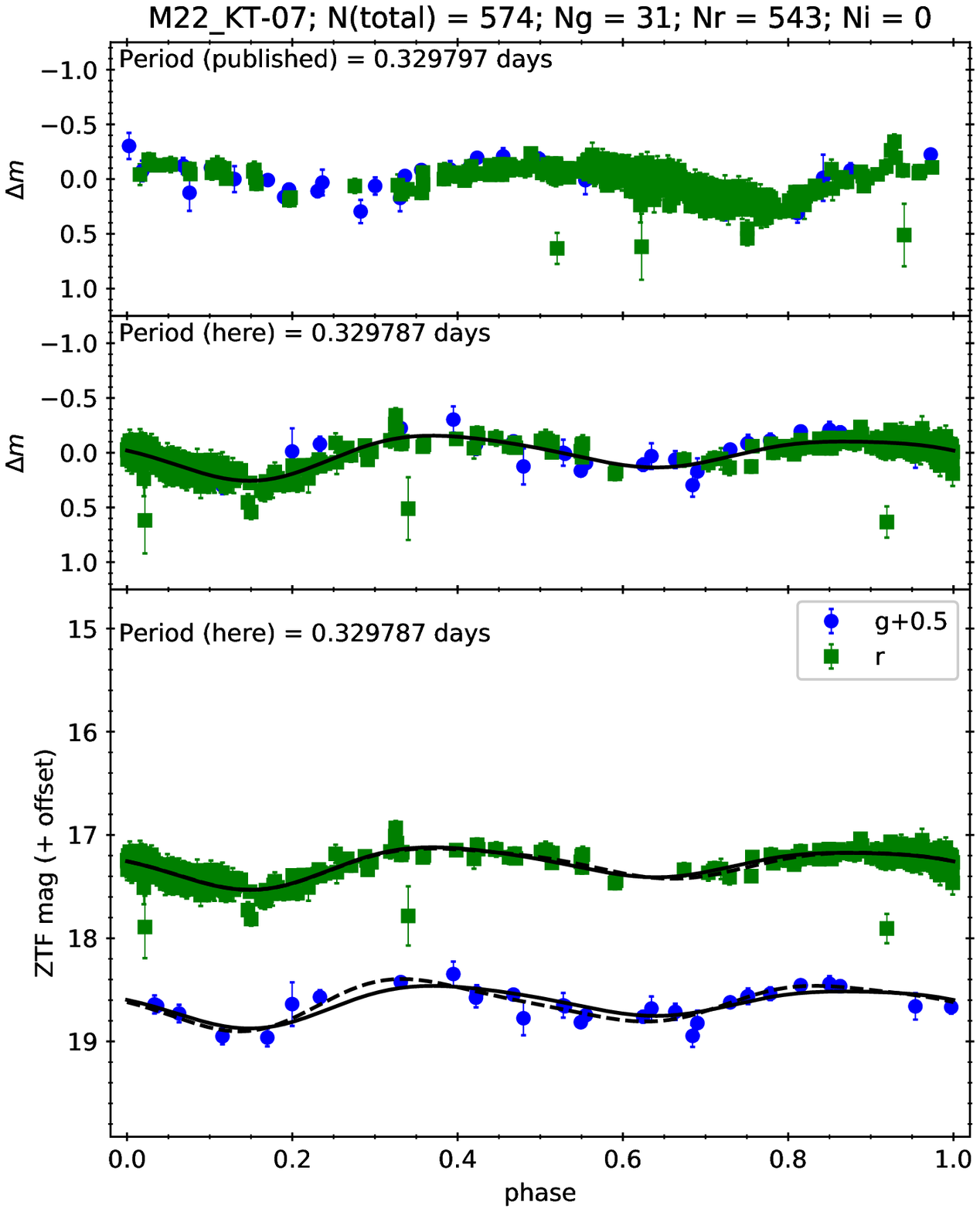} \\
    \includegraphics[scale=0.34]{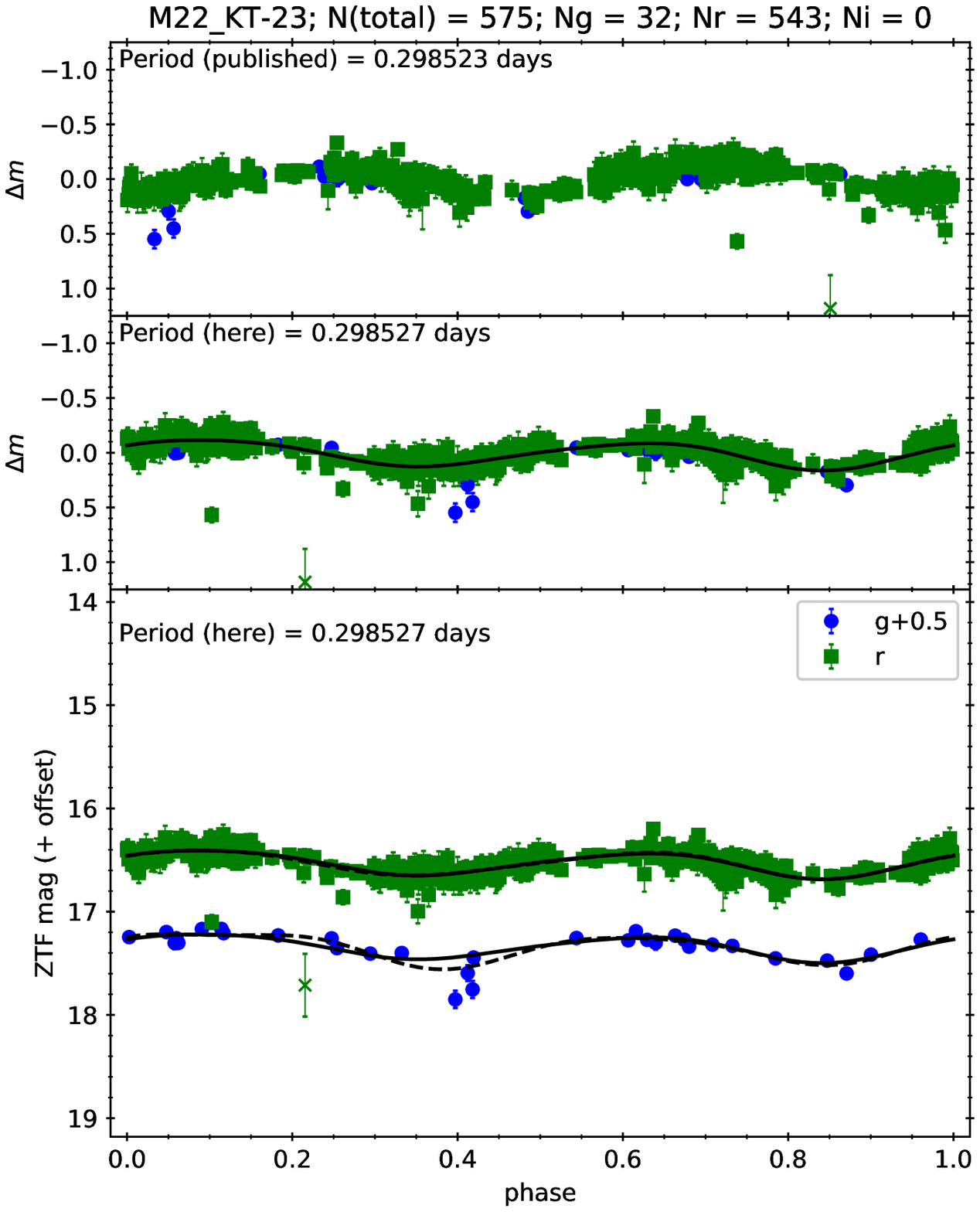} &  \includegraphics[scale=0.34]{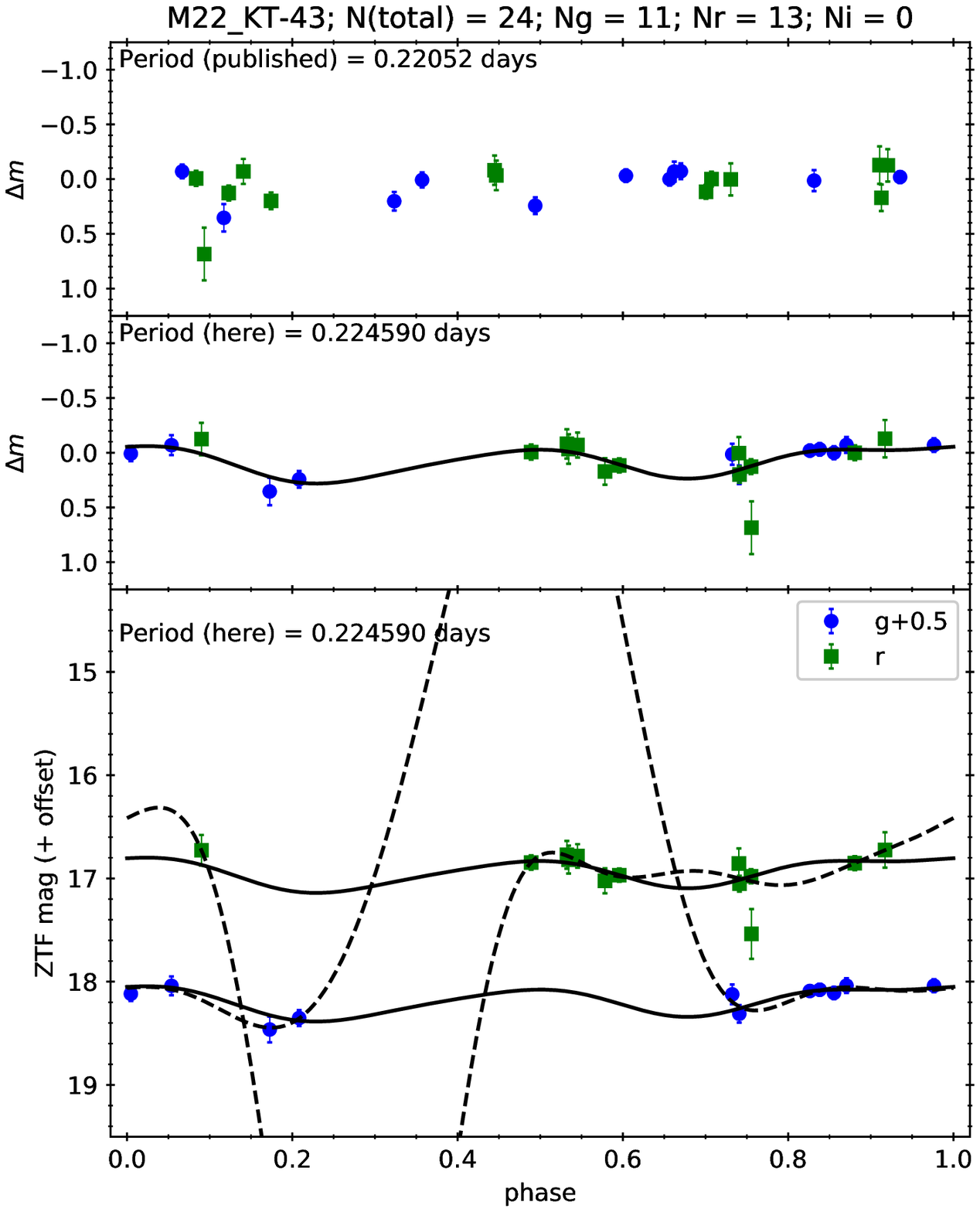} &  \includegraphics[scale=0.34]{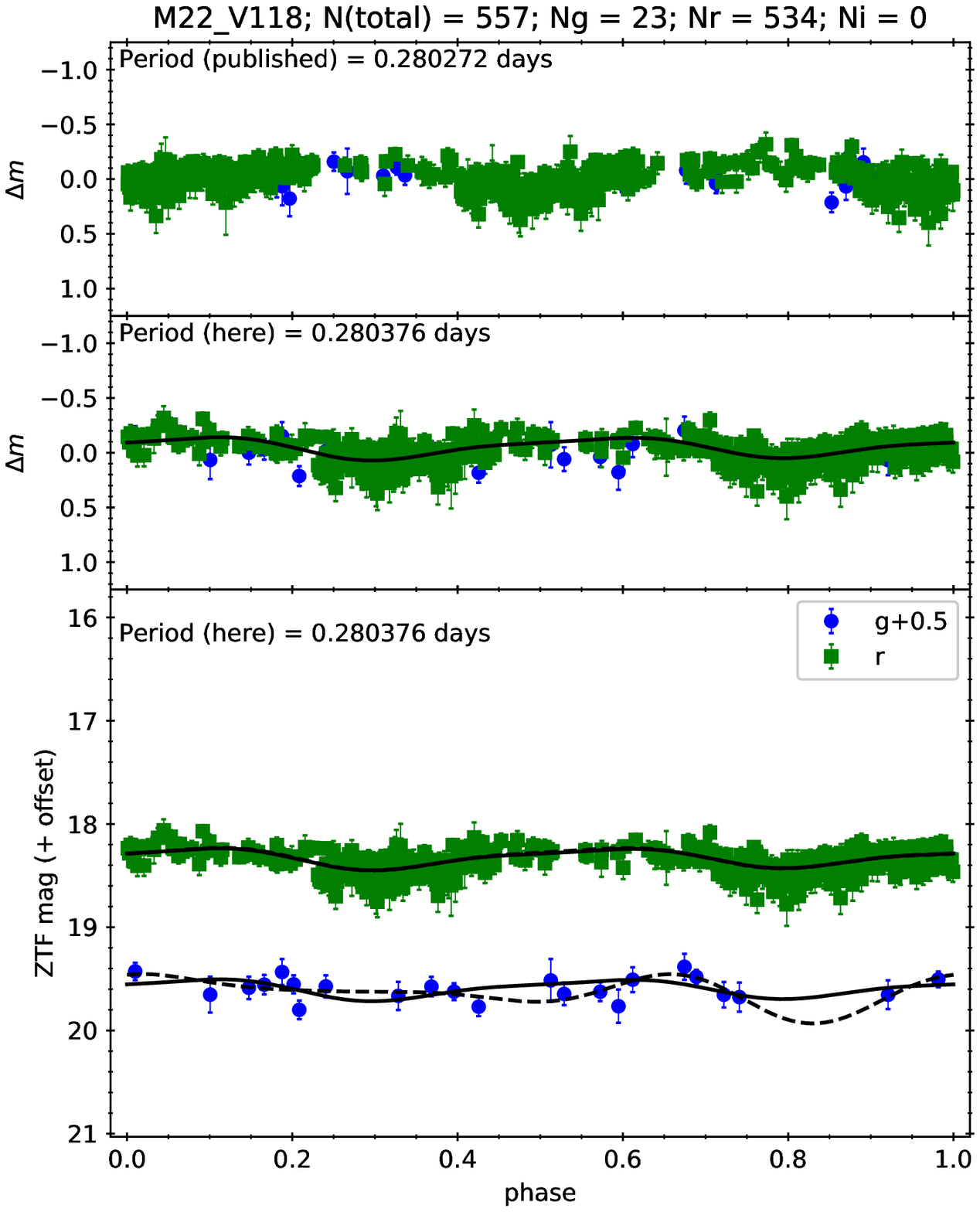} 
  \end{tabular}
  \caption{ZTF light curves folded with published periods (top panels in each sub-figures) and periods found in this work (middle panels in each sub-figures). The solid curves in the middle panels of each sub-figures are the template light curves by fitting a $4^{\mathrm{th}}$-order Fourier expansion to the combined light curves. These template light curves where then shifted vertically to fit the individual $gr(i)$-band light curves displayed in the bottom panels of each sub-figures. For comparison, the dashed curves in the bottom panels are the fitted $4^{\mathrm{th}}$-order Fourier expansion to individual $gr(i)$-band light curves. The blue circles, green squares and red triangles represent the $g$-, $r$- and $i$-band data, respectively. Crosses are the rejected outliers of the light curves.}\label{fig_lcA}
\end{figure*}

In Figure \ref{fig_lcA}, we present the folded ZTF light curves for all of the 30 CB listed in Table \ref{tab_cb}.

\begin{figure*}
  %\epsscale{2}
  \figurenum{16}
  \begin{tabular}{ccc}
    \includegraphics[scale=0.34]{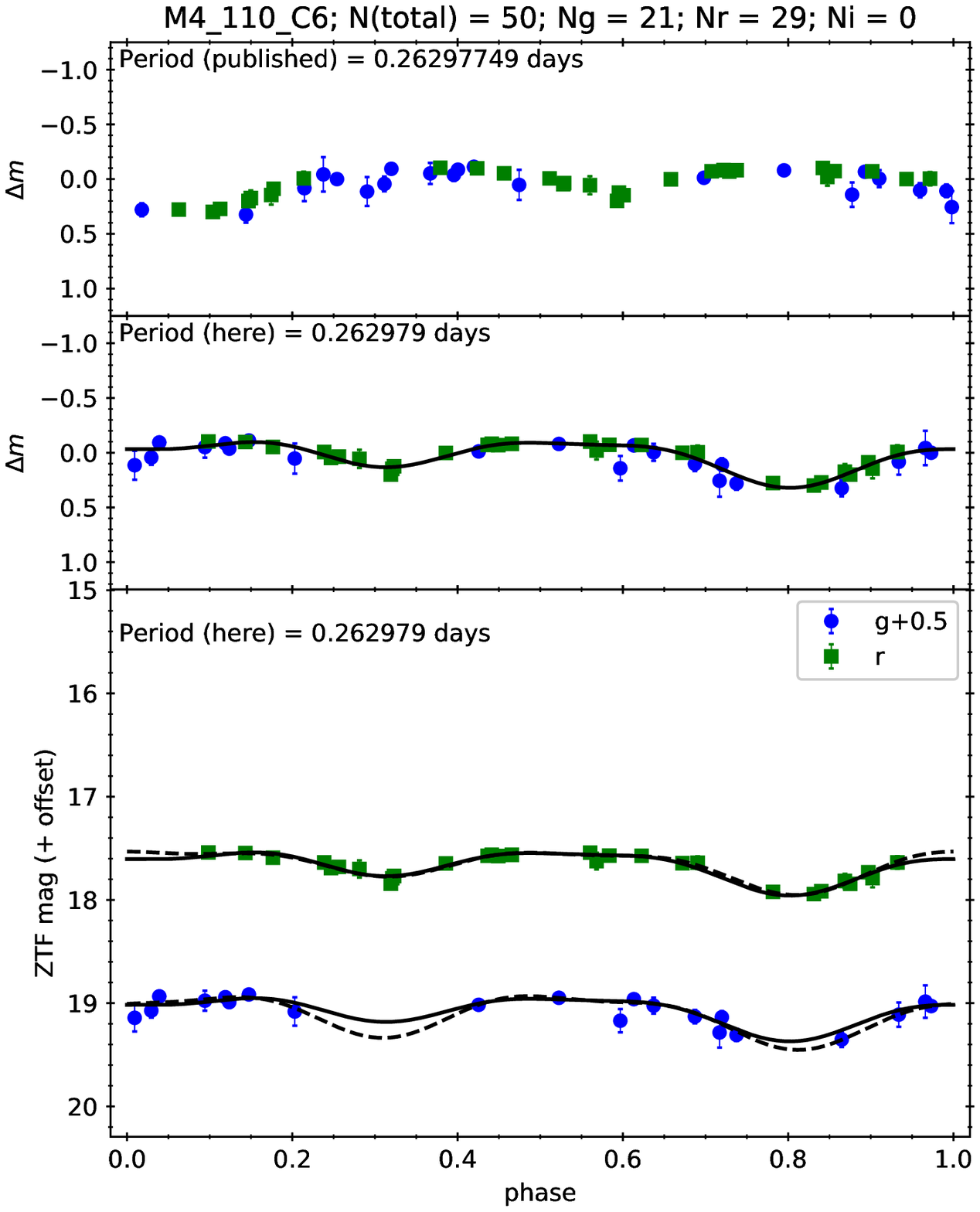} &  \includegraphics[scale=0.34]{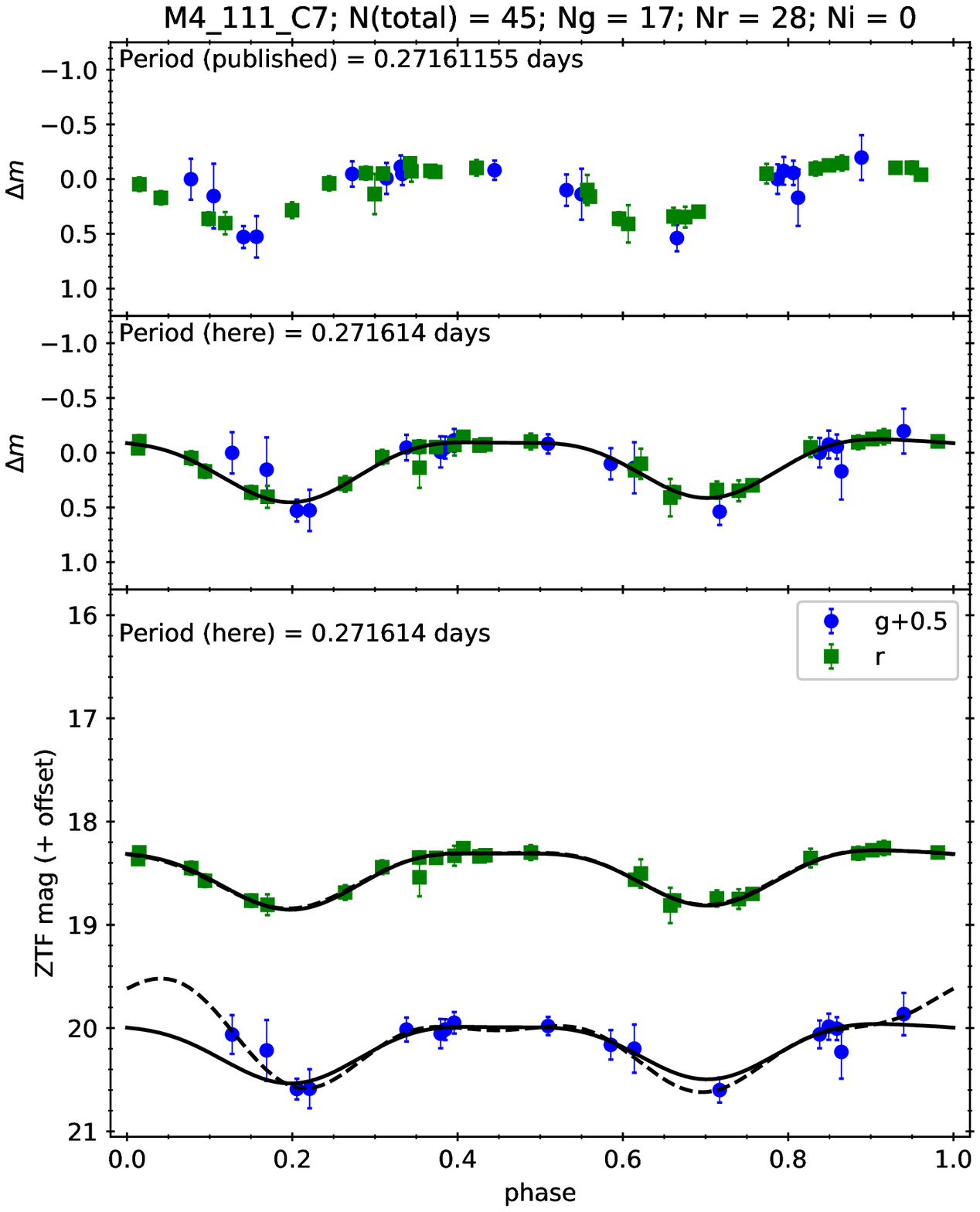} &  \includegraphics[scale=0.34]{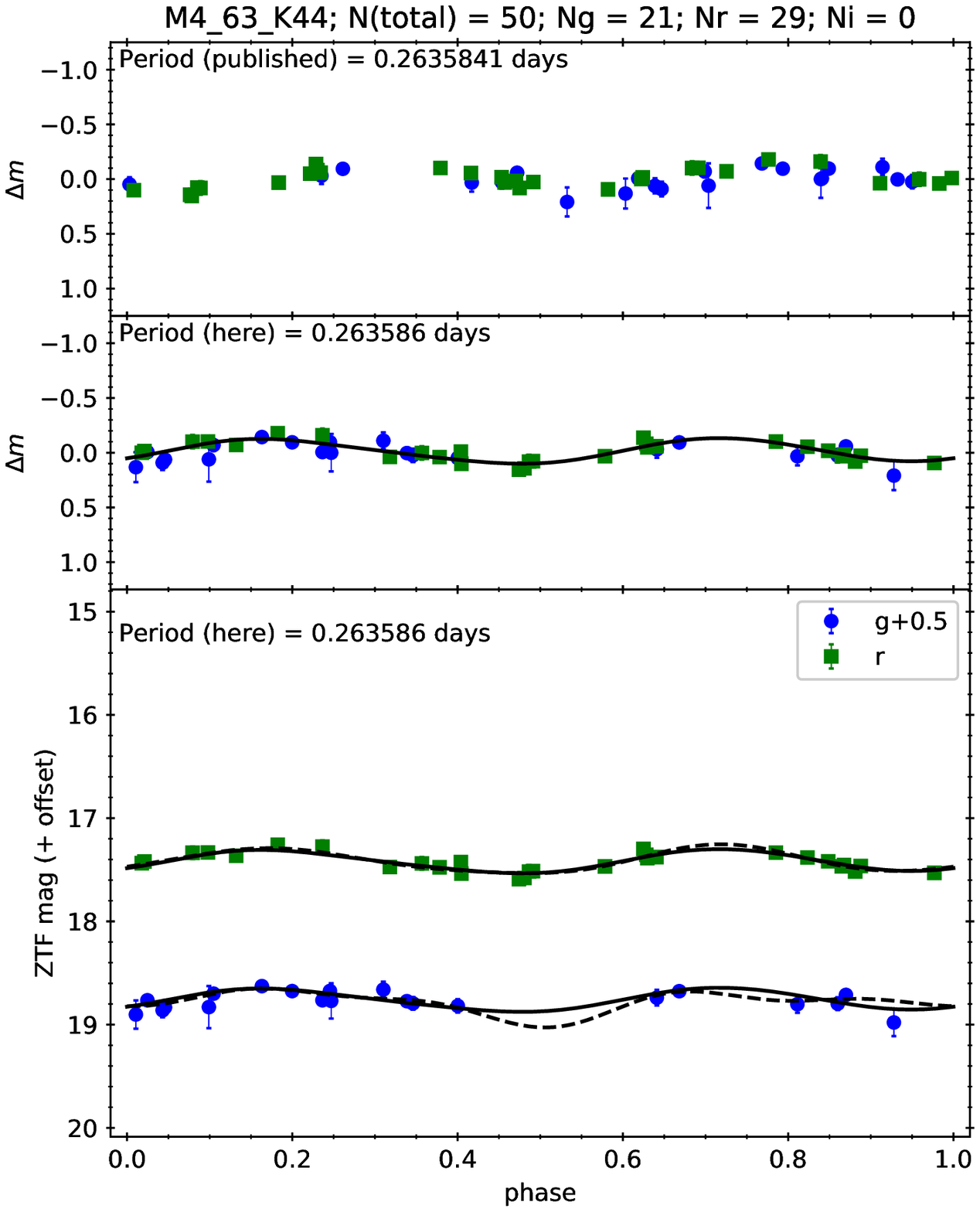} \\
    \includegraphics[scale=0.34]{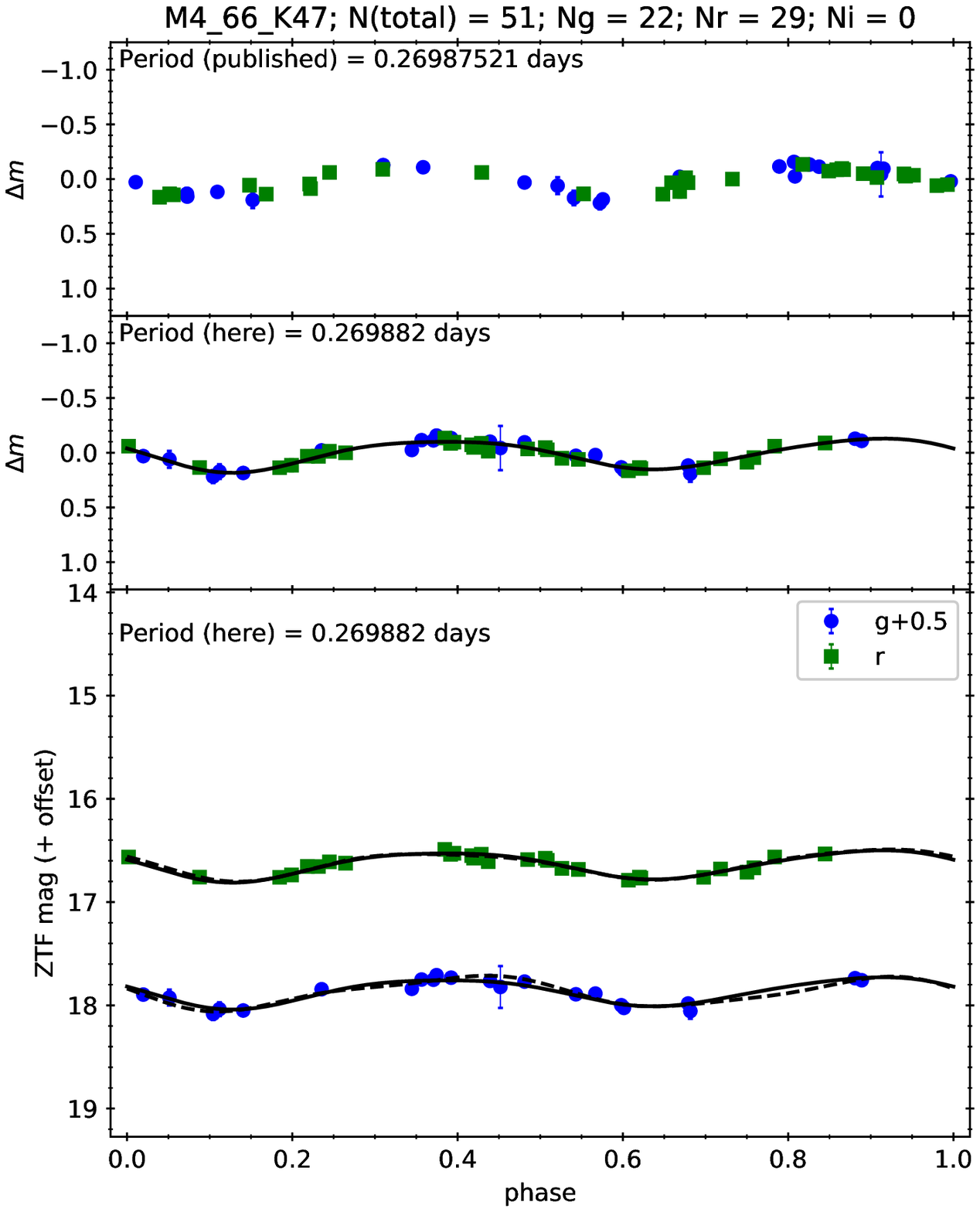} &  \includegraphics[scale=0.34]{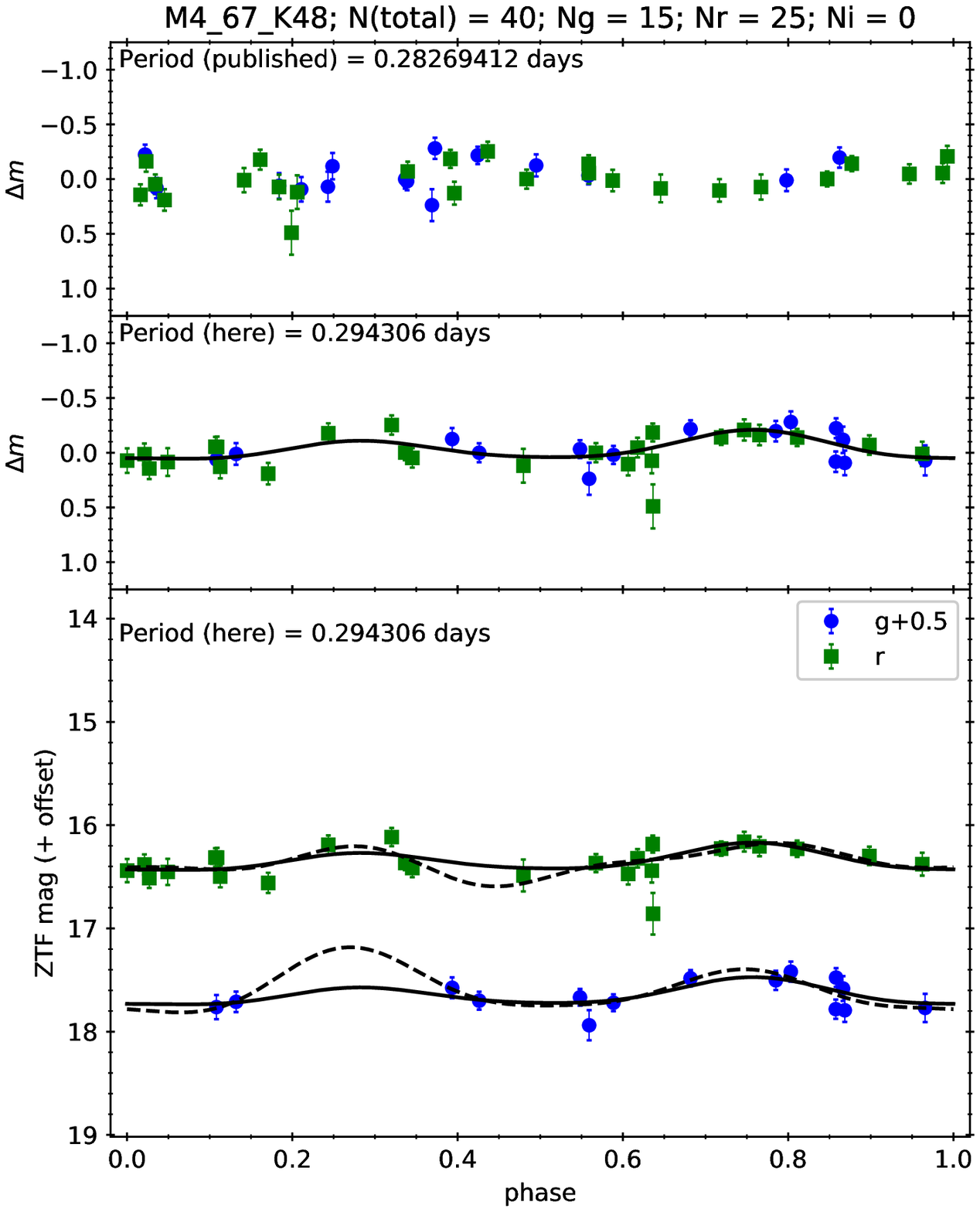} &  \includegraphics[scale=0.34]{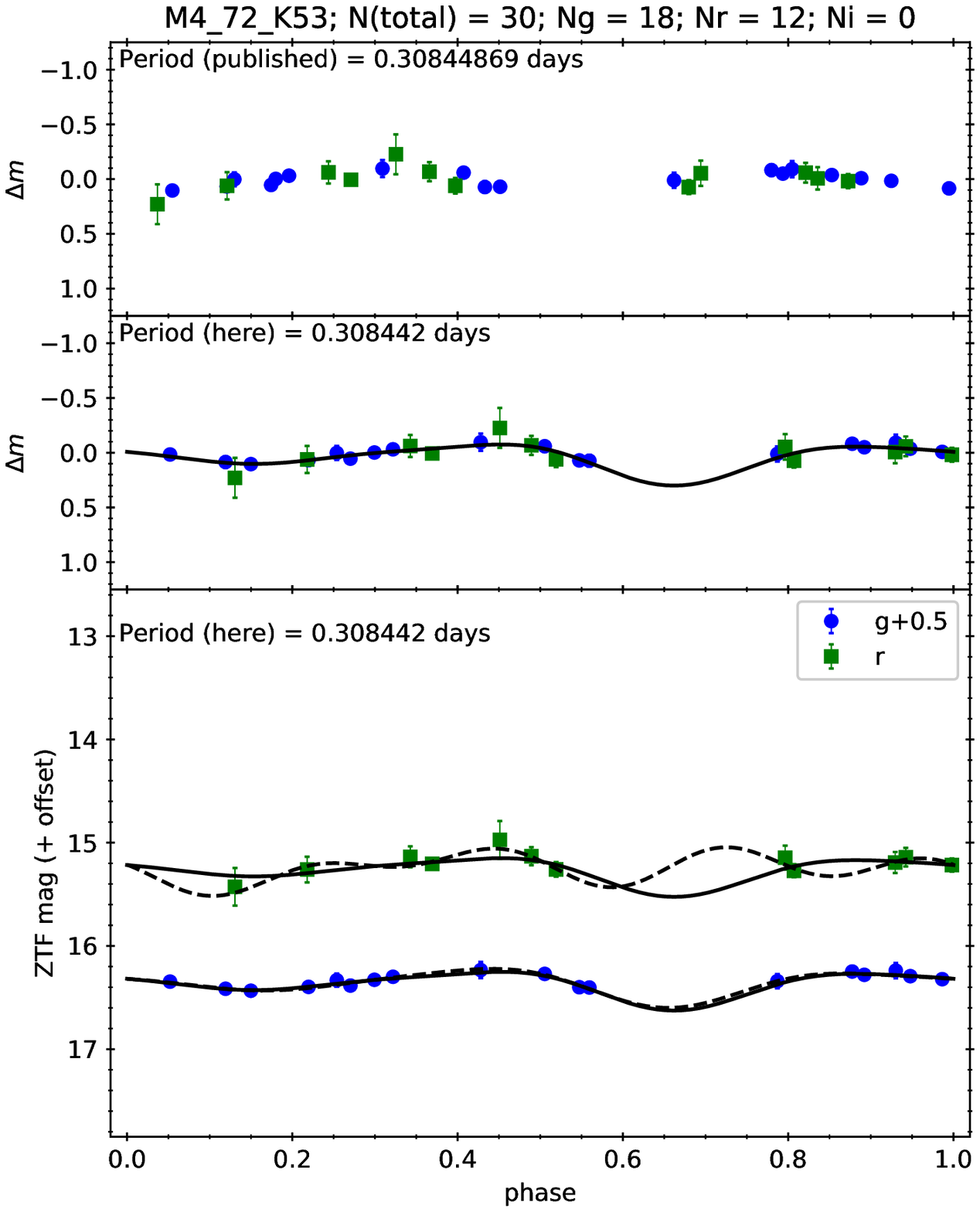} \\
    \includegraphics[scale=0.34]{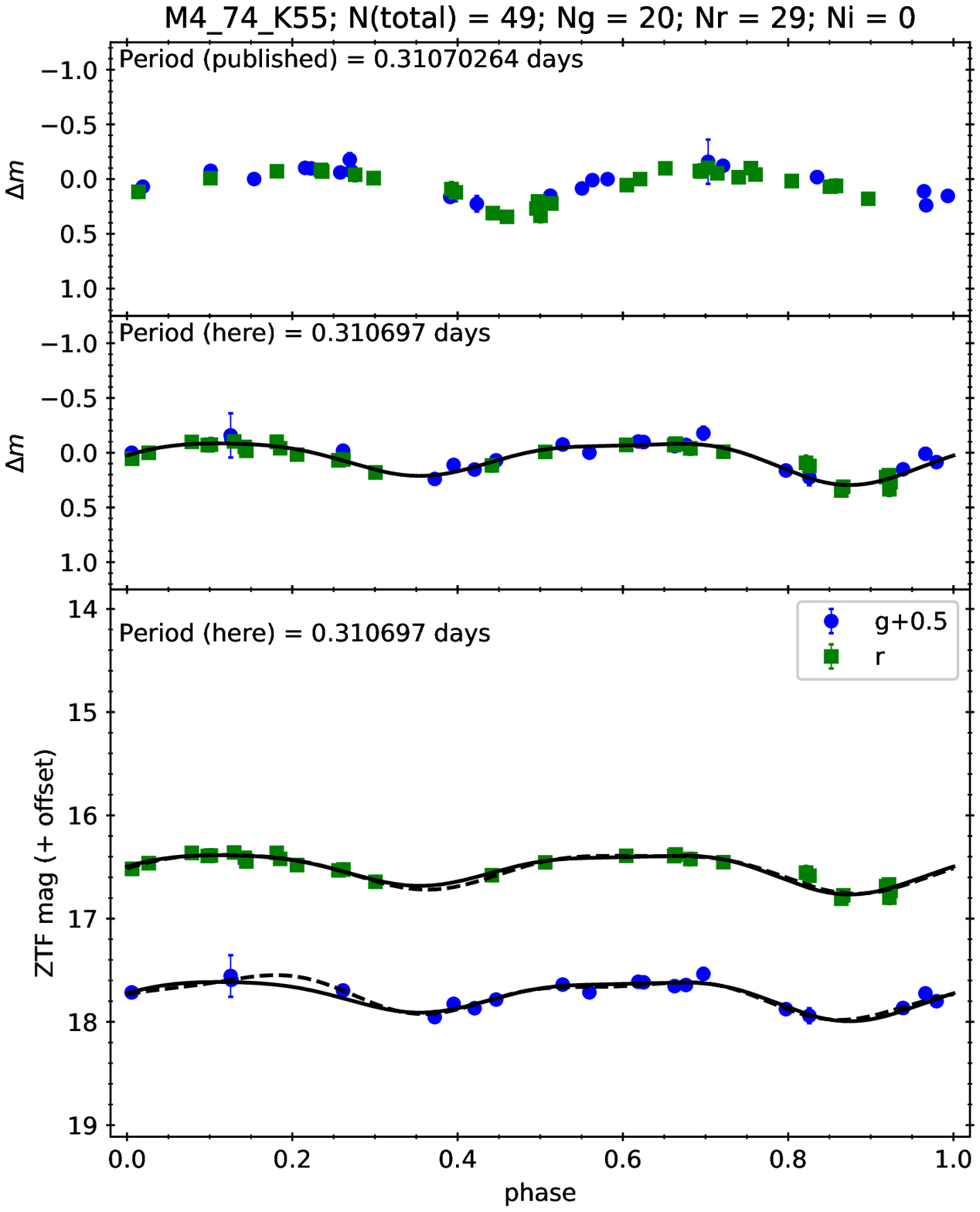} &  \includegraphics[scale=0.34]{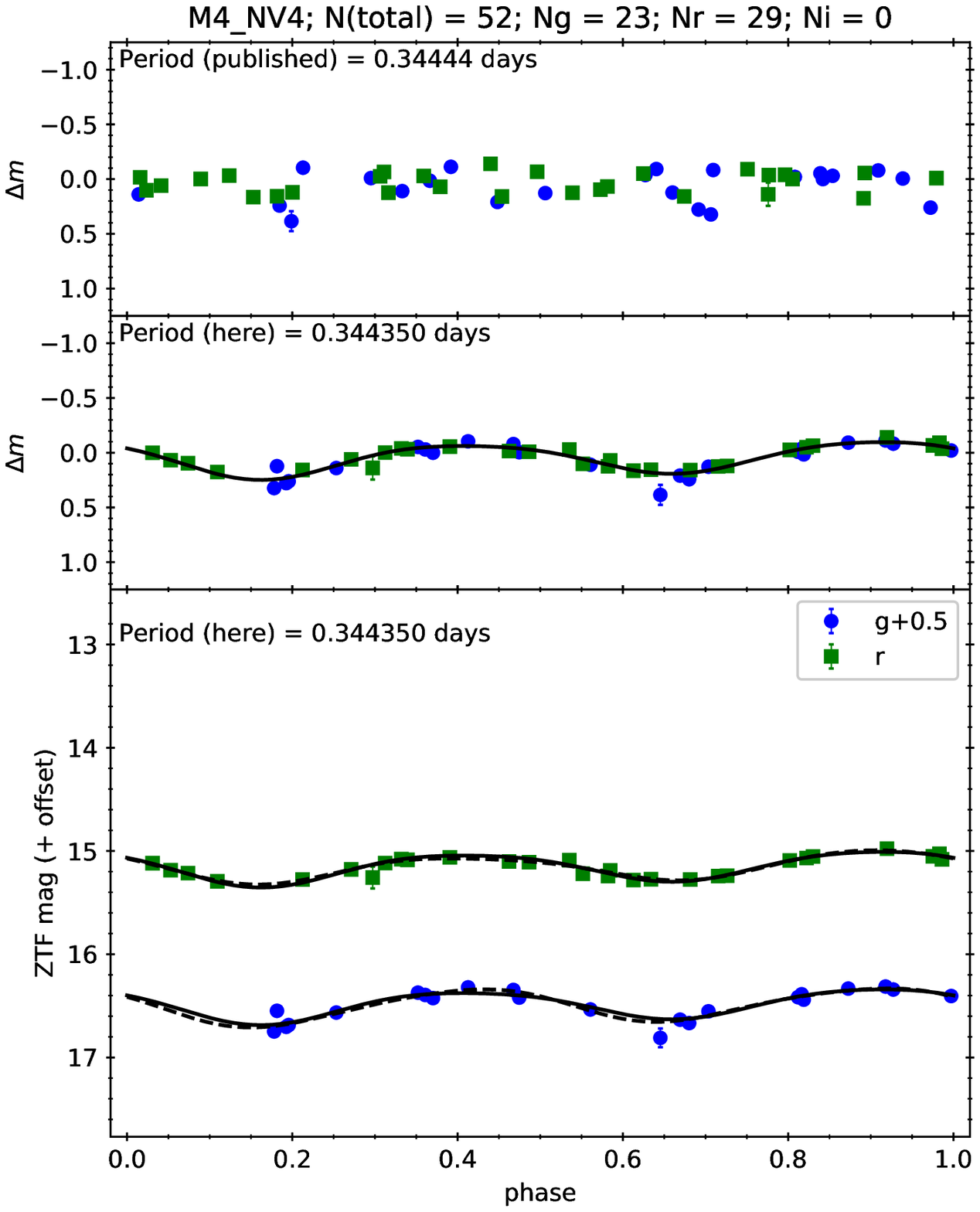} &  \includegraphics[scale=0.34]{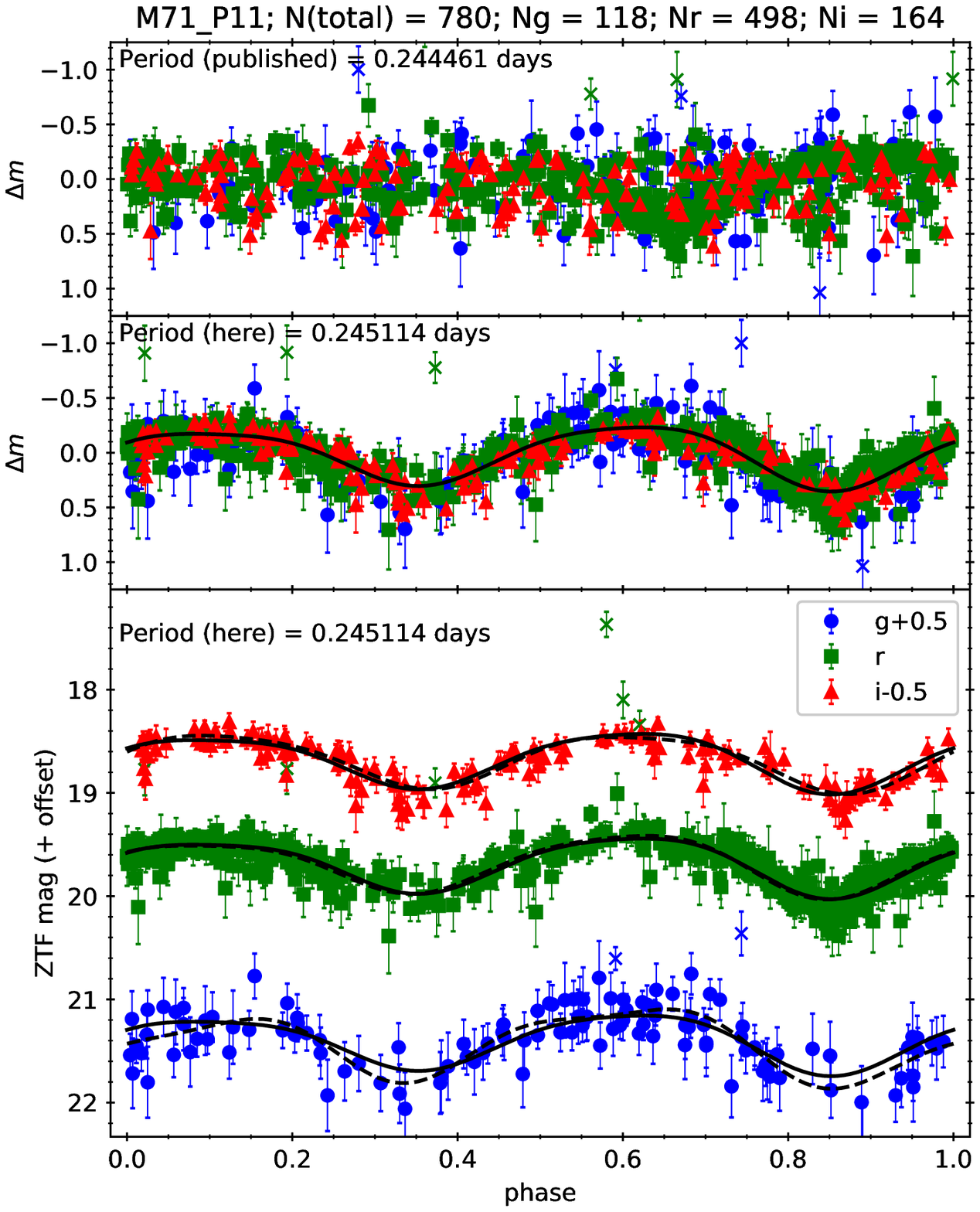} 
  \end{tabular}
  \caption{{\it Continue}.}
\end{figure*}

\begin{figure*}
  %\epsscale{2}
  \figurenum{16}
  \begin{tabular}{ccc}
    \includegraphics[scale=0.34]{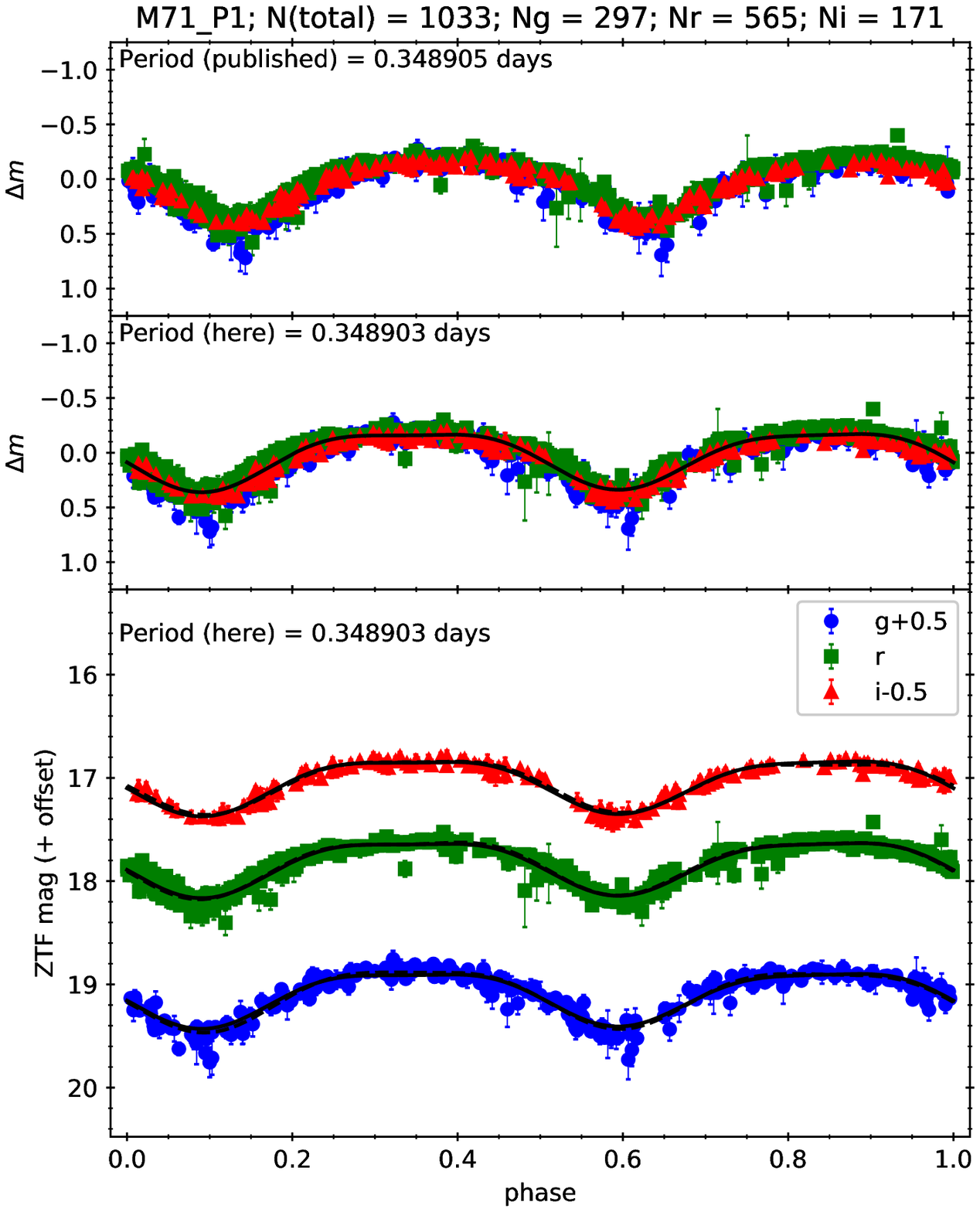} &  \includegraphics[scale=0.34]{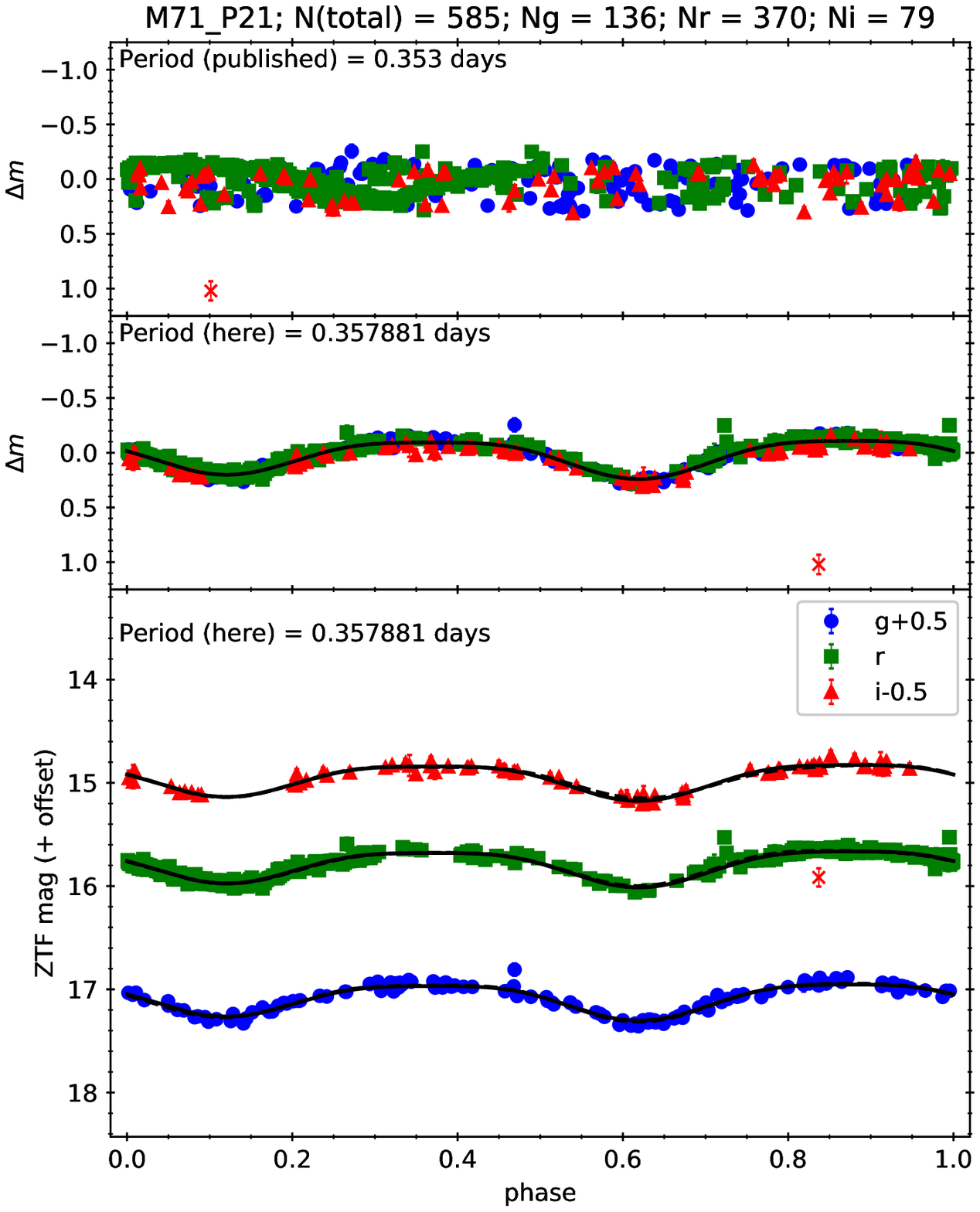} &  \includegraphics[scale=0.34]{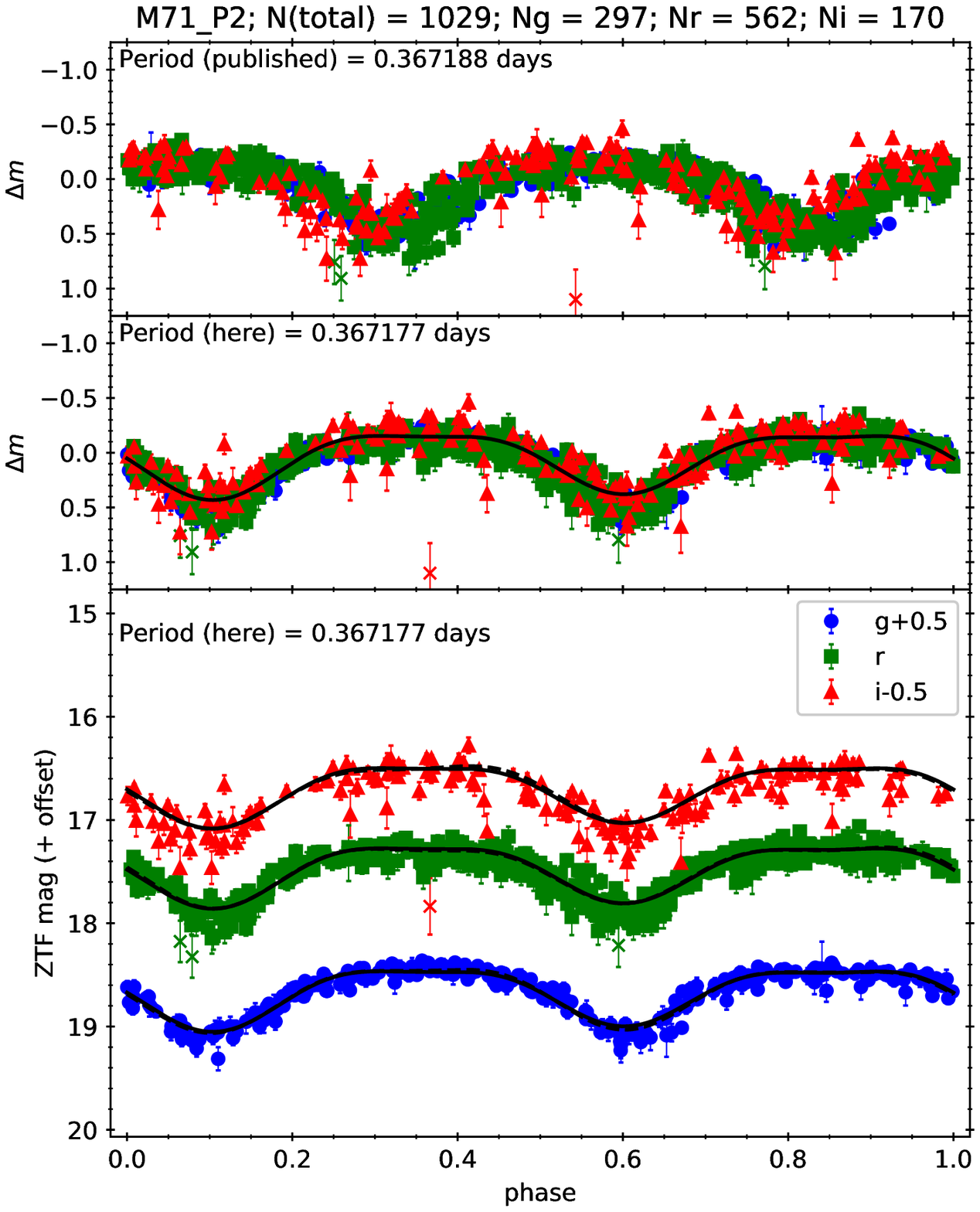} \\
    \includegraphics[scale=0.34]{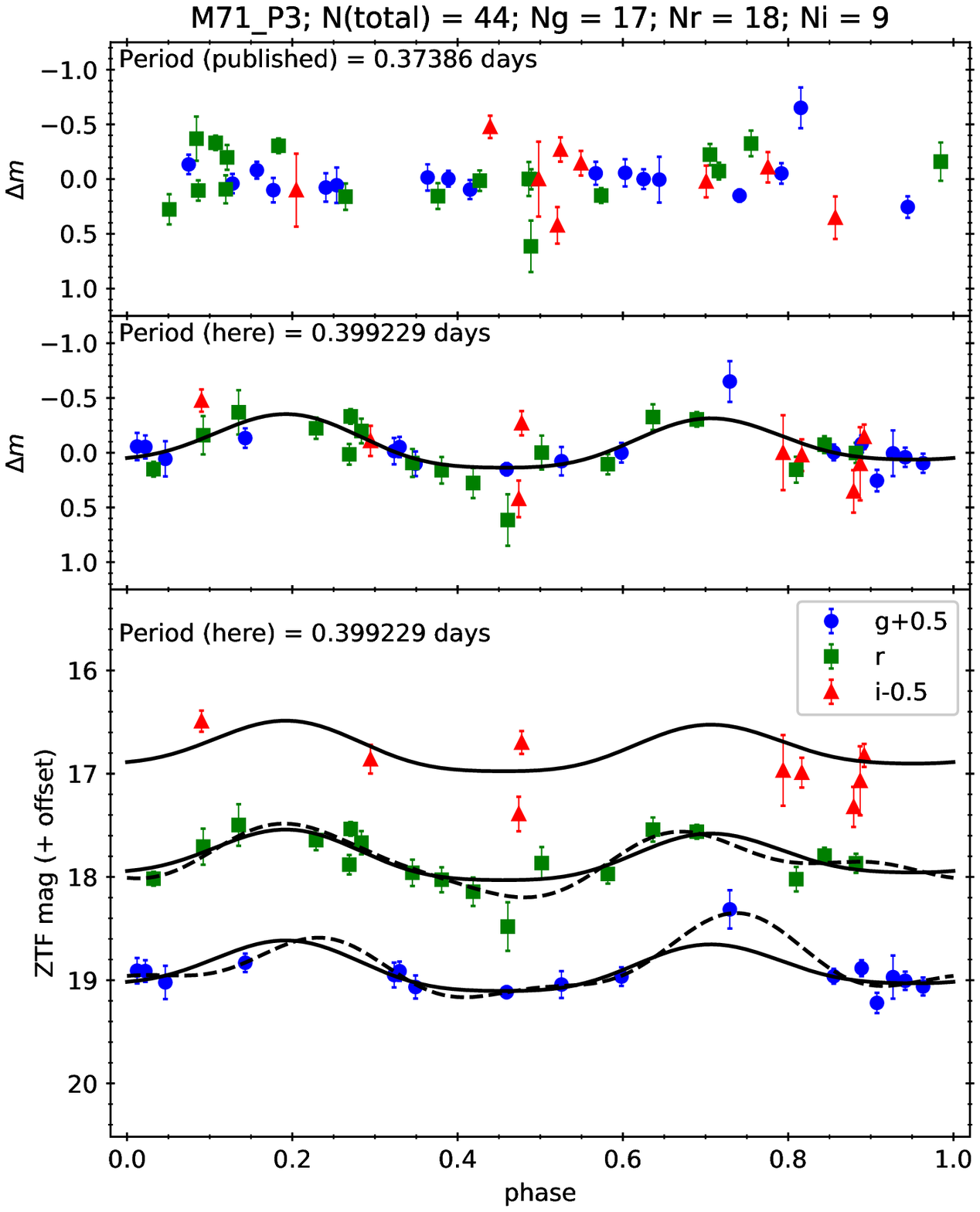} &  \includegraphics[scale=0.34]{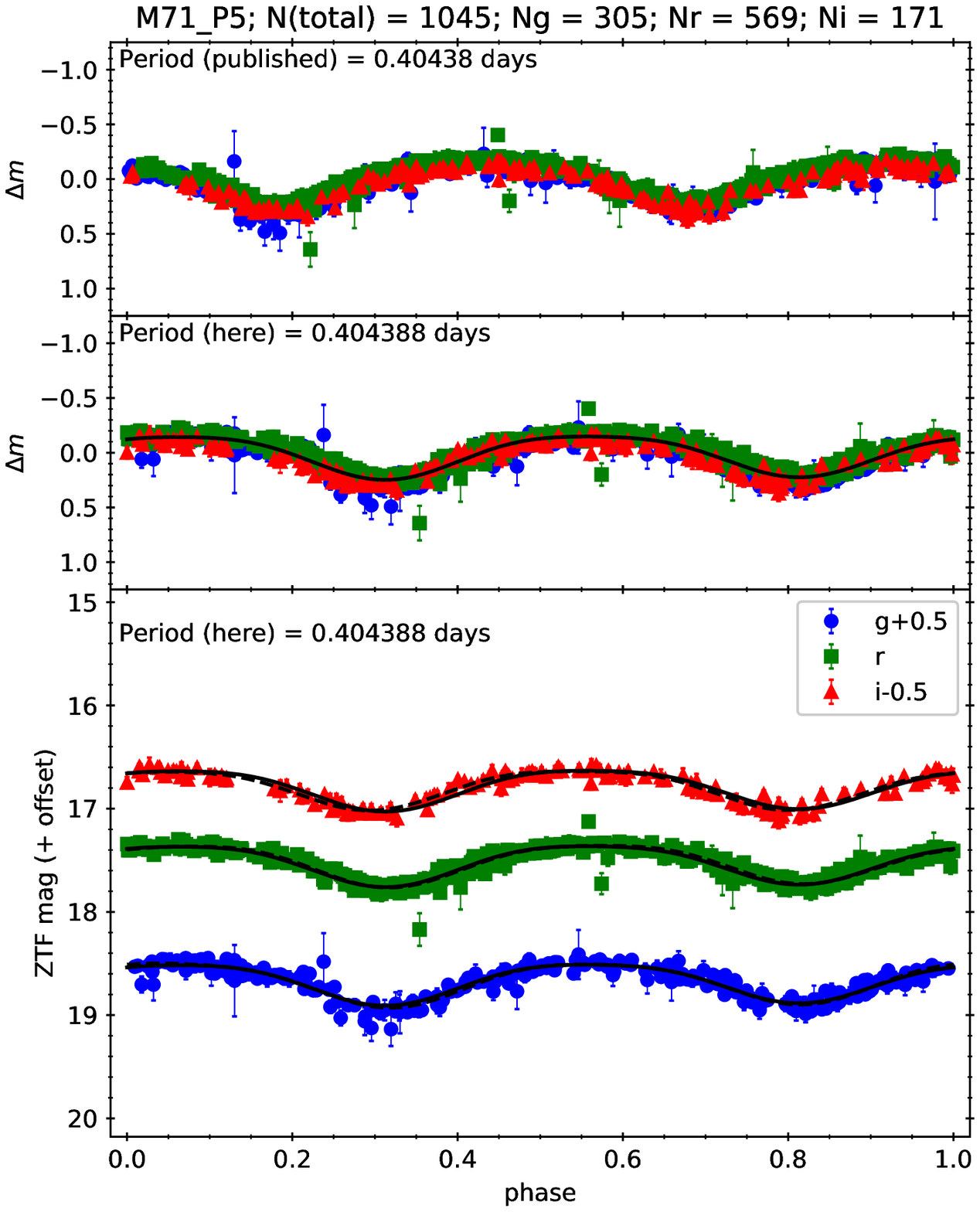} &  \includegraphics[scale=0.34]{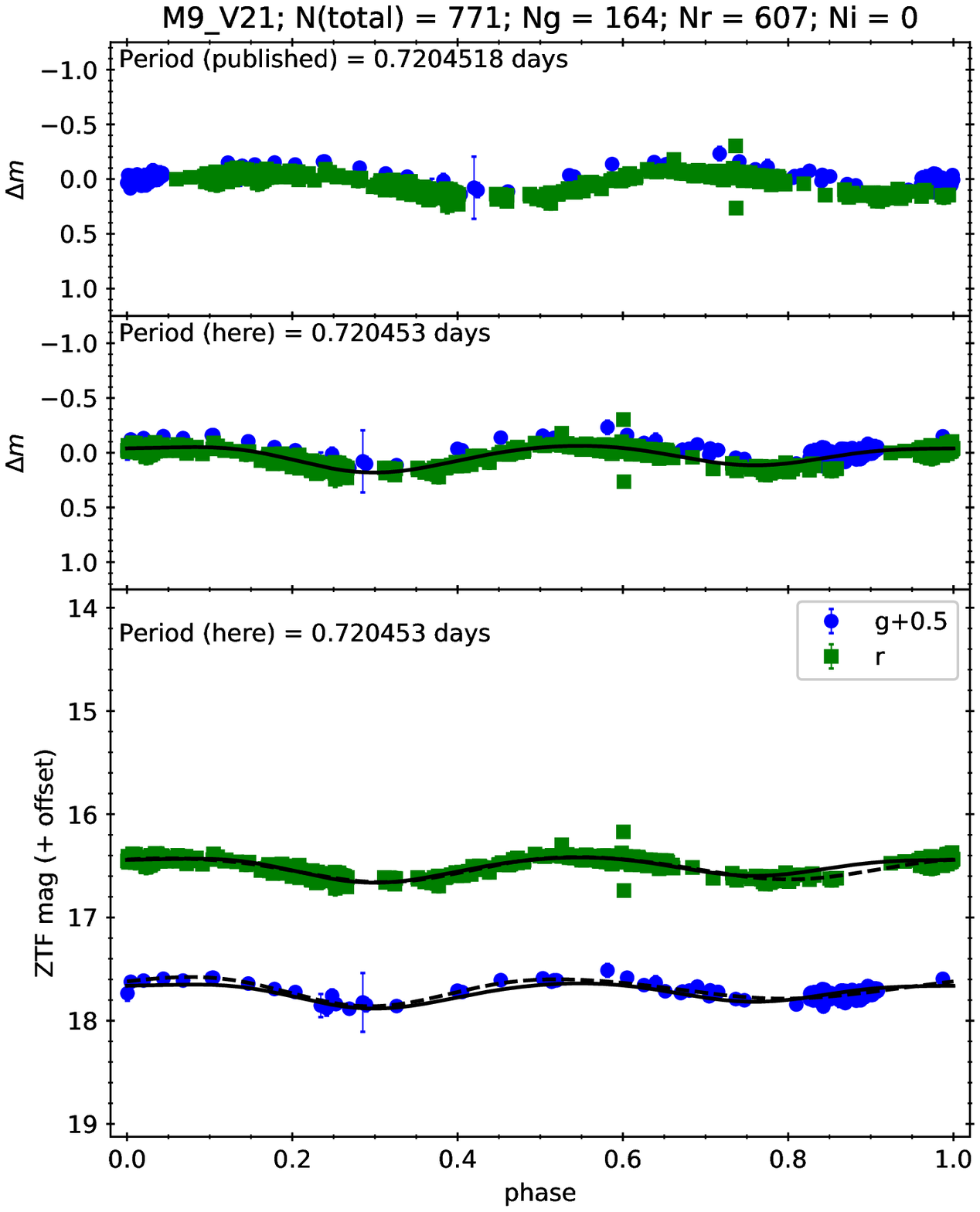} \\
    \includegraphics[scale=0.34]{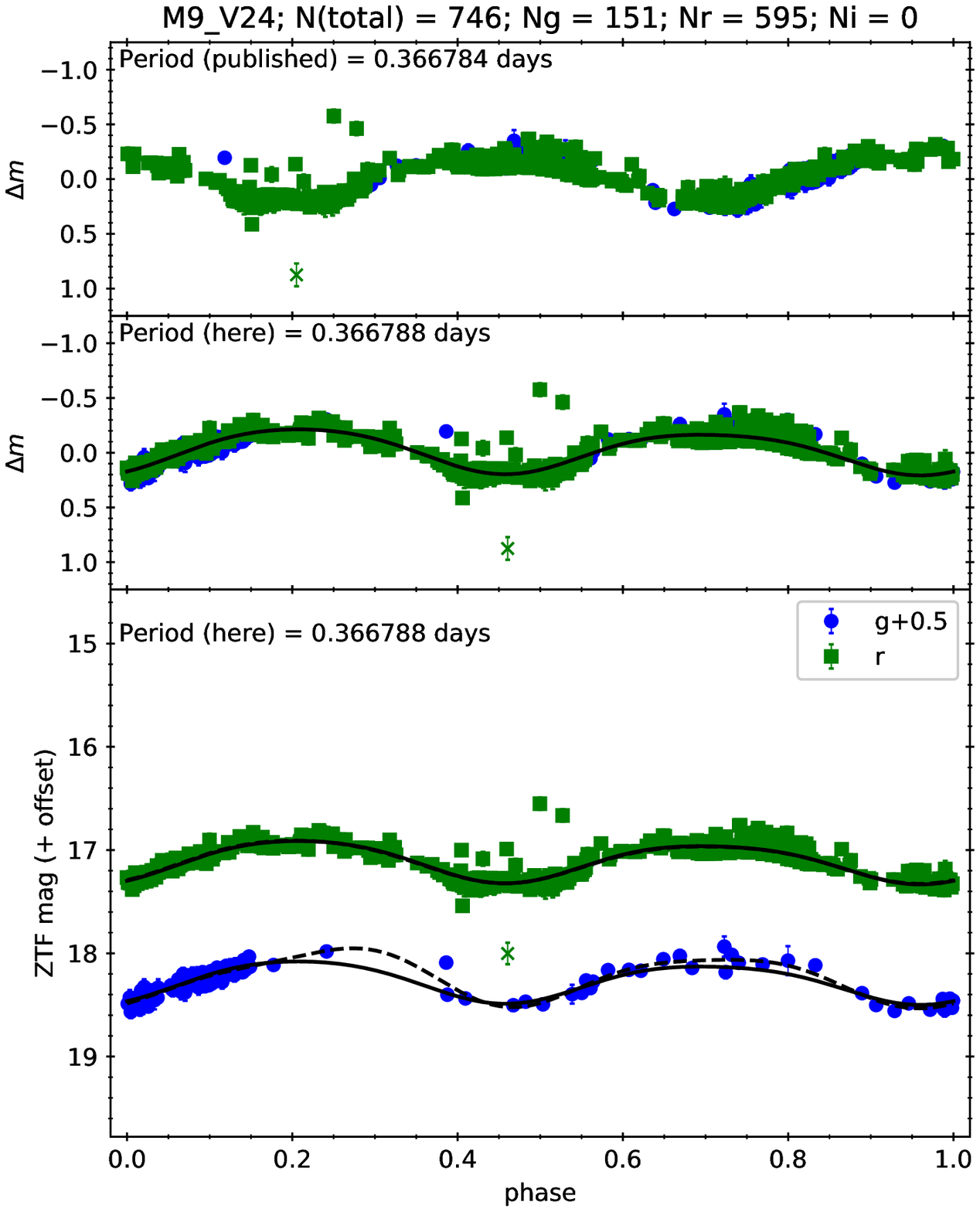} &  \includegraphics[scale=0.34]{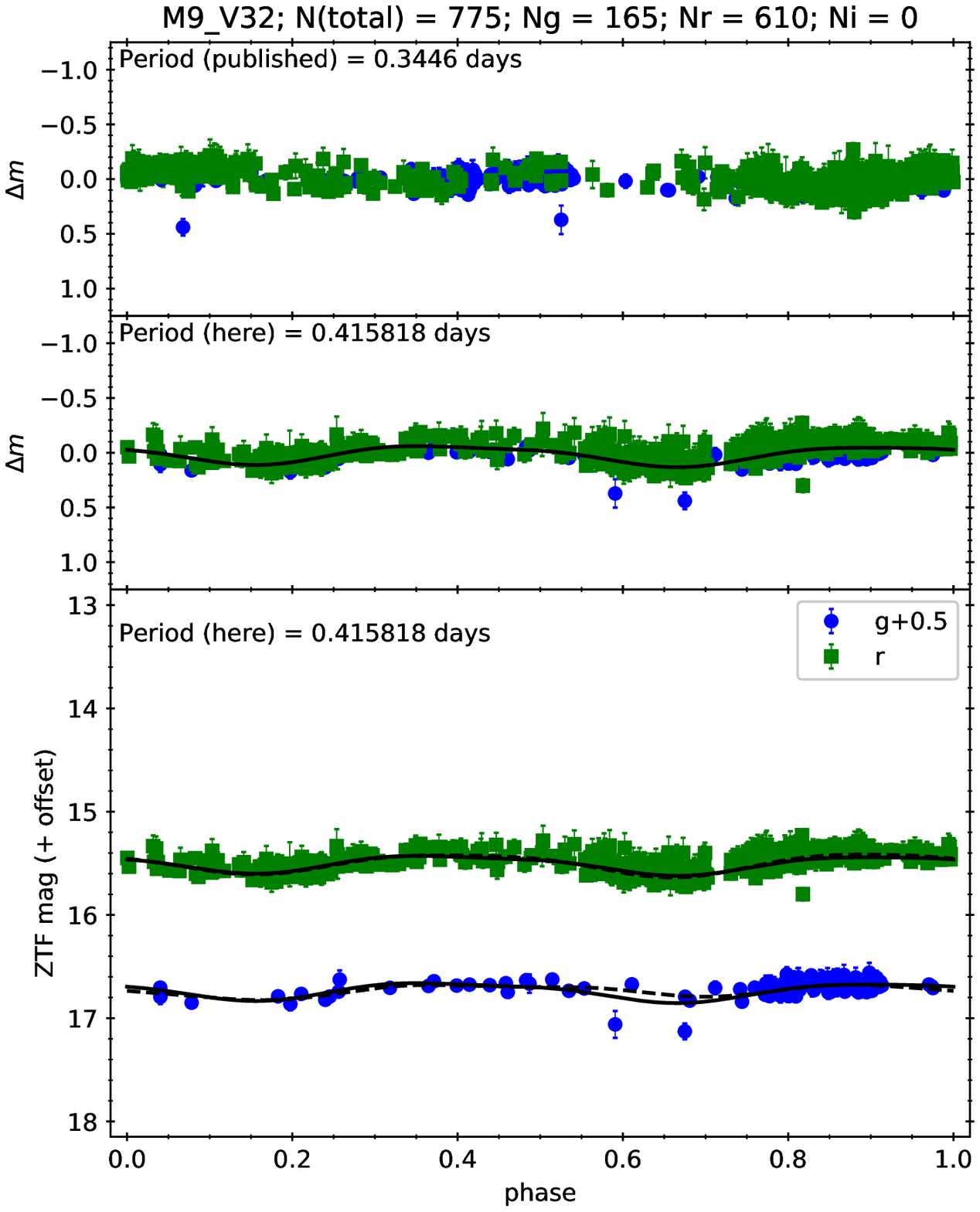} &  \includegraphics[scale=0.34]{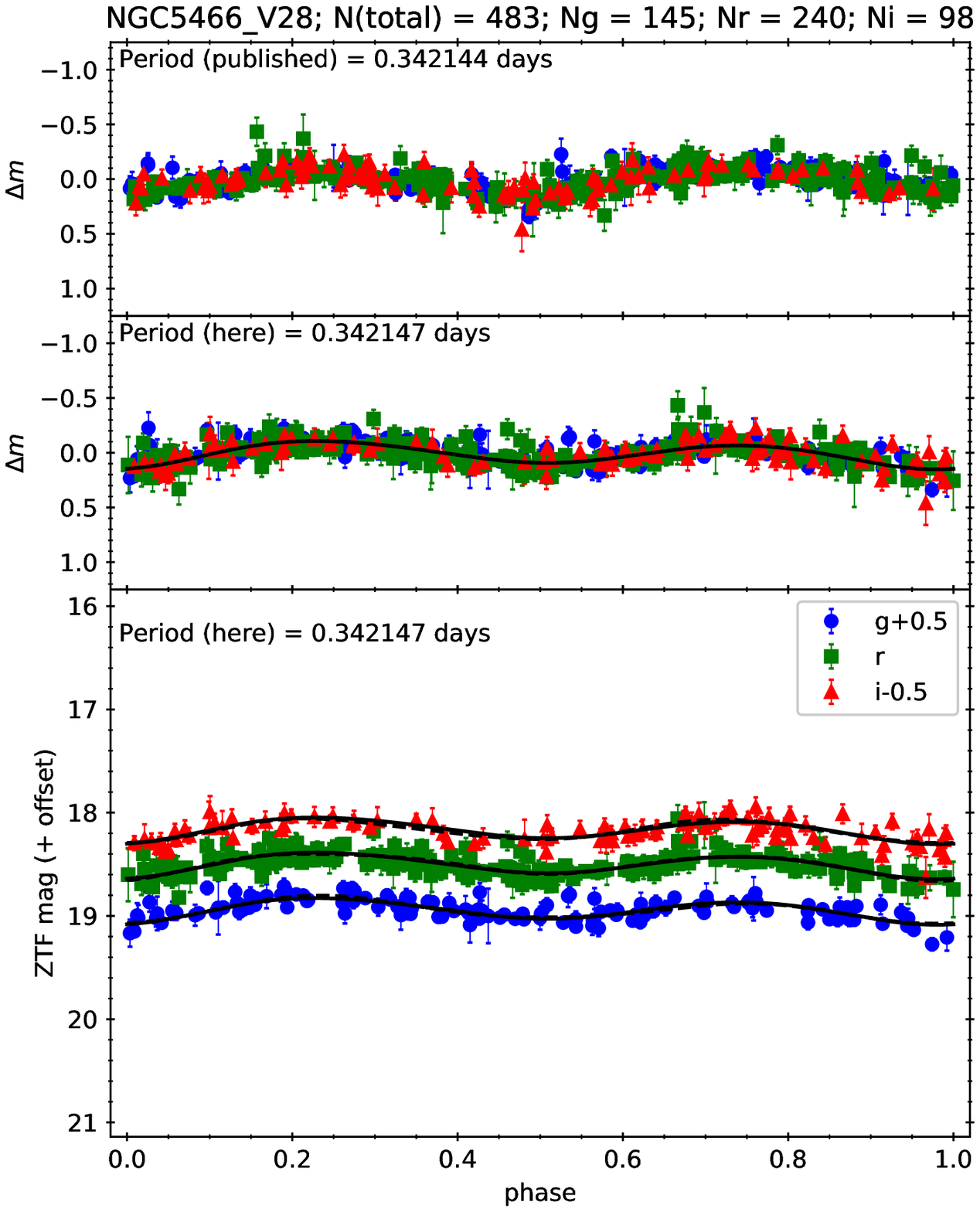} 
  \end{tabular}
  \caption{{\it Continue}.}
\end{figure*}

\begin{figure*}
  %\epsscale{2}
  \figurenum{16}
  \begin{tabular}{ccc}
    \includegraphics[scale=0.34]{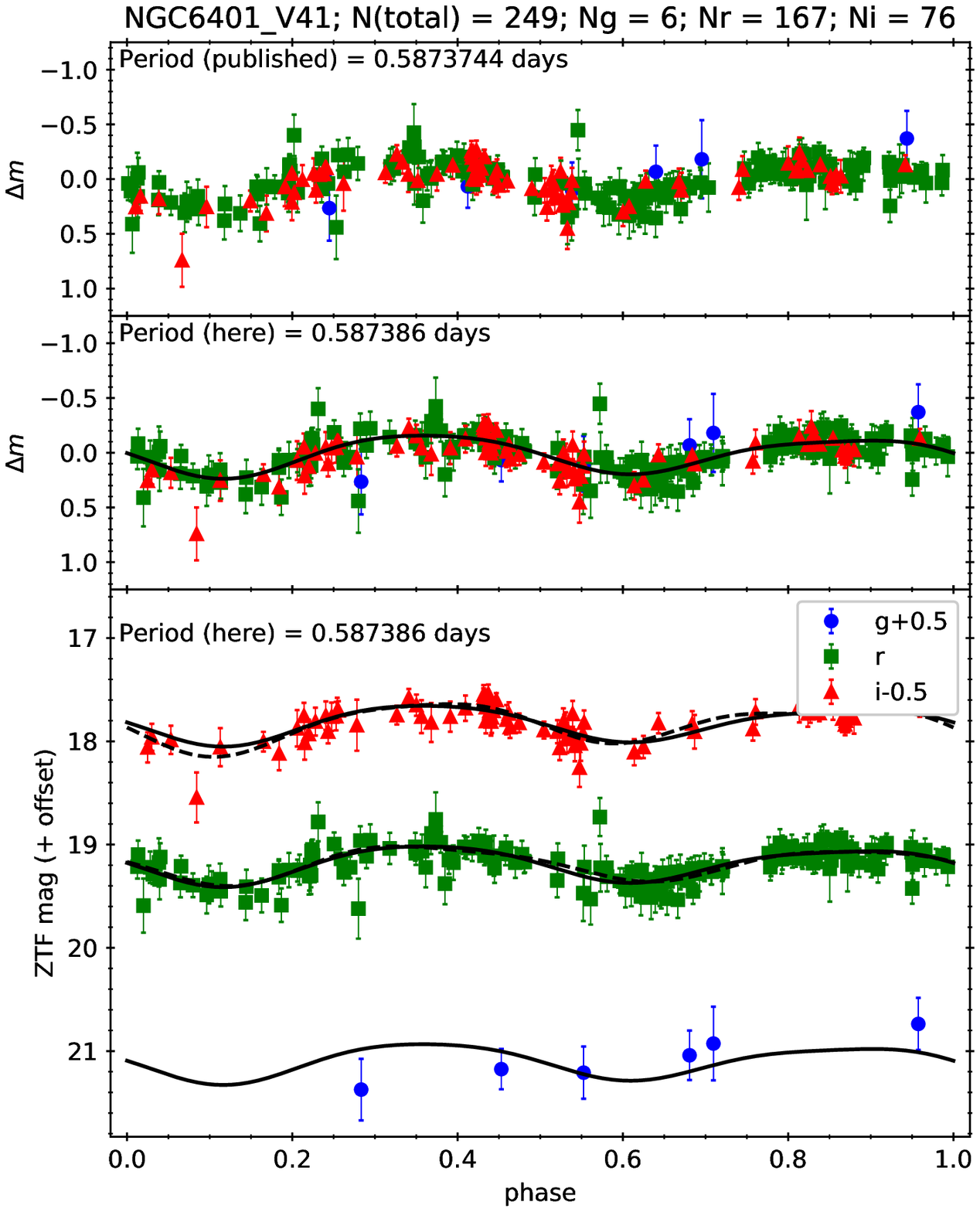} &  \includegraphics[scale=0.34]{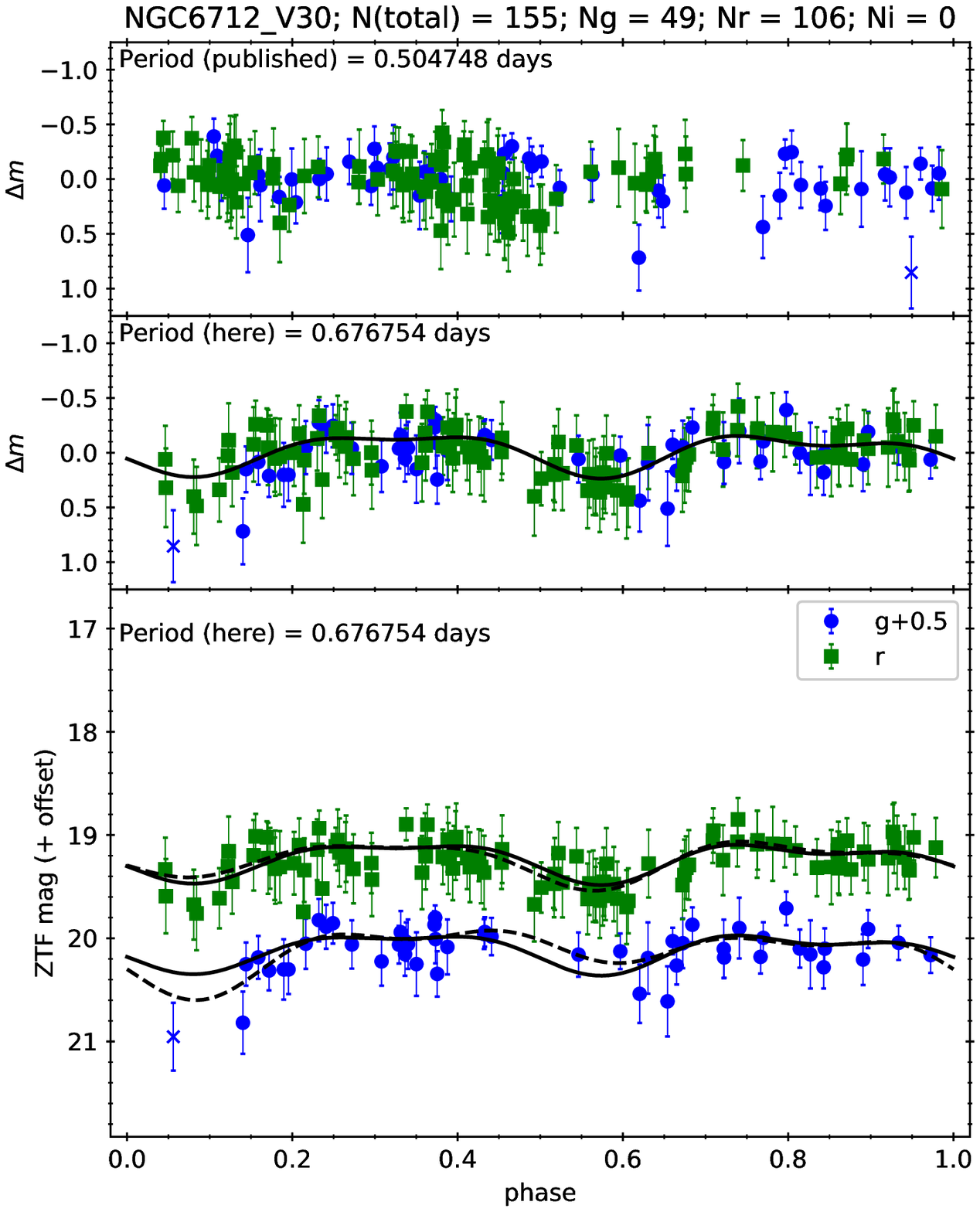} &  \includegraphics[scale=0.34]{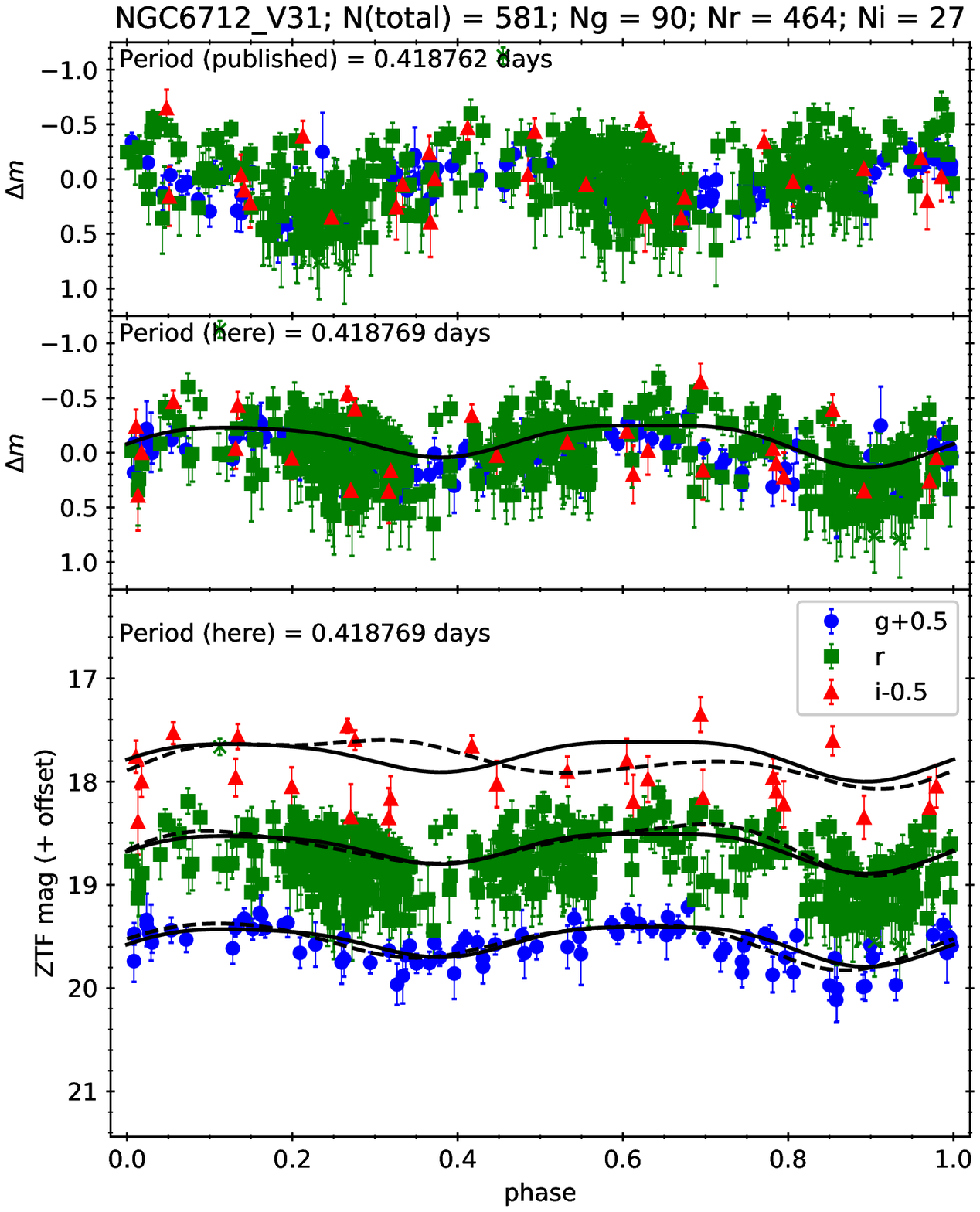} 
  \end{tabular}
  \caption{{\it Continue}.}
\end{figure*}

% ===============================================
%               REFERENCE
% ===============================================

\end{document}